\documentclass[12pt,preprint]{aastex}
\begin{document}
\title{Jet Properties of Compact Steep-Spectrum Sources and an Eddington-Ratio-Driven Unification Scheme of Jet Radiation in Active Galactic Nuclei}

\author{Jin Zhang\altaffilmark{1}, Hai-Ming Zhang\altaffilmark{2}, Ying-Ying Gan\altaffilmark{3}, Ting-Feng Yi\altaffilmark{4}, Jun-Feng Wang\altaffilmark{5}, En-Wei Liang\altaffilmark{3}}
\altaffiltext{1}{Key Laboratory of Space Astronomy and Technology, National Astronomical Observatories, Chinese Academy of Sciences, Beijing 100012, People's Republic of China; jinzhang@bao.ac.cn}
\altaffiltext{2}{School of Astronomy and Space Science, Nanjing University, Nanjing 210023, China}
\altaffiltext{3}{Guangxi Key Laboratory for Relativistic Astrophysics, School of Physical Science and Technology, Guangxi University, Nanning 530004, People's Republic of China; lew@gxu.edu.cn}
\altaffiltext{4}{Department of Physics, Yunnan Normal University, Kunming 650500, People's Republic of China}
\altaffiltext{5}{Department of Astronomy, Xiamen University, Xiamen, Fujian 361005, People's Republic of China; jfwang@xmu.edu.cn}
\begin{abstract}

Compact steep-spectrum sources (CSSs) likely represent a population of young radio-loud active galactic nuclei (AGNs) and have been identified as $\gamma$-ray emitting sources. We present a comprehensive analysis of their $\gamma$-ray emission observed with {\em Fermi}/LAT and establish their broadband spectral energy distributions (SEDs). We derive their jet properties by the SED fits with a two-zone leptonic model for radiations from the compact core and large-scale extended region, and explore the possible signature of a unification picture of jet radiation among subclasses of AGNs. We show that the observed $\gamma$-rays of CSSs with significant variability are contributed by the radiation of their compact cores via the inverse Compton process of the torus photons. The derived power-law distribution index of the radiating electrons is $p_1\sim1.5-1.8$, magnetic field strength is $B\sim0.15-0.6$ G, and Doppler boosting factor is $\delta\sim2.8-8.9$. Assuming that the jet is composed of $e^{\pm}$ pairs, the compact cores of CSSs are magnetized and have a high radiation efficiency, similar to that of flat spectrum radio quasars. The $\gamma$-ray emitting CSSs on average have higher Eddington ratio and black hole mass than those non-GeV-detected CSSs, and they follow the correlation between the jet power in units of Eddington luminosity ($P^{e^{\pm}}_{\rm jet}/L_{\rm Edd}$) and Eddington ratio ($R_{\rm Edd}$) with other sub-classes of AGNs, $P^{e^{\pm}}_{\rm jet}/L_{\rm Edd}\propto R^{0.52\pm 0.03}_{\rm Edd}$, indicating that $R_{\rm Edd}$ would be a key physical driver for the unification scheme of AGN jet radiation.

\end{abstract}


\section{Introduction}           
\label{sect:intro}

The compact steep-spectrum sources (CSSs) as a subclass of active galactic nuclei (AGNs) are characterized by a compact size (angular size $\leq1\arcsec-2\arcsec$) and a steep high frequency spectrum at the radio band ($\alpha\geq0.5$, $F_{\nu}\propto\nu^{-\alpha}$; Fanti et al. 1990). The radio morphology of CSSs is typically characterized by fully developed radio lobes and by a small linear size ($<15$ kpc), and CSSs make up a significant fraction ($\sim30\%$) of sources at 5 GHz (O'Dea 1998 for a review). The other radio properties, for example, low degree of polarization, complex morphology, and high surface brightness were reported by van Breugel et al. (1984). Fanti et al. (1990) suggested that most of CSSs ($\geq70\%$) are likely to be intrinsically small other than a projection effect, and represent an early stage of radio source evolution. They estimated the age of CSSs to be $\leq5\times10^{6}$ years from their occurrence rate, which is consistent with the dynamical timescale. The lack of the synchrotron break frequency in the spectrum of lobe component at mm-wavelength also supports that CSSs are young sources with age of $\leq10^{5}$ years (Kameno et al. 1995). However, some CSSs exhibit distorted morphology, which is suggested to be attributed to the interaction with dense clouds in their environment (e.g., van Breugel et al. 1984; Wilkinson et al. 1984, 1991; Saikia et al. 1995). The core radiation of CSSs contributes a small fraction of total flux ($\leq36\%$) in the radio band and a large fraction of the radio emission is dominated by the extended lobes (Kameno et al. 1995). With the very long baseline interferometry (VLBI) observation results for 18 CSSs at 22 and 43 GHz, a correlation between variability at 22 GHz and spectral index at mm-wavelengths is observed, which can be explained by a two-component model with a flat-spectrum core and steep-spectrum lobes (Kameno et al. 1995).

It is generally thought that CSSs may eventually evolve into radio sources with large-scale jets, i.e., Fanaroff-Riley (FR) I and FR II radio galaxies (RGs) (O'Dea 1998; Polatidis \& Conway 2003; Randall et al. 2011). According to the unification model for radio-loud (RL) AGNs, FR I and FR II RGs are the parent populations of blazars with large viewing angles and small Doppler factors (Urry \& Padovani 1995). Connection between CSSs and narrow-line Seyfert 1 galaxies (NLS1s) was reported (Oshlack et al. 2001; Komossa et al. 2006; Yuan et al. 2008; Caccianiga et al. 2014; Gu et al. 2015). They may share the same radio properties (Komossa et al. 2006; Caccianiga et al. 2014). Furthermore, Wu (2009) reported that the central black hole (BH) masses and Eddington ratios of CSSs are similar to these of NLS1s. CSSs may be the parent population of RL-NLS1s (Caccianiga et al. 2014; Berton et al. 2016; Liao \& Gu et al. 2020).

The detection of $\gamma$-ray emission in RL-NLS1s by \emph{Fermi}/LAT is convincing evidence for the existence of relativistic jets in this class of AGNs (Abdo et al. 2009; D'Ammando et al. 2012; Yao et al. 2015, 2019; Paliya et al. 2018; Yang et al. 2018; Paliya 2019). Evidence for interactions between kpc-scale jets and ambient gas in CSSs is revealed (O'Dea 1998 for a review), and superluminal motion is also observed in a few CSSs (Cotton et al. 1997 for 3C 138; Taylor et al. 1995 for 3C 216; Gawro\'{n}ski \& Kus 2006 for 3C 309.1), indicating that the beaming effect may impact on some orientation-dependent properties of CSSs (Saikia et al. 1995). High energy radiations of young AGNs have been  theoretically explored based on the relativistic jet models. It was suggested that GeV $\gamma$-ray emission may be produced from their cocoons/radio lobes via inverse Compton (IC) processes (Stawarz et al. 2008; see also Migliori et al. 2014) or by thermal bremsstrahlung radiation mechanism (Kino et al. 2007, 2009). Hadronic processes in mini radio lobes on pc-scale may also result in two bumps at $\sim$1 MeV and $\sim$1 GeV in the broadband spectral energy distributions (SEDs) of these AGNs (Kino \& Asano 2011), and the IC emission in mini shells on pc-scale may produce detectable very high energy (VHE, $>$100 GeV) $\gamma$-ray emission (Ito et al. 2011; Kino et al. 2013).

{\em Fermi}/LAT has discovered $\gamma$-ray emission from some CSSs as reported in the latest Fourth Fermi LAT source catalog (4FGL, Abdollahi et al. 2020), which covers 8 years of \emph{Fermi}/LAT data in the energy range from 50 MeV to 1 TeV. Different from the other $\gamma$-ray emitting AGNs, the core radiation of CSSs is very weak and even no significant core is observed in some CSSs. These $\gamma$-ray emitting CSSs could be good candidates to investigate the jet properties among the young AGNs. The radiation mechanism and the properties of emitting regions of their $\gamma$-ray emission are still uncertain. They would be crucial for revealing the nature of this subclass of AGNs and even the unification of different AGN subclasses. In this paper, we comprehensively analyze the data observed by \emph{Fermi}/LAT for these $\gamma$-ray emitting CSSs in Section 3, including the long-term light curves, the spectra, the counts maps, and the variability index. We also compile their broadband SEDs and fit them with a two-zone leptonic model in Section 4, then obtain their jet parameters and investigate the jet properties. Comparisons of jet properties between $\gamma$-ray emitting CSSs and other kinds of $\gamma$-ray emitting AGNs are presented in Section 5 to explore the intrinsic unification among these $\gamma$-ray emitting AGNs. Discussion and conclusions are given in Section 6.

\section{GeV-selected CSSs}

Five CSSs (3C 138, 3C 216, 3C 286, 3C 380, 3C 309.1) are included in 4FGL (Abdollahi et al. 2020). An et al. (2016) reported that the $\gamma$-ray emitting source PKS 0202+149 (4C 15.05) is also a CSS. Therefore, six CSSs are included in 4FGL. However, O'Dea (1998) reported that 3C 216 and 3C 380 may be not really CSSs and their small angular extent is due to projection effects (see also Barthel et al. 1988; Wilkinson et al. 1991; van Breugel et al. 1992; Taylor et al. 1995). Although the nature of 3C 216 and 3C 380 is still controversial, we present data and theoretical analysis for all six CSSs for completeness. We do not include them in the discussion on CSSs, and present some discussion for them based on our analysis (see Section 6). We describe these sources in the following.

\emph{\textbf{3C 138}} located at $z=0.759$ (PKS 0518+16, Spinrad et al. 1985). It has a steep spectrum of $\alpha=0.65$ straight up to 22 GHz with a turn-over at about 100 MHz (Fanti et al. 1990; Shen et al. 2001). The VLBI observations at 15 and 22 GHz show that its main jet consists of several knots extending about 400 mas in a position angle of 65$^{\circ}$ and a weak counter-jet about 250 mas is presented in the opposite direction (Cotton et al. 1997 and references therein). The VLBI image at 5 GHz also exhibits two notable emission regions, i.e., the core and the eastern lobe, which are separated by 400 mas at a position angle of 70$^{\circ}$, and there are several discretely lower surface brightness regions between them, but no counter-jet was significantly detected at 5 GHz (Shen et al. 2001). Due to the flattest spectral index and the highest brightness temperature, component A is suggested as a nucleus component (Shen et al. 2005). The superluminal motion with an apparent velocity of 9.7$c$ was reported by Cotton et al. (1997), and later was suggested to be 3.3$c$ (Shen et al. 2001).

\emph{\textbf{3C 216}} also identified with the quasar 0906+430 at $z=0.67$ (Smith \& Spinrad 1980). The VLBI observations indicate the superluminal motion of $\sim4c$ and a small viewing angle of $\theta<20^{\circ}$ (Taylor et al. 1995 and references therein). A steep spectrum of $\alpha=0.79$ at low frequencies is observed, but it flattens above 5 GHz with $\alpha=0.29$ (Taylor et al. 1995). The VLBI Space Observatory Programme observations reveal the pc-scale structure of 3C 216 that can be well described by compact jet models (Paragi et al. 2000). Using the measured brightness temperature Paragi et al. (2000) estimated a lower-limit of the Doppler boosting factor of $\delta\sim3$ for the core-jet with the viewing angle less than 19$^{\circ}$, and thus they concluded that the observed small projected size of 3C 216 is probably caused by both interaction and projection effects. It displays high optical polarization and variability, similar to typical blazars (Angel \& Stockman 1980). The bright core-lobe features of 3C 216 extend over 2.${\arcsec}$5 (Pearson et al. 1985) and they are embedded in a faint diffuse radio halo with a diameter $\sim7\arcsec$ (Barthel et al. 1988; Taylor et al. 1995).

\emph{\textbf{3C 286}} also named B1328+307, at redshift of $z=0.849$ (Cohen et al. 1977). The steep radio spectrum has a spectral index of $\alpha=0.61$ between 1.4 and 50 GHz and then a turnover at about 300 MHz, below which it is flat till $\sim$75 MHz (An et al. 2017). The source displays a primary core and a second lobe $\sim$2.6 arcsec to the south-west (Spencer et al. 1989; An et al. 2017). The radio emission at 15 GHz of this source is rather stable and no significant variation is observed in the past 10 years. The high polarization of the source has been detected in the radio band (Akujor \& Garrington 1995). A compact bright nucleus associated with the radio core is also detected by the Hubble Space Telescope (deVries et al. 1997). Two compact components with comparable flux densities in the inner 10-mas region are resolved by the VLBI and the more compact component showing an inverted spectrum with a turnover between 5 and 8 GHz may infer the core (An et al. 2017). A jet speed of $\sim0.5 c$ and an inclination angle of $\sim48^{\circ}$ are derived for 3C 286. The optical spectrum observed with the SDSS-BOSS clearly indicates that 3C 286 can be classified as a NLS1 (Berton et al. 2017).

\emph{\textbf{3C 309.1}} located at $z=0.904$ (S5 1458+71, Burbidge \& Burbidge 1969). The overall extent of 3C 390.1 is $2\arcsec$ and it has a steep spectrum of $\alpha=0.69$ between 1 and 10 GHz (Wilkinson 1972). Its extended structures have a steep spectrum with $\alpha=0.94\pm0.08$ between 100 MHz and 15 GHz and exhibit a large rotation in position angle (Kus et al. 1981). Forbes et al. (1990) reported that 3C 309.1 is surrounded by very massive cooling flows, however, this is not consistent with the small rotation measures (Aaron et al. 1997). The image at 15 GHz observed with the VLBA shows a compact core and a extended component 20 mas to the south with some extended diffuse emission. The total fluxes at 8 and 14.5 GHz display obvious increase during 1990 to 2000, and the corresponding spectral index also varies (Aller et al. 2003). A relativistic motion of the new blob nearby the core with apparent velocity of $7.0\pm0.5c$ was observed (Gawro\'{n}ski \& Kus 2006).

\emph{\textbf{3C 380}} located at $z=0.692$ (TXS 1828+487, Wilkinson et al. 1991). It has a steep radio spectrum with $\alpha=0.7$ between 300 MHz and 5 GHz and the extent extends to $\sim1\arcsec$ (Readhead \& Wilkinson 1980). The superluminal motion in the outer regions of the jet is observed and corresponds to an apparent velocity $\sim8c$, indicating that the outer part of the VLBI-scale jet is within $10^{\circ}$ of the line-of-sight. However, the absence of strong variability at radio and optical bands may imply an intrinsic bend near the core-jet, and the core-jet points $\sim30^{\circ}$ away while the overall source axis is within $10^{\circ}$ of the line-of-sight (Wilkinson et al. 1991). Using the multi-epoch VLBI observations, Polatidis \& Wilkinson (1998) revealed a curved pc-scale jet with complex substructure and an apparent acceleration from the core to $\sim100$ pc. They also suggested that 3C 380 most likely is a powerful FR II RG seen approximately end-on. With the space-VLBI observations the apparent superluminal motions are observed, however, no acceleration in pc-scale and changes of the position angle were confirmed (Kameno et al. 2000). One-sided jet is observed in both pc-scale and kpc-scale for 3C 380 (Gabuzda et al. 2014).

\emph{\textbf{4C 15.05}} also known as NRAO 91 and PKS 0202+149. On the basis of [O~{\scriptsize III}] $\lambda3727$ and [Ne~{\scriptsize I}] $\lambda$3833 lines, it was estimated to be located at $z=0.833$ (Stickel et al. 1996), but a smaller redshift of $z=0.405$ was reported by Perlman et al. (1998). Recently, Jones et al. (2018) suggested that the neutral hydrogen absorption feature of this source agrees very well with the value of $z=0.833$. Its mean spectral index between 400 MHz to 8 GHz is $\alpha=0.33$ (Herbig \& Readhead 1992), which is slightly steeper than that of blazars (Fan et al. 2010; Pei et al. 2019). This source displays the structure of a core and double lobes with the total projected size of $\sim$1.3 kpc. A core-jet structure in pc-scale extends the projected size of $\sim$25 pc at a position angle perpendicular to the kpc-scale structure (An et al. 2016). The significant apparent superluminal motion of $\sim16 c$ is detected (An et al. 2016). 4C 15.05 had been identified as a $\gamma$-ray emitting AGN with EGRET (von Montigny et al. 1995).

\section{\emph{Fermi}/LAT Data Analysis}

The \emph{Fermi}/LAT has provided a powerful tool for monitoring AGNs at $\gamma$-ray energy band (Ackermann et al. 2015), which is sensitive to photon energies greater than 20 MeV. For this work, the data of sources were taken from the Fermi Science Support Center (FSSC) covering the period from 2008 August 4 to 2019 July 16 (MET 239557417--584949858), approximately 11 years. the data analysis was performed with the publicly available software \textit{fermitools} (ver. 1.0.0). Using standard data quality selection criteria ``$(DATA\_QUAL > 0)  \&\& (LAT\_CONFIG == 1)$", the events with energies from 100 MeV to 300 GeV are considered. In order to reduce the contamination from the $\gamma$-ray Earth limb, the maximum zenith angle is set to be 90$\degr$. Data within a $14\degr \times14\degr$ radius of interest (ROI) centered on the source position are binned in 12 logarithmically spaced bins in energy and a spatial bin of 0.1$\degr$ per pixel is used. The \textit{$P8R3\_SOURCE\_V2$} set of instrument response functions (IRFs) is used. For the background model, we include the diffuse Galactic interstellar emission (IEM, $gll\_iem\_v07.fits$) and isotropic emission (``$iso\_P8R3\_SOURCE\_V2\_v1.txt$") templates released by FSSC\footnote{http://fermi.gsfc.nasa.gov/ssc/data/access/lat/BackgroundModels.html}, as well as the individual $\gamma$-ray sources listed in the 4FGL (Abdollahi et al. 2020). The normalization and spectral parameters of the discrete $\gamma$-ray sources within 8$\degr$ in the background model were kept free. The Galactic emission and the isotropic component were also kept the normalization free during the data analysis. We use the Maximum Likelihood test statistic (TS) to estimate the significance of $\gamma$-ray signals, which is defined by TS$=2(\ln\mathcal{L}_{1}-\ln\mathcal{L}_{0})$, where $\mathcal{L}_{0}$ is the likelihood of background without the point source (null hypothesis) and $\mathcal{L}_{1}$ is the likelihood of the background including the point source.

\begin{figure}
 \centering
   \includegraphics[angle=0,scale=0.3]{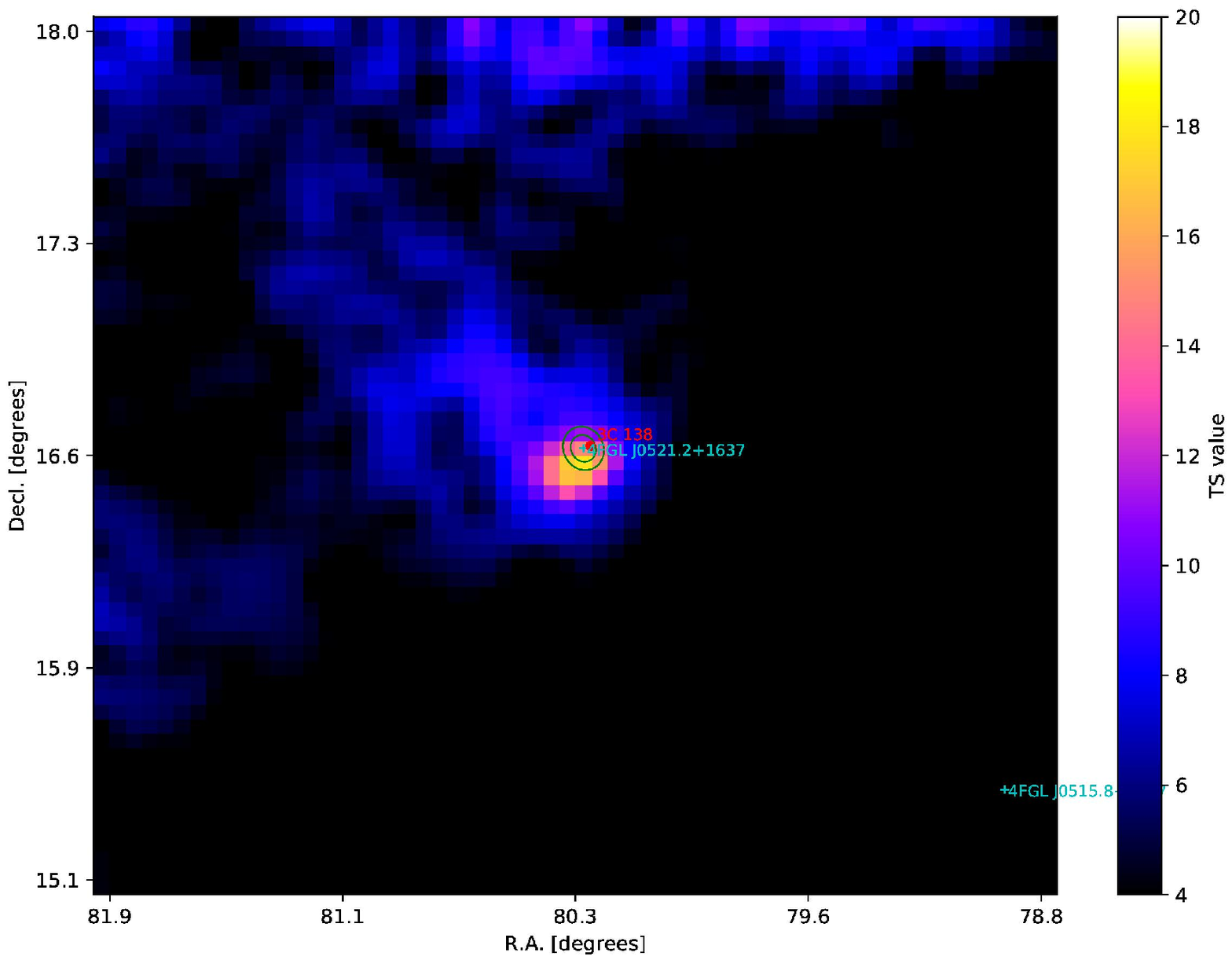}
   \includegraphics[angle=0,scale=0.3]{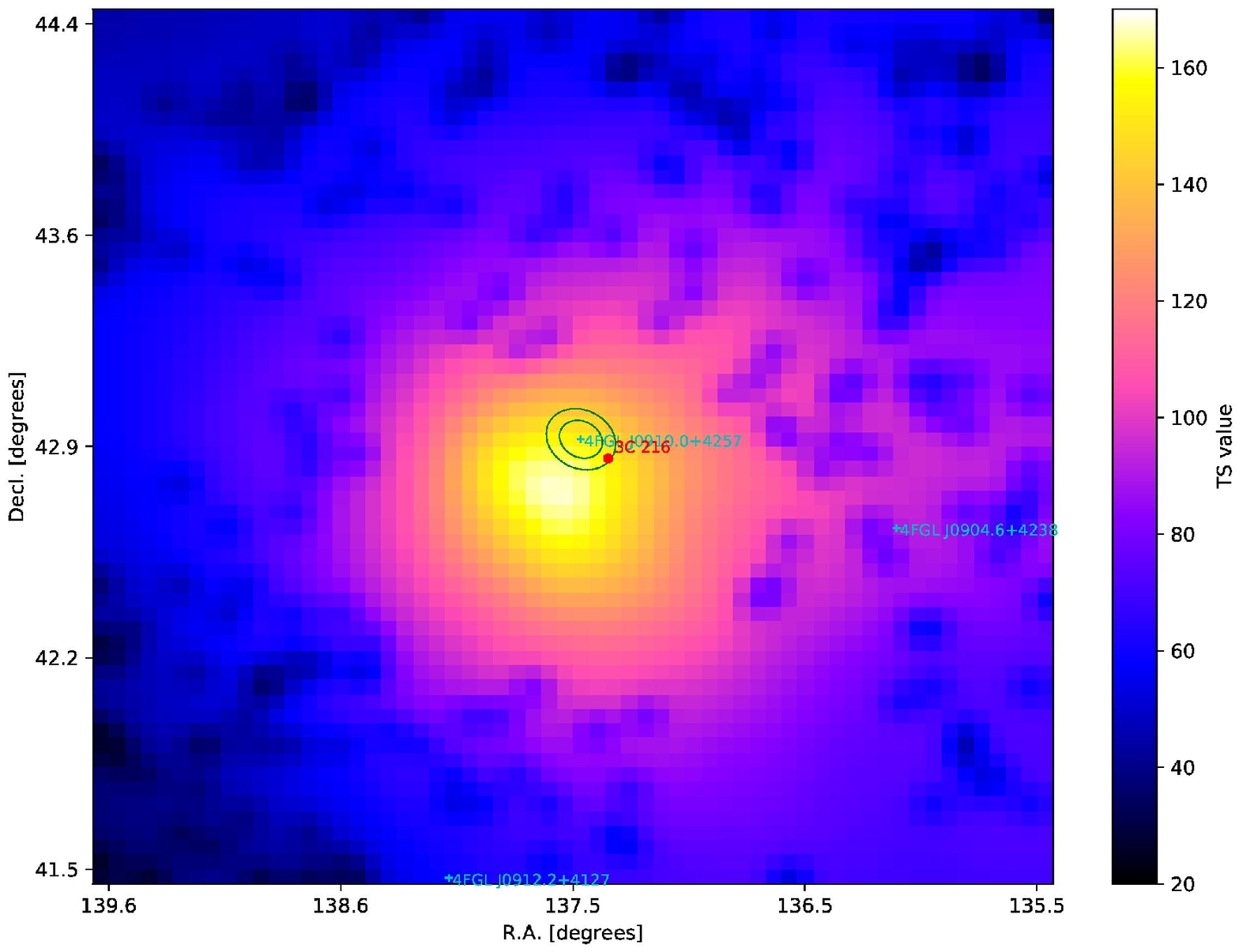}
   \includegraphics[angle=0,scale=0.3]{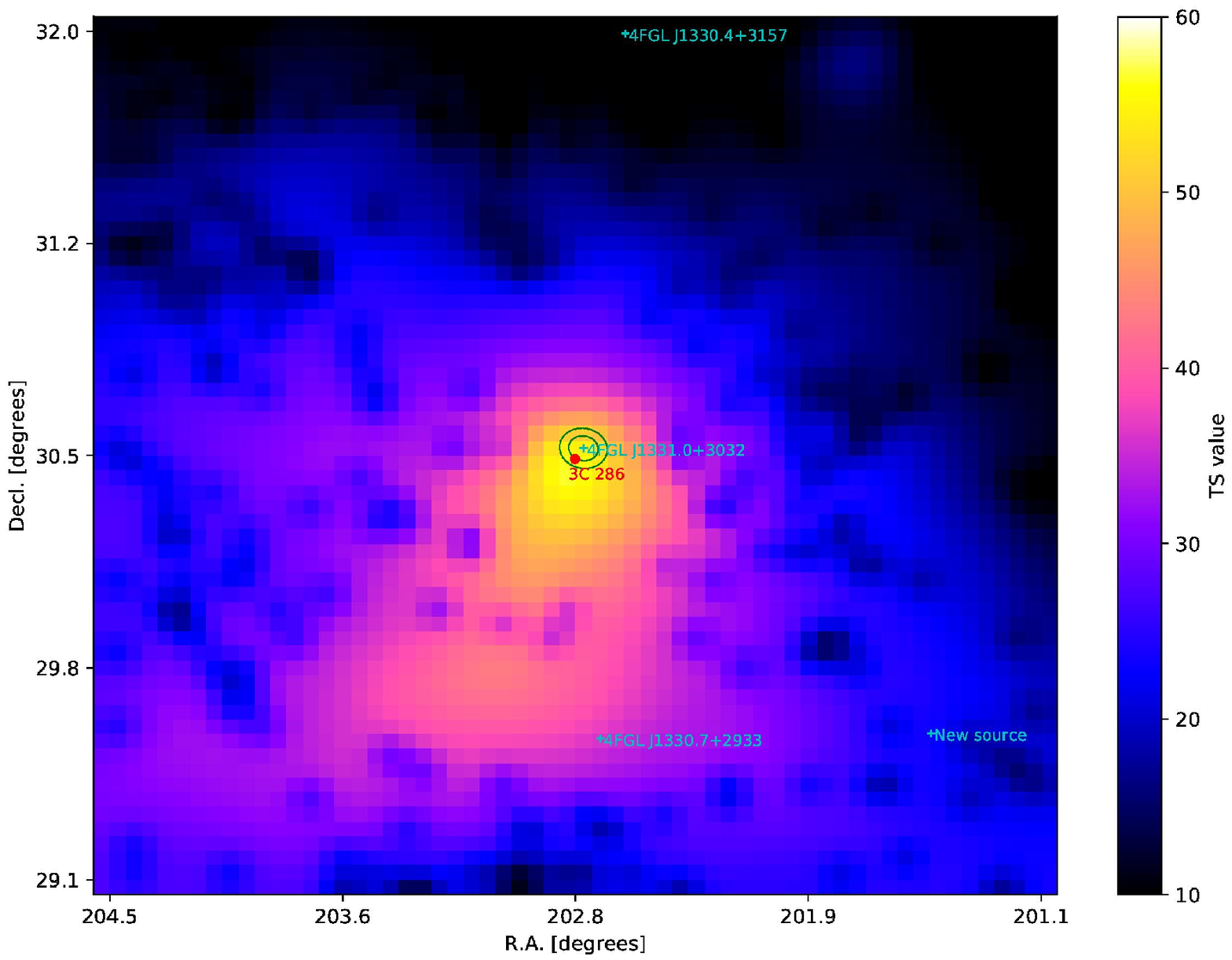}
\caption{TS maps of the six CSSs. The light blue crosses represent the positions of the 4FGL point sources and the corresponding $68\%$ and $95\%$ containments are shown as green contours. The red points represent the positions of these CSSs (taken from Abdollahi et al. 2020). Note that a new background source (New source) is needed to add in the background model of 3C 286 and the derived TS value for this new source is 41.2 by assuming a power-law spectrum. The maps are created with a pixel size of 0.05 and smoothed by Gaussian kernel ($\sigma=0.35\degr$).}\label{TSmap}
\end{figure}

Note that the 4FGL point sources are based on the 8-year survey data, here we make the new background source test using the package \textit{gtfindsrc}. Only a new background source that has TS$\sim$41.2 ($>5\sigma$) with a power-law spectrum was found in the region of $3\degr\times3\degr$ centered on 3C 286. As illustrated in Figure \ref{TSmap}, the six CSSs are located within the $95\%$ containment of the associating 4FGL point sources, confirming that these CSSs are spatially associated with these 4FGL point sources. The information of these CSSs and associating 4FGL point sources is given in Table 1.

The photon spectra of CSSs are well described by a power-law spectral function, i.e., $dN(E)/dE = N_{0}(E/E_{0})^{-\Gamma_{\gamma}}$, where $N(E)$ is the photon distribution as a function of energy and $\Gamma_{\gamma}$ is the photon spectral index. The spectral model of 4C 15.05 (4FGL J0204.8+1513) is a log-normal function in 4FGL. Note that a spectrum should be significantly curved if TS$_{\rm curv}>9$ (3$\sigma$ significance; Abdollahi et al. 2020), where TS$_{\rm curv}=2\log(\mathcal{L}_{\rm log-normal}/\mathcal{L}_{\rm power-law})$. However, we find that TS$_{\rm curv}\sim1$ for 4C 15.05. Therefore, its photon spectrum is taken as the power-law spectral model in our analysis.

\begin{figure}
 \centering
   \includegraphics[angle=0,scale=0.31]{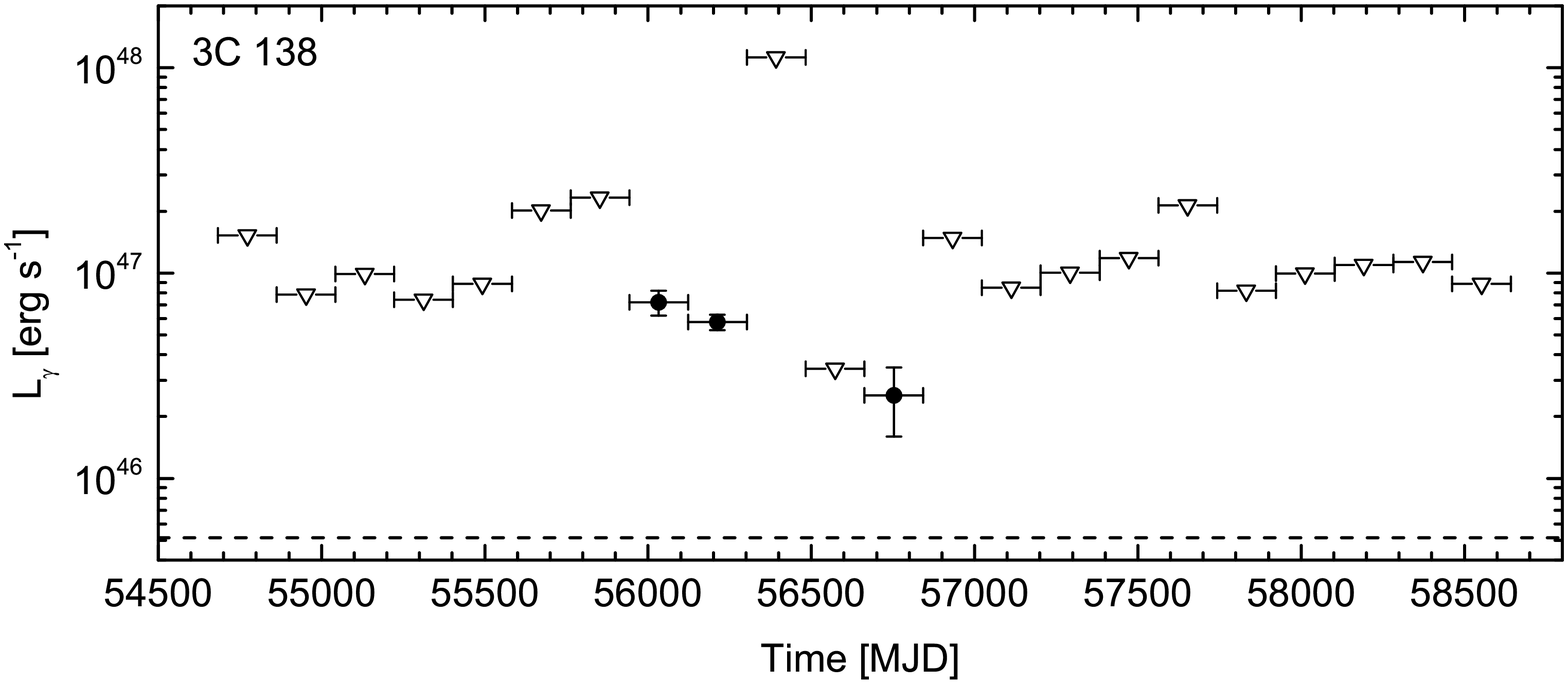}
   \includegraphics[angle=0,scale=0.27]{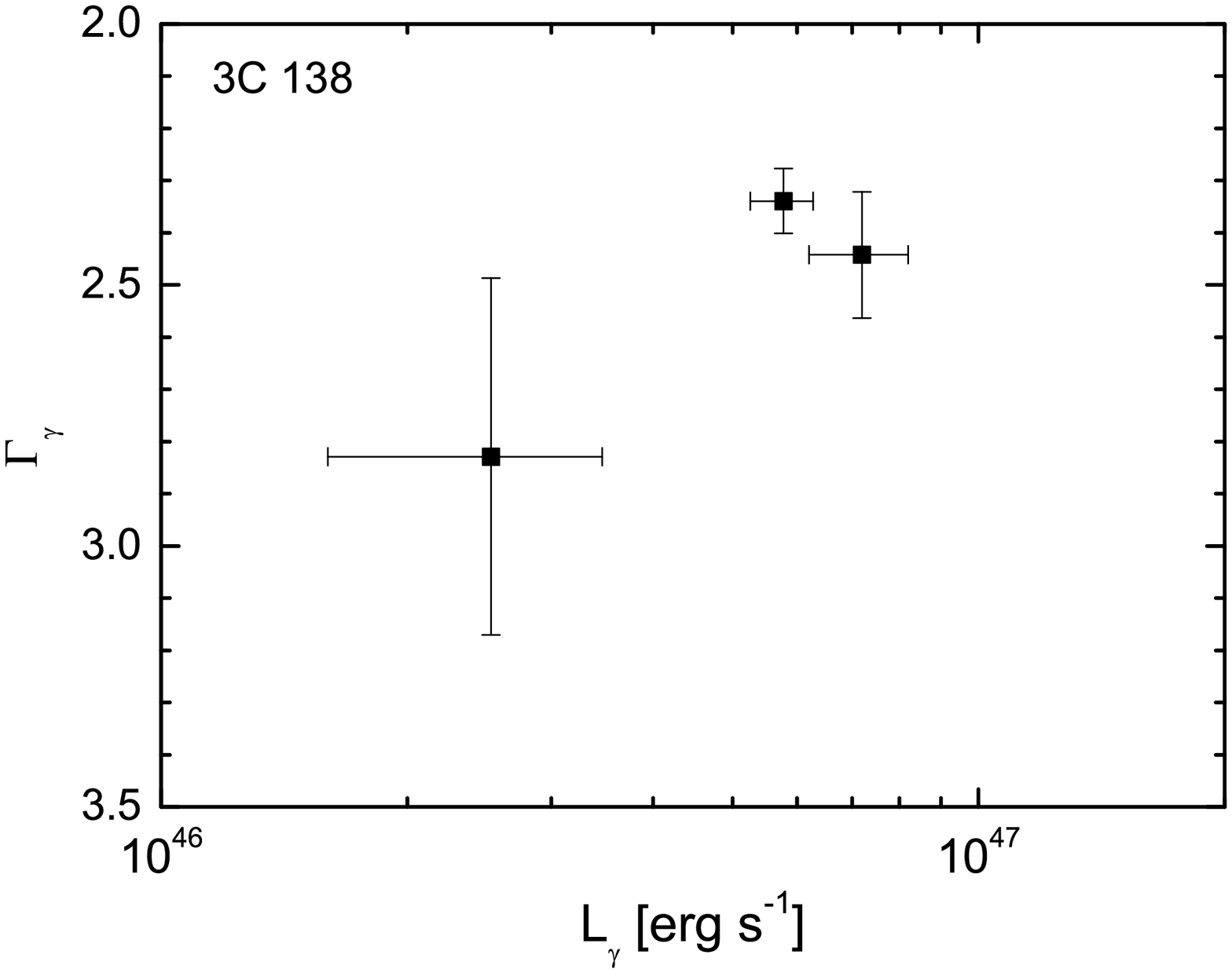}
   \includegraphics[angle=0,scale=0.31]{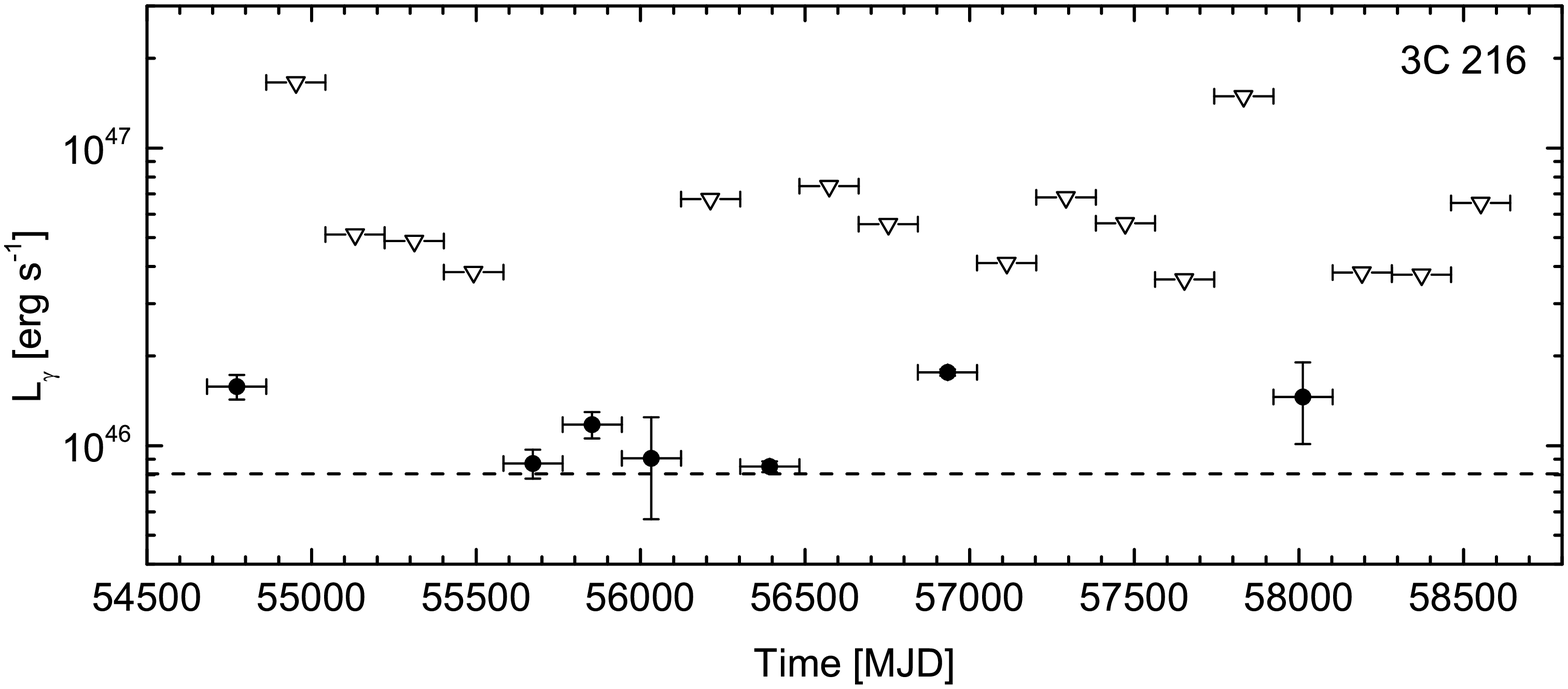}
   \includegraphics[angle=0,scale=0.27]{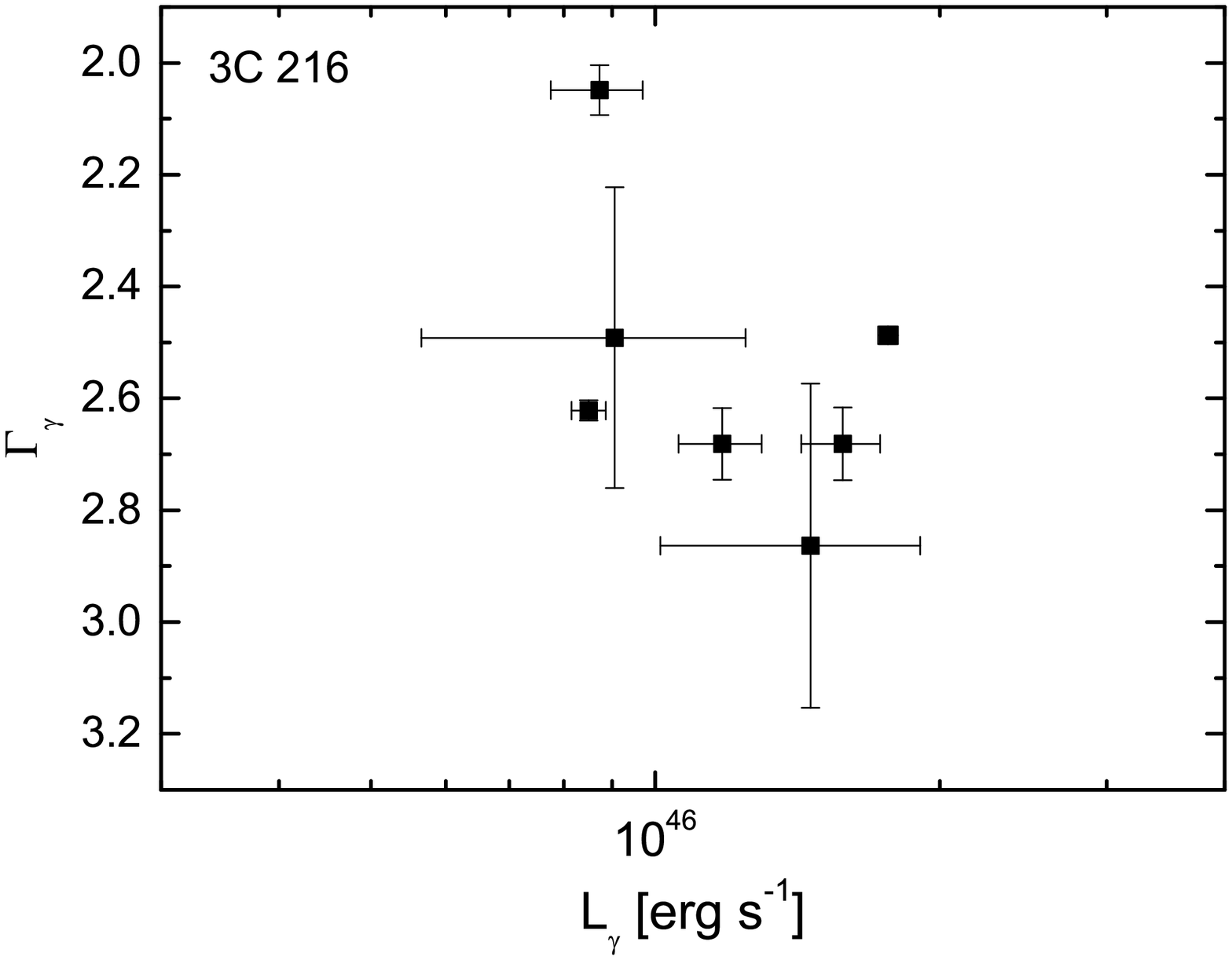}
   \includegraphics[angle=0,scale=0.31]{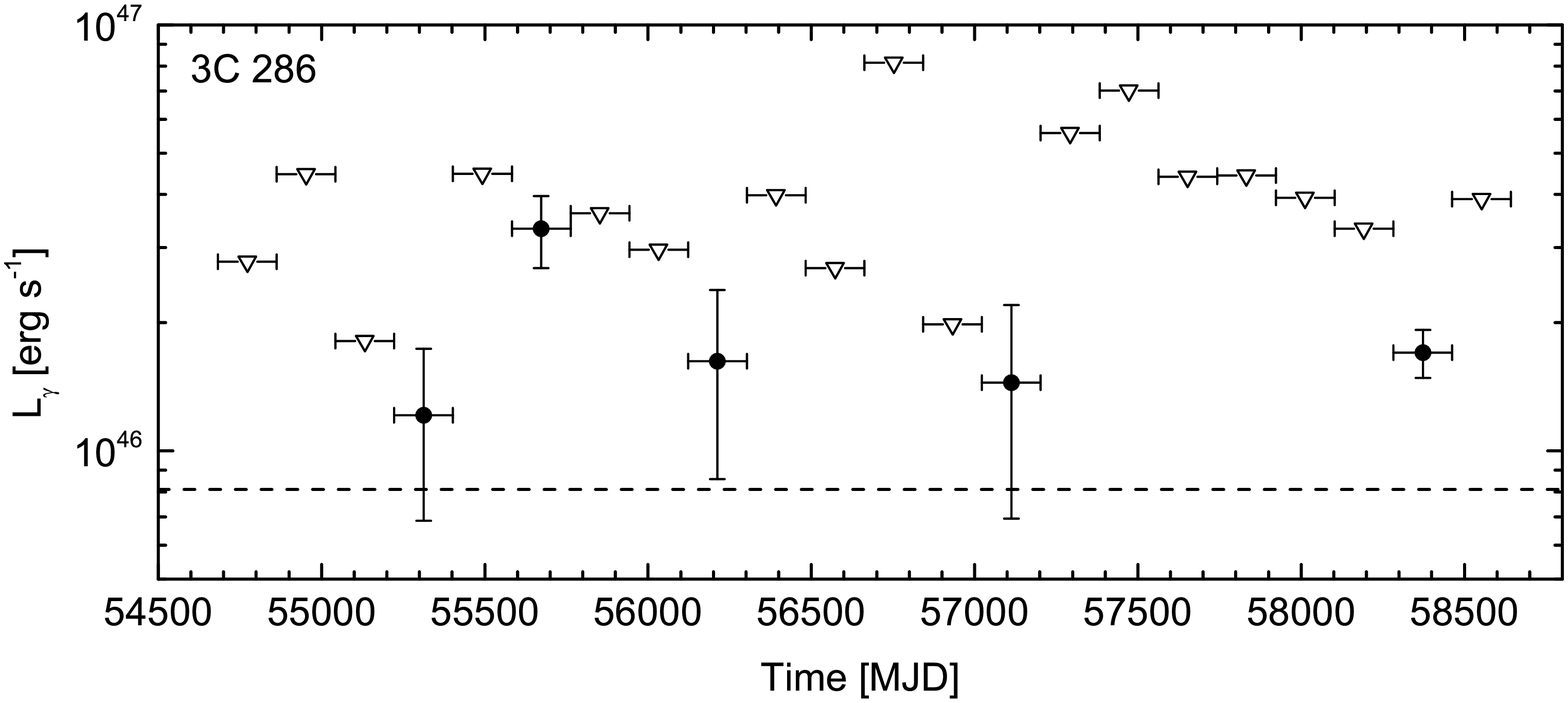}
   \includegraphics[angle=0,scale=0.27]{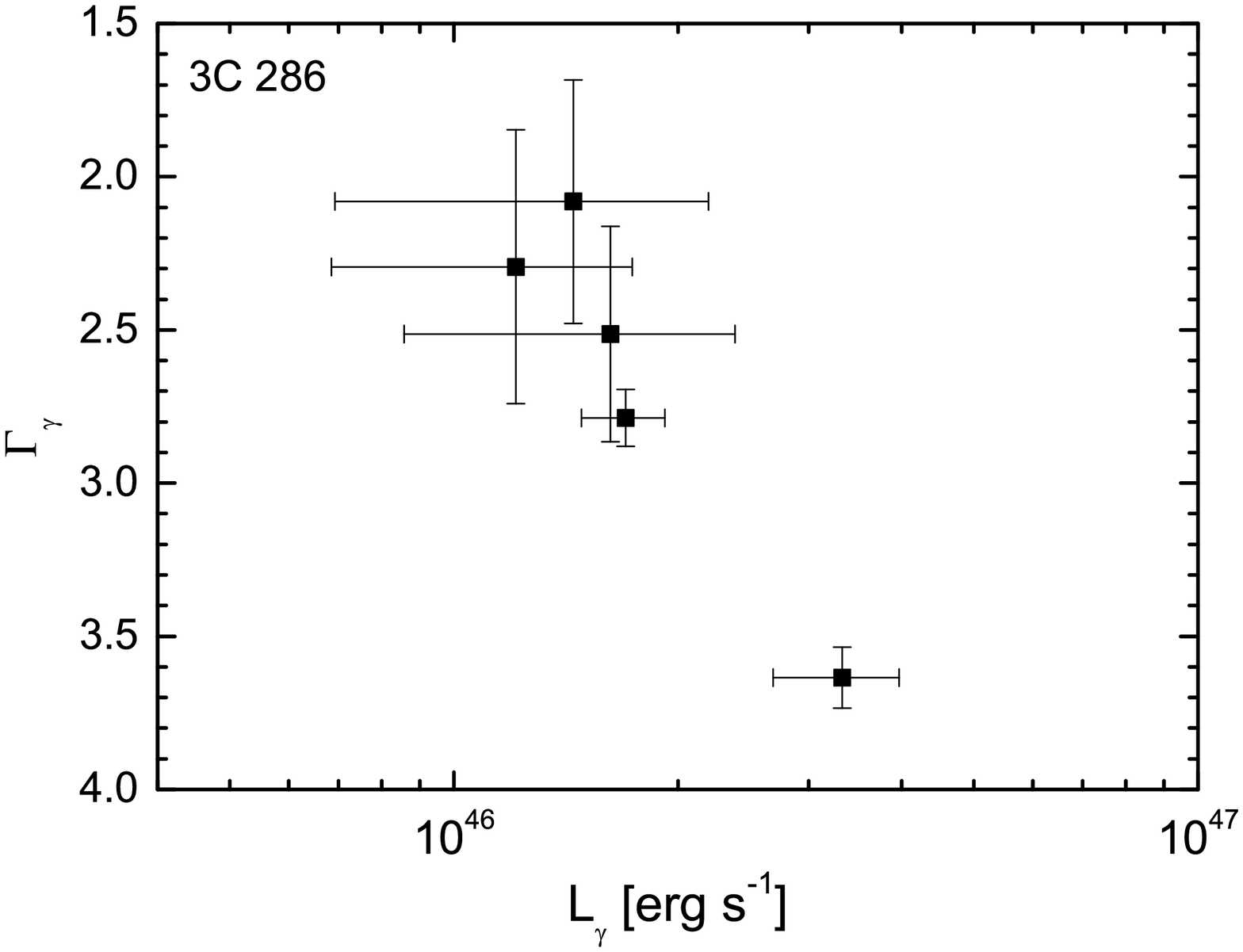}
\caption{\emph{Left-panels}: Long-term $\gamma$-ray light curves observed by the \emph{Fermi}/LAT in time bins of 180 days. The opened triangles indicate TS$<$9 for this time bin. The horizontal dashed lines represent the $\sim$11-year average luminosity of sources observed by \emph{Fermi}/LAT. \emph{Left-panels}: photon spectral index ($\Gamma_{\gamma}$) as a function of $\gamma$-ray luminosity ($L_{\gamma}$).}\label{LC}
\end{figure}

The light curves are obtained by the derived fluxes with the power-law fits. The $\gamma$-ray light curves of these CSSs in time bins of 180-day with TS$\geq$9 (approximatively corresponds to $\sim3\sigma$ detection, Mattox et al. 1996) are presented in Figure \ref{LC}. If TS$<9$, an upper-limit is presented (95$\%$ confidence level). The average luminosity of the past $\sim$11 years for sources is also given in the figure. Except for 3C 380 and 4C 15.05, only several time bins in the long-term light curves of these CSSs satisfy TS$\geq$9, indicating that on average this kind of AGNs has weak $\gamma$-ray emission comparing with blazars. The detection data points spread on both sides of the average luminosity for 3C 380 and 4C 15.05, however, almost all the detection data points of other four CSSs have higher luminosity than the average luminosity. We roughly estimate the variability amplitude of these sources with $F_{\max}/F_{\min}$, where $F_{\max}$ and $F_{\min}$ are respectively the highest and lowest fluxes in the long-term $\gamma$-ray light curves (excluding the time bin data with TS$<$9). The derived largest and smallest values of $F_{\max}/F_{\min}$ are 7.5 for 3C 309.1 and 2.1 for 3C 216, respectively. We can observe that these $\gamma$-ray emission CSSs generally do not show the fast and large flares like blazars.

Likelihood-based statistic is also the most common method to quantify variability (Nolan et al. 2012; Abdollahi et al. 2020; Peng et al. 2019; Xi et al. 2020). To gauge the variability of sources, we follow the definition in 2FGL (Nolan et al. 2012) and compute the variability index (TS$_{\rm var}$) as
\begin{equation}
\rm TS_{\rm var} =  2\sum_{i=1}^N [ log(\mathcal{L}_{i}(F_i)) - log(\mathcal{L}_{i}(F_{\rm glob}))],
\end{equation}
where $F_i$ is the fitting flux for bin $i$, $\mathcal{L}_{i}(F_i)$ is the likelihood corresponding to bin $i$,  and $F_{\rm glob}$ is the best fit flux for the glob time by assuming a constant flux. Since we generate the light curves in time bins of 180-day, the source is considered to be probably variable if TS$_{\rm var}>45.82$, where TS$_{\rm var}=45.82$ corresponds to $3\sigma$ confidence level in a $\chi^2_{N-1}$(TS$_{\rm var}$) distribution with $N-1=21$ degrees of freedom, $N$ is the number of time bins. Five among six CSSs are variable sources with this criterion, i.e., TS$_{\rm var}=128.1$ for 3C 138, TS$_{\rm var}=64.3$ for 3C 286, TS$_{\rm var}=167.0$ for 3C 309.1, TS$_{\rm var}=149.1$ for 3C 380, and TS$_{\rm var}=211.4$ for 4C 15.05. Only 3C 216 does not show as a variable source with TS$_{\rm var}=11.8$.

The photon spectral index ($\Gamma_{\gamma}$) as a function of luminosity ($L_{\gamma}$) is also illustrated in Figure \ref{LC}. The distinct spectral variations are observed in these CSSs. The largest change of $\Gamma_{\gamma}$ is presented in 3C 286 and 4C 15.05, i.e., from 2.08$\pm$0.40 to 3.63$\pm$0.10 and from 1.85$\pm$0.01 to 2.90$\pm$0.02, respectively. Although no significant flux variation is observed in 3C 216, its photon spectral index also shows variations, from 2.05$\pm$0.04 to 2.86$\pm$0.29. Only 3C 138 seems to show the behavior of ``harder when brighter", which has been seen in the $\gamma$-ray emitting blazars (e.g., Zhang et al. 2013, 2018a, 2020). And the tendency of ``steeper when brighter" is displayed in 3C 286 and 3C 309.1. The Pearson correlation analysis yields a correlation coefficient of $r=0.95$ and a chance probability of $p=0.01$ for 3C 286 and $r=0.57$, $p=0.19$ for 3C 309.1, respectively. No correlated tendency between $\Gamma_{\gamma}$ and $L_{\gamma}$ is presented in 3C 216 and 3C 380. For 4C 15.05, the relation between $\Gamma_{\gamma}$ and $L_{\gamma}$ seems to change from anti-correlation into correlation. Excluding the four data points with $L_{\gamma}>5\times10^{46}$ erg s$^{-1}$, the Pearson correlation analysis yields $r=0.45$ and $p=0.09$ for 4C 15.05. Recently, a transition from softer-when-brighter to harder-when-brighter is also observed in the monthly $\gamma$-ray flux--index plot of 3C 273 (Kim et al. 2020). This transition may be due to a balance between acceleration and cooling of relativistic particles (Kim et al. 2020) or a shift of the inverse Compton peak in the SED (Shah et al. 2019).

\begin{figure}
 \centering
   \includegraphics[angle=0,scale=0.4]{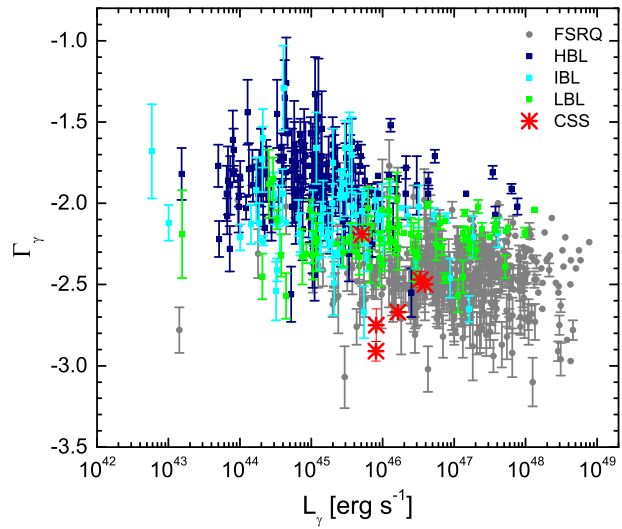}
\caption{The $\sim$11-year average $\Gamma_{\gamma}$ observed by \emph{Fermi}/LAT as a function of the corresponding $L_{\gamma}$ for the six CSSs. The blazar data are taken from Ackermann et al. (2015). }\label{Gamma-L}
\end{figure}

As illustrated in Figure \ref{Gamma-L}, we also show the Fermi blazars in the $L_{\gamma}$--$\Gamma_{\gamma}$ plane, where the blazar data are taken from Ackermann et al. (2015) and belong to the clean sample with confirmed redshift, including 414 flat spectrum radio quasars (FSRQs), 162 high-frequency-peaked BL Lacs (HBLs), 69 intermediate-frequency-peaked BL Lacs (IBLs), and 68 low-frequency-peaked BL Lacs (LBLs). The radiation properties of the six CSSs in the GeV band are analogous to these of FSRQs, however, they on average have steeper spectra and lower $L_{\gamma}$ than FSRQs.

\section{Broadband SED Modeling}

\begin{figure}
 \centering
   \includegraphics[angle=0,scale=0.25]{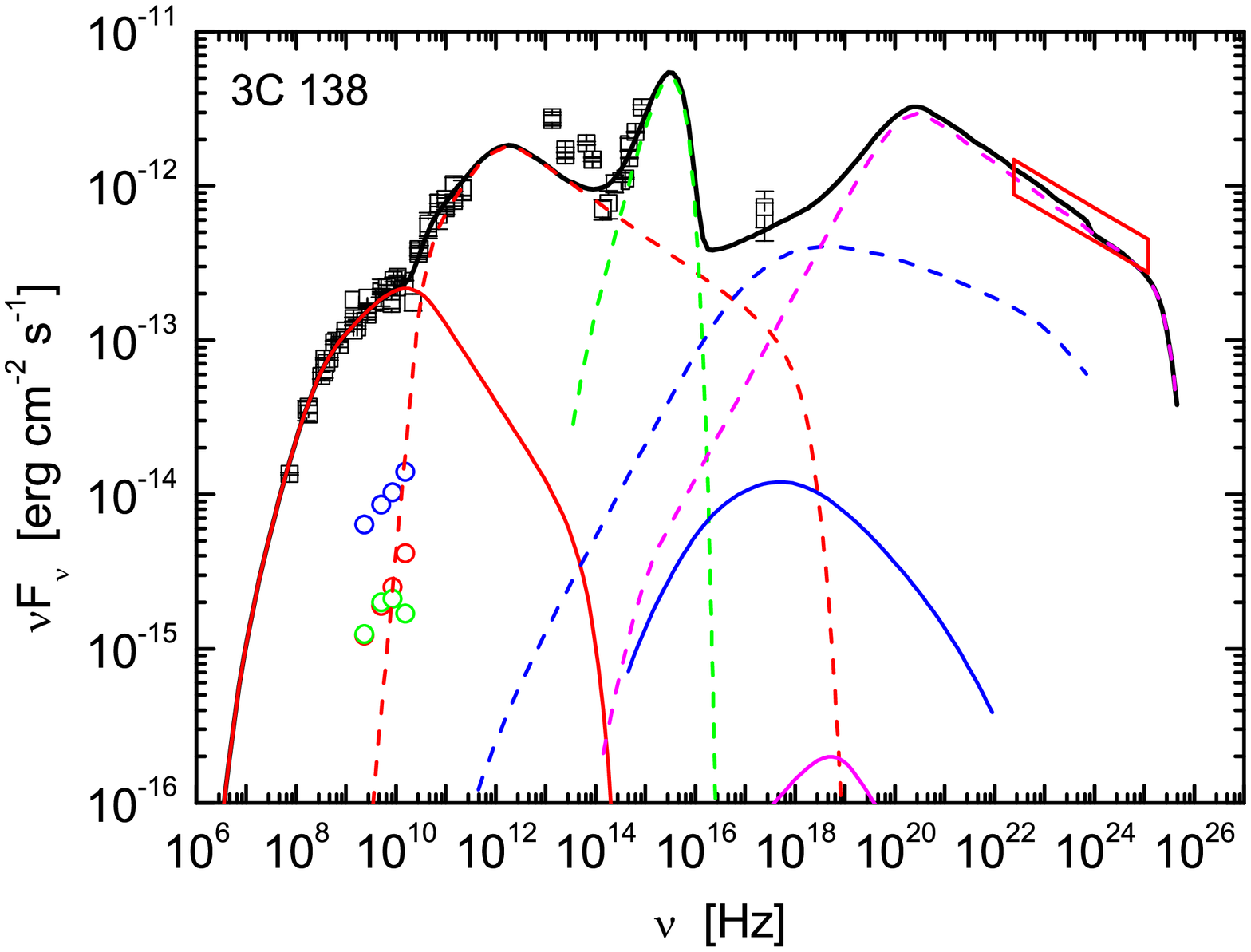}
   \includegraphics[angle=0,scale=0.25]{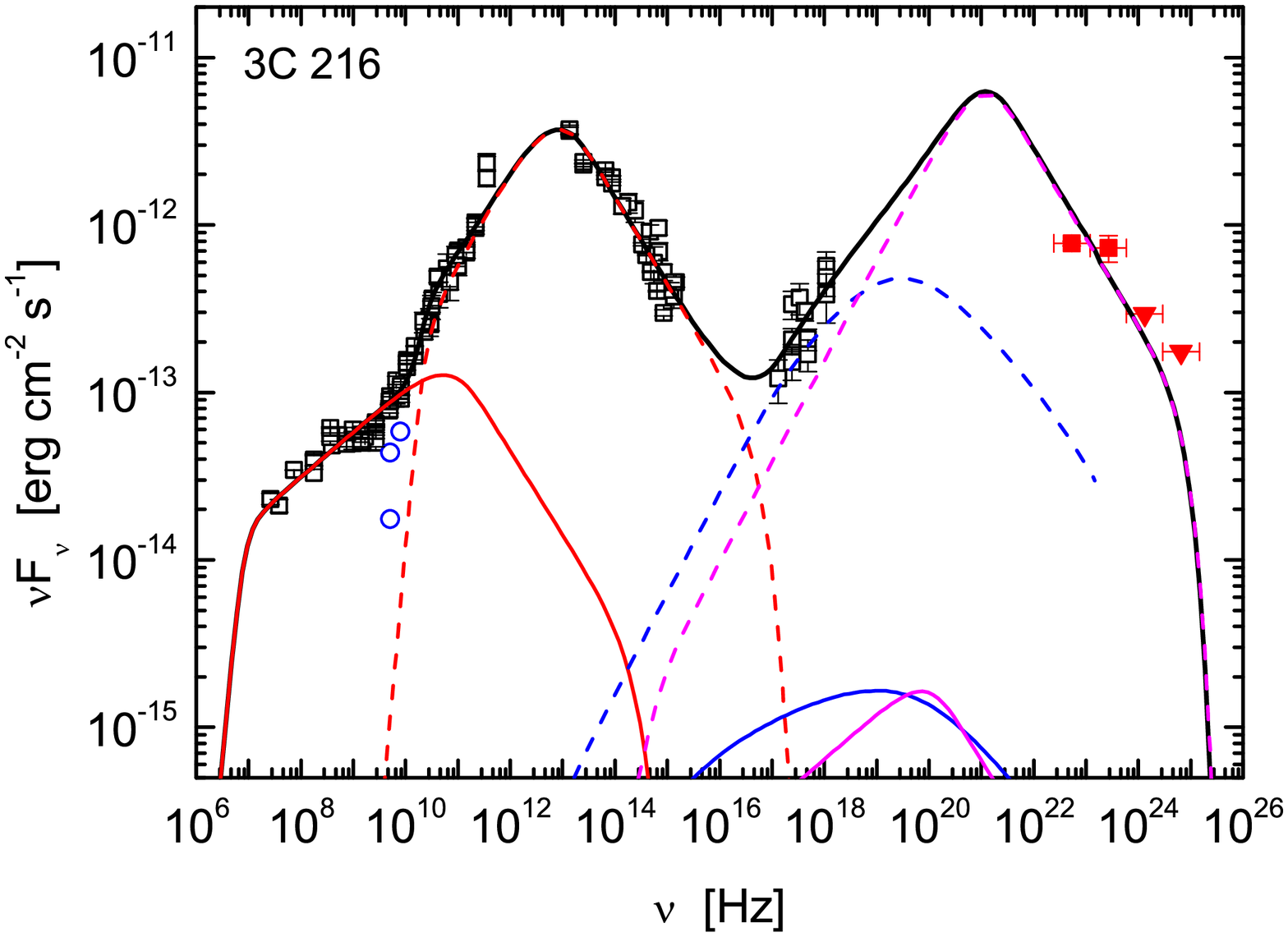}
   \includegraphics[angle=0,scale=0.25]{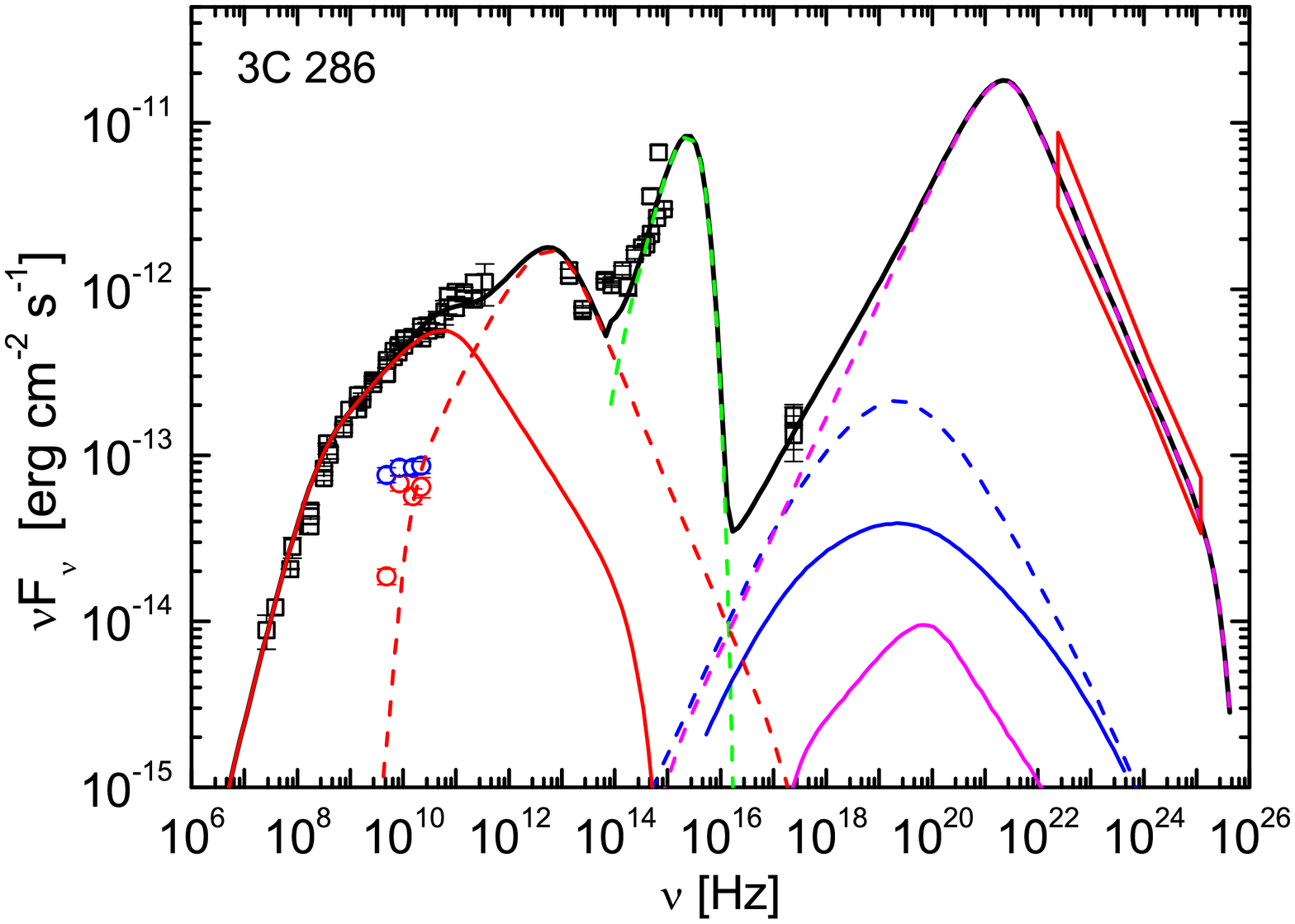}
   \includegraphics[angle=0,scale=0.25]{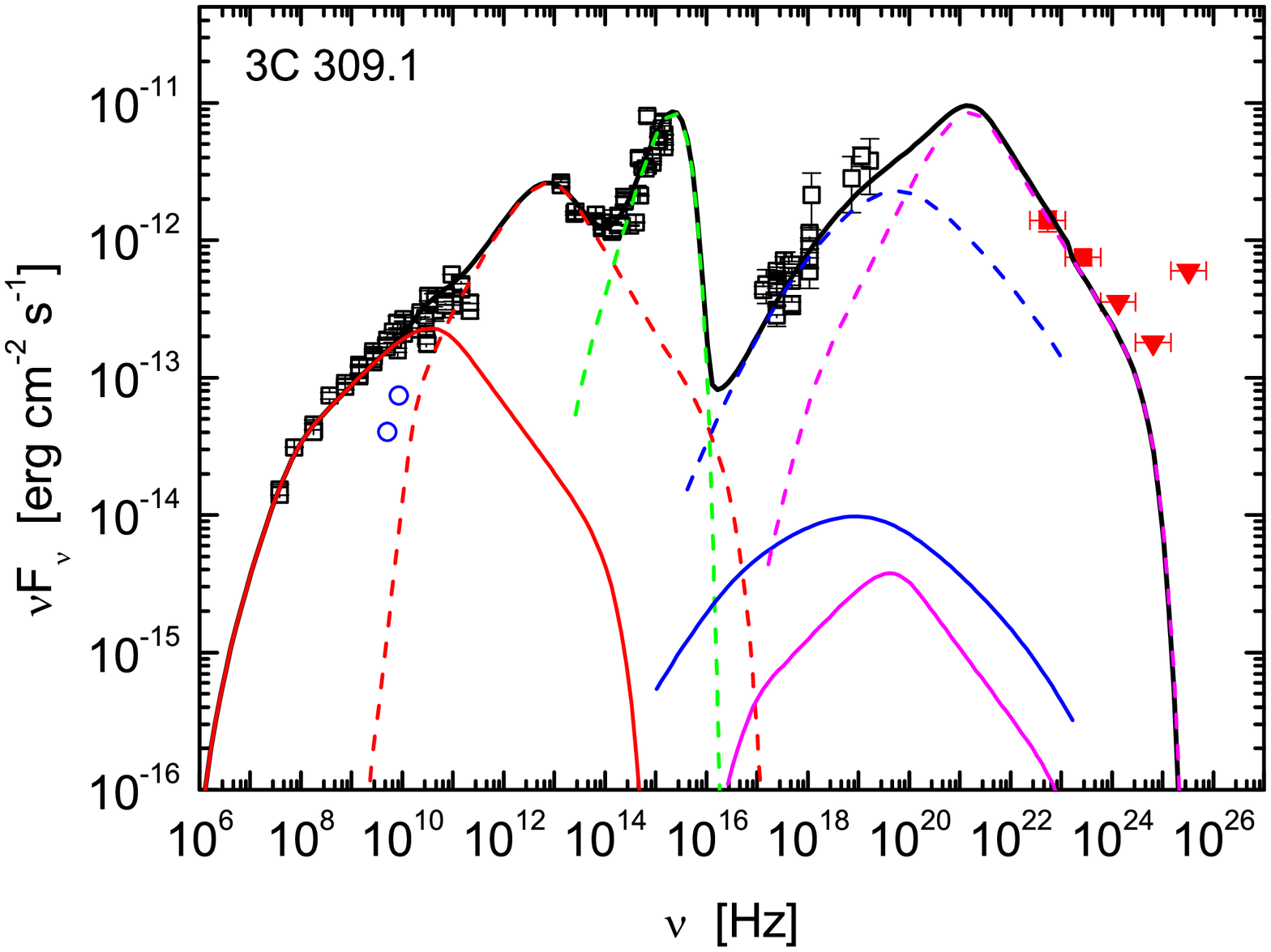}
   \includegraphics[angle=0,scale=0.25]{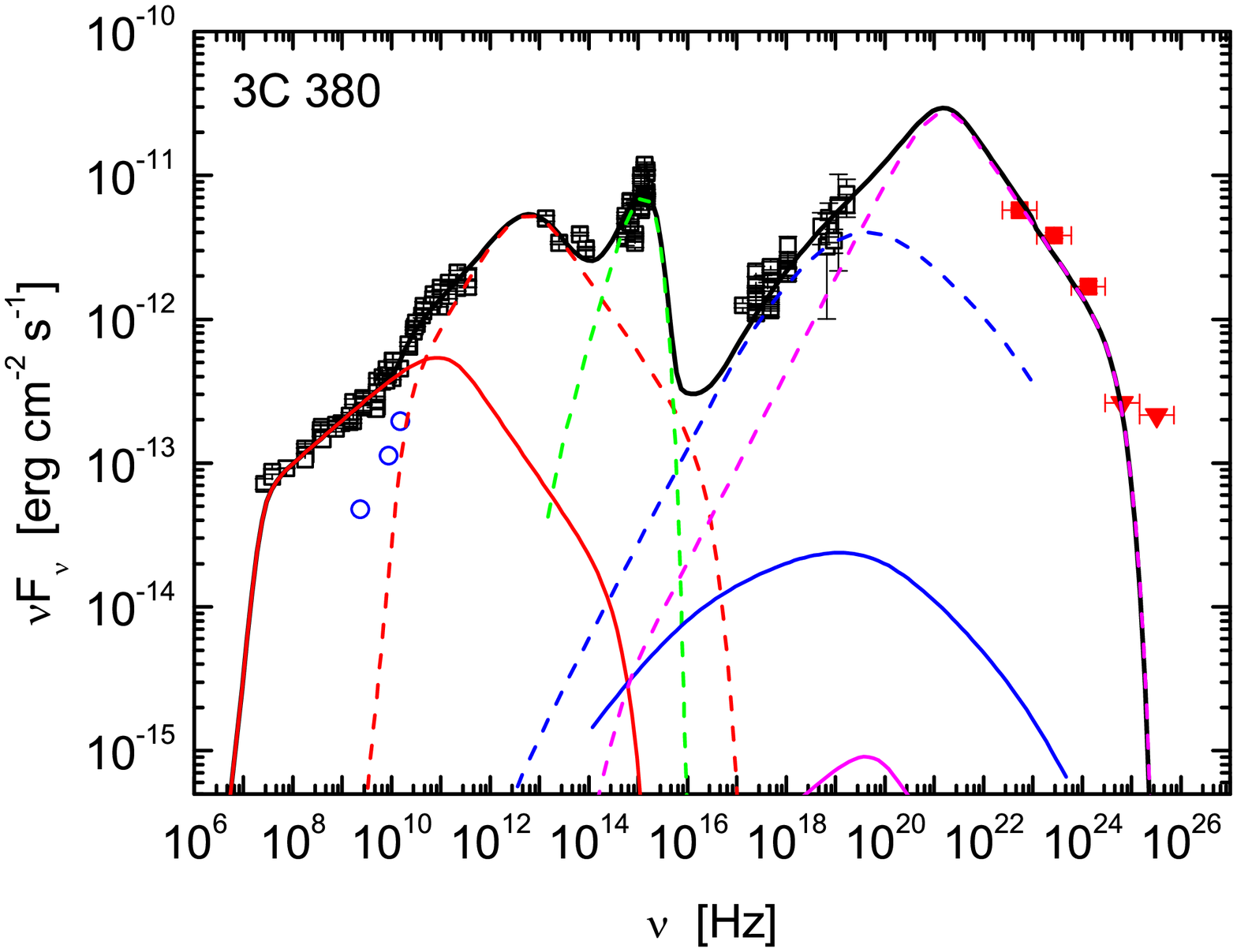}
   \includegraphics[angle=0,scale=0.25]{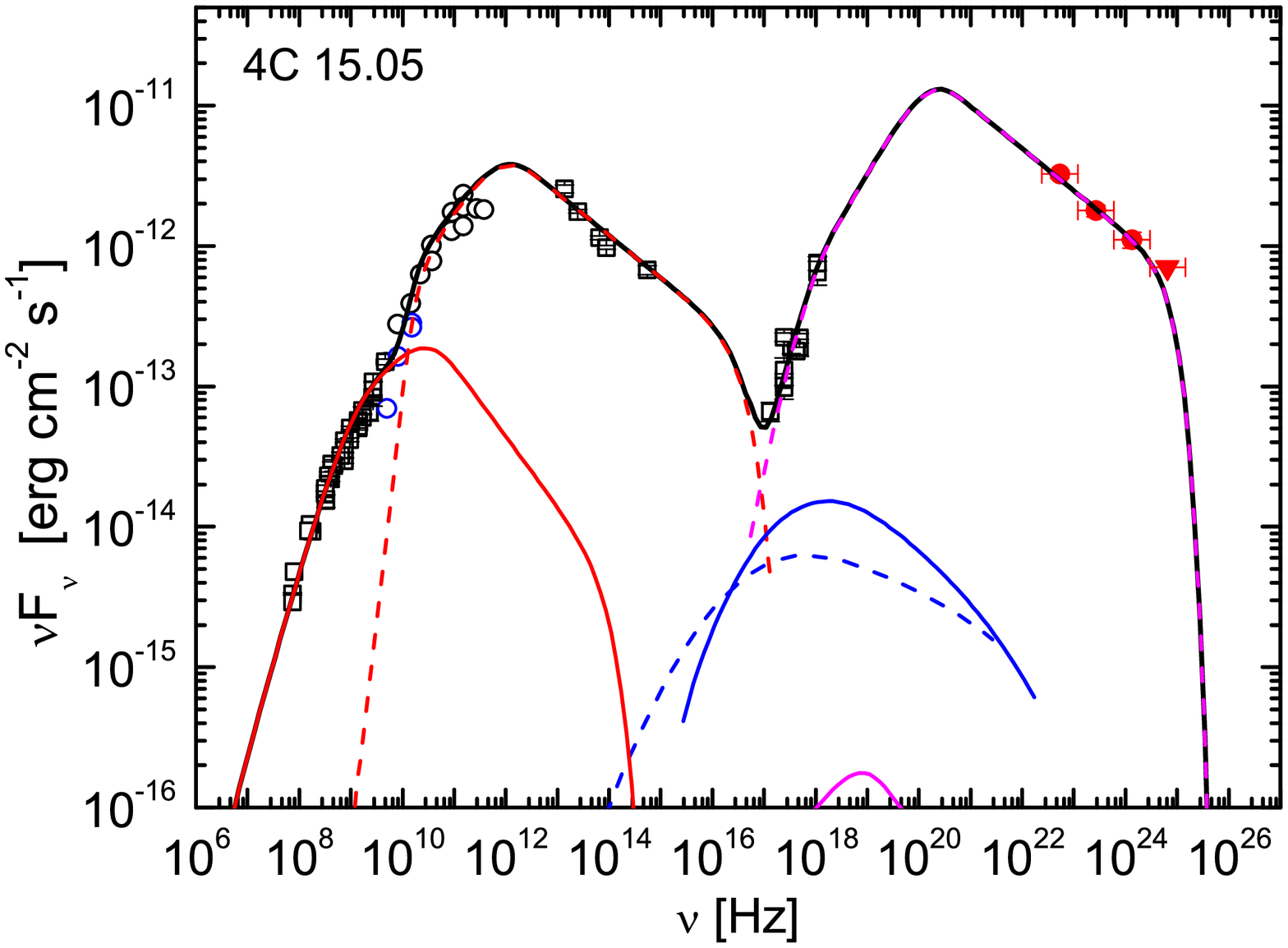}
\caption{Observed SEDs with model fitting. The data marked as opened black squares are taken from the ASI Science Data Center (ASDC). For 4C 15.05, the opened black circles in the radio band are from Bloom et al. (1994). The red solid symbols at the $\gamma$-ray band indicate the average spectrum of the \emph{Fermi}/LAT observations, where the down-triangles represent upper-limits. The opened blue circles in the panels of 3C 216, 3C 309.1, 3C 380, and 4C 15.05 represent the core fluxes, which are taken from the NASA/IPAC Extragalactic Database (NED). The opened blue, red, and green circles in the panel of 3C 138 respectively represent the fluxes from components A, B and C (Shen et al. 2005). The opened blue and red circles in the panel of 3C 286 respectively represent the fluxes from components C1 and C2 (An et al. 2017). The thick black solid lines are the sum of emission from each component; synchrotron radiation (red lines), accretion disk (green lines), SSC process (blue lines), and EC process (magenta lines), and among them the dashed lines and solid lines respectively represent the radiation from the core and the extended region.  }\label{SED}
\end{figure}

With the {\em Fermi}/LAT data and multi-wavelength data compiled from the literature and ASI Science Data Center (ASDC)\footnote{https://tools.ssdc.asi.it/SED/}, we establish the broadband SEDs of the six CSSs, as displayed in Figure \ref{SED}. The SEDs in $\nu>10$ GHz are apparently similar to the SEDs of FSRQs and NLS1s (e.g., Zhang et al. 2014; Sun et al. 2015). The variability in $\gamma$-ray band shown in Figure \ref{LC} and the significant position offset observations reported for 3C 216 (Paragi et al. 2000), 3C 138 (Shen et al. 2001), and 3C 309.1 (Ros \& Lobanov 2001), together with the apparent superluminal motions in some CSSs, likely indicate that the $\gamma$-rays would be from their compact core-jets, which still have the relativistic bulk motion. The radio emission in $\sim$0.01--10 GHz in the SEDs should not be dominated by the emission of the compact core because of a significant synchrotron-self-absorption effect on the radio emission from the core region. It would be attributed to the emission from the large-scale extended regions, similar to large-scale hotspots and knots (e.g., Zhang et al. 2010, 2018b). Therefore, we employ a two-zone leptonic radiation model to fit the constructed SEDs.

\subsection{The Compact Core Region}

The core region is assumed as a homogenous sphere with radius $R$, magnetic field strength $B$, and Doppler factor $\delta$, where $\delta=1/\Gamma(1-\beta\cos\theta)$, $\Gamma$ is the bulk Lorentz factor, $\theta$ is the viewing angle, and $\beta=1/\sqrt{1-\Gamma^{2}}$. The electron population distribution is taken as a broken power-law, i.e.,
\begin{equation}
N(\gamma )= N_{0}\left\{ \begin{array}{ll}
\gamma ^{-p_1}  &  \mbox{ $\gamma_{\rm min}\leq\gamma \leq \gamma _{\rm b}$}, \\
\gamma _{\rm b}^{p_2-p_1} \gamma ^{-p_2}  &  \mbox{ $\gamma _{\rm b} <\gamma <\gamma _{\rm max} $.}
\end{array}
\right.
\end{equation}

The synchrotron (syn), synchrotron-self-Compton (SSC), IC scattering of external field photons processes of the relativistic electrons are considered to model the broadband SEDs. The blue bump of the thermal emission from the accretion disk is prominent for 3C 138, 3C 286, 3C 309.1 and 3C 380, as shown in Figure \ref{SED}. We use the standard accretion disk spectrum (Davis \& Laor 2011) to explain this thermal emission (see also Zhang et al. 2015). The inner ($R_{\rm in}$) and outer ($R_{\rm out}$) radii and inclination to the line of sight ($i$) of the accretion disk are taken as $R_{\rm in}=R_{\rm s}$ (Krolik \& Hawley 2002)\footnote{The inner radiative edge of the accretion disk may be at
the marginally stable orbit radius and outside the Schwarzschild radius. However, $R_{\rm in}=4.5R_{\rm s}$ is taken for 3C 286 since the small $R_{\rm in}$ would result in very high disk luminosity and Eddington ratio.}, $R_{\rm out}=700R_{\rm s}$, and $\cos i=1$, where $R_{\rm s}$ is the Schwarzschild radius. The black hole mass ($M_{\rm BH}$) as listed in Table 3 is also fixed, and then we vary the Eddington ratio to model the emission from the accretion disk.

As displayed in Figure \ref{LC}, the $\gamma$-ray light curves in time bins of 180 days show slightly variability for the sources, and thus we use the time-scale of 180 days to constrain the size of the radiation region for the compact core. If $R=\delta c\Delta t/(1+z)\sim4.7\times10^{17}\delta/(1+z)$, where $\Delta t=180$ days, the energy dissipation region should be outside the broad-line regions (BLRs) of these CSSs. In this case, the photons from torus provide the seed photons for the IC process (IC/torus) of the relativistic electrons in the compact core. The energy density of the torus photon field in the comoving frame is $U^{'}_{\rm IR}=3\times10^{-4}\Gamma^2$ erg cm$^{-3}$ and the spectrum of the torus can be approximated by a blackbody with a peak in the comoving frame at $3\times10^{13}\Gamma$ Hz (e.g., Cleary et al. 2007; Kang et al. 2014).

Parameters $p_1$ and $p_2$ are fixed and derived from the spectral indices of the observed SEDs, i.e., in the radio (or X-ray) band and the GeV $\gamma$-ray (or IR-optical) band, which are obtained by fitting the observed data with a power-law function (see also Zhang et al. 2012, 2013). $\gamma_{\rm min}$ is fixed as $\gamma_{\rm min}=1$ or constrained with the observed SEDs via a method reported in Tavecchio et al. (2000). $\gamma_{\rm max}$ is also poorly constrained by the last observation point in the GeV energy band, or taken as 10$^5$. We fix the value of viewing angle and vary the value of $\Gamma$ (obtaining the corresponding $\delta$), $B$, $N_0$ and $\gamma_{\rm b}$ to fit the broadband SEDs of the core region.

\subsection{The Large-scale Extended Region}

The radio emission below $10^{10}$ Hz in SEDs should be dominated by the radiation of extended region at large-scale. The extended region is also assumed to be a homogenous sphere and the radius is roughly derived from the angular radius at the radio band as listed in Table 2. Even at the pc-scale, only small motion is detected for some CSSs by the VLBI observations. Hence we do not consider the relativistic effect of the extended regions in large-scale and assume $\delta=\Gamma=1$ during SED modeling. The electron distribution is also taken as Eq. (2). The cosmic microwave background (CMB) provides the seed photons of IC process (IC/CMB) and the CMB energy density in the comoving frame is $U^{'}_{\rm CMB}=\frac{4}{3}\Gamma^2(1+z)^4U_{\rm CMB}$ (Dermer \& Schlickeiser 1994; Georganopoulos et al. 2006), where $U_{\rm CMB}=4.2\times10^{-13}$ erg cm$^{-3}$. The syn+SSC+IC/CMB model under the equipartition condition ($U_{B}=U_{\rm e}$) is used to reproduce the radiation of extended region, where $U_{B}$ and $U_{\rm e}$ are the energy densities of magnetic fields and electrons.

$p_1$ is fixed and derived by fitting the radio spectrum with a power-law function. $p_2$ cannot be constrained and is fixed as $p_2=4$. $\gamma_{\rm min}$ is fixed as $\gamma_{\rm min}=100$ or is taken larger values to match the SEDs. $\gamma_{\rm max}$ is also poorly constrained and taken as $\gamma_{\rm max}=50\gamma_{\rm b}$. As illustrated in Figure \ref{SED}, it seems that there is a break around at $10^{10}$ Hz in the broadband SEDs of these CSSs, and the emission below this break may be dominated by the extended regions. We thus assume that the synchrotron radiation peak of the extended regions in large-scale is around $10^{10}$ Hz, which is used to constrain the values of $\gamma_{\rm b}$. We adjust the free parameters of $\gamma_{\rm b}$ and $N_0$ to fit the SEDs below $10^{10}$ Hz of the six CSSs.

\subsection{Results}

We fit the SEDs with the two-zone leptonic model by considering the Klein-Nishina effect and the absorption of high-energy $\gamma$-ray photons by extragalactic background light (Franceschini et al. 2008). Note that the observed SEDs are contributed by the radiations of both core and extended region. The model parameters are poorly constrained, and we therefore only search for the parameter sets that can represent the SEDs. The uncertainty of the parameters cannot be constrained in such analysis. The results are illustrated in Figure \ref{SED}. Under the equipartition condition and assuming $\delta=\Gamma=1$, the predicted IC fluxes by model from the extended regions are much lower than the \emph{Fermi}/LAT observation data, which implies $\gamma$-ray emission of these CSSs should be from the radiation of their compact cores. The model fitting parameters are shown in Tables 2 and 3.

\begin{figure}
 \centering
   \includegraphics[angle=0,scale=0.25]{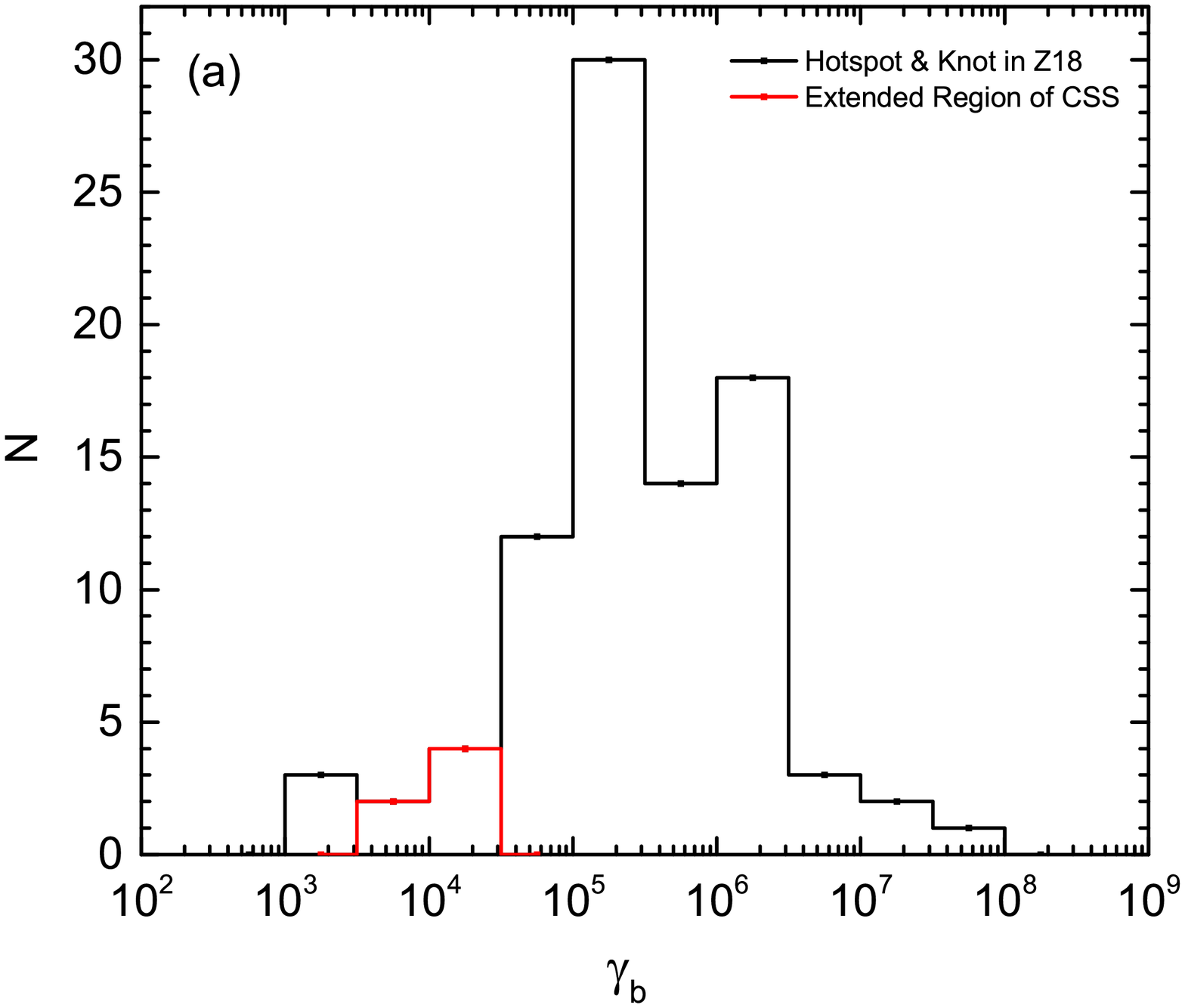}
   \includegraphics[angle=0,scale=0.25]{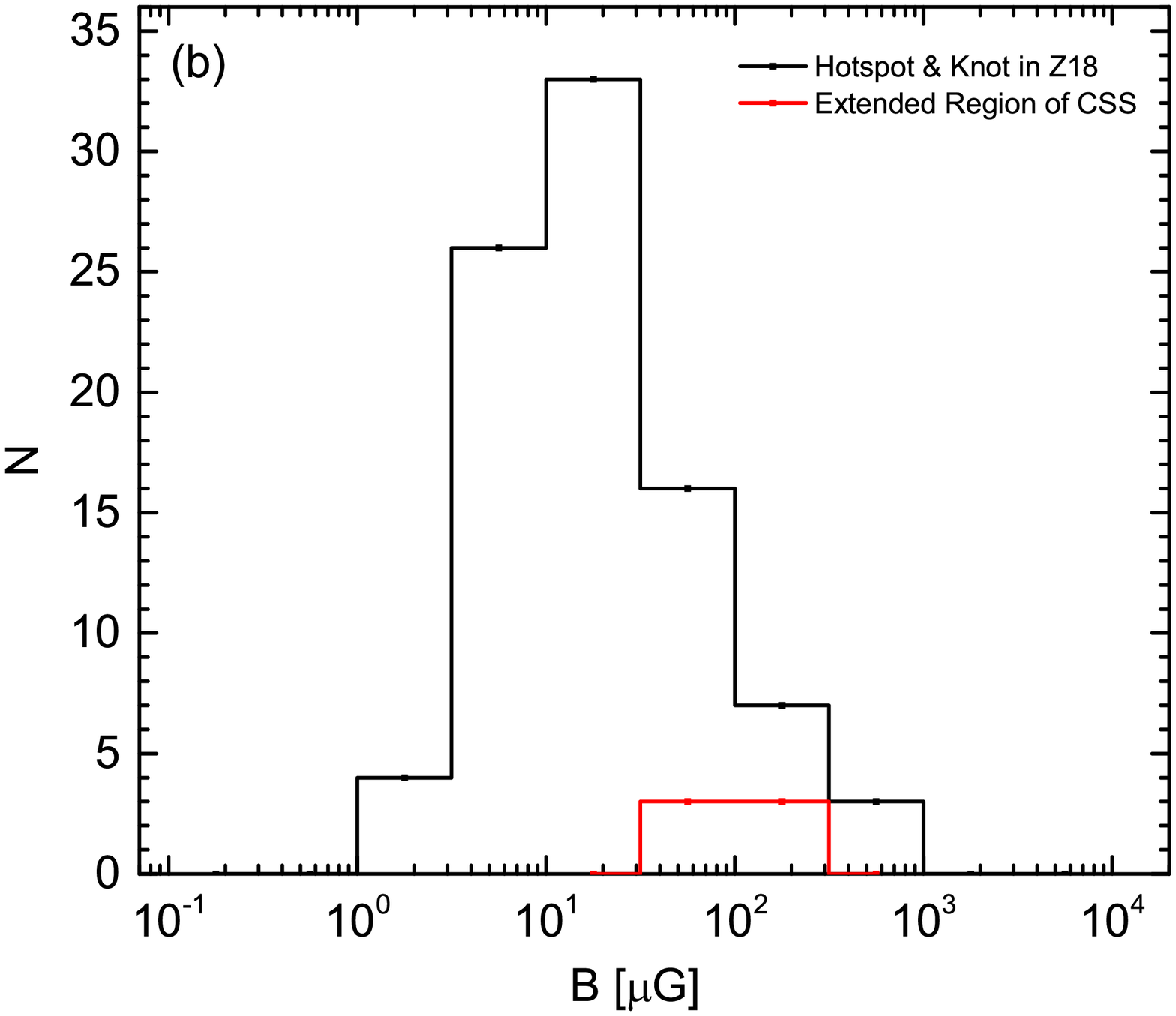}
   \includegraphics[angle=0,scale=0.25]{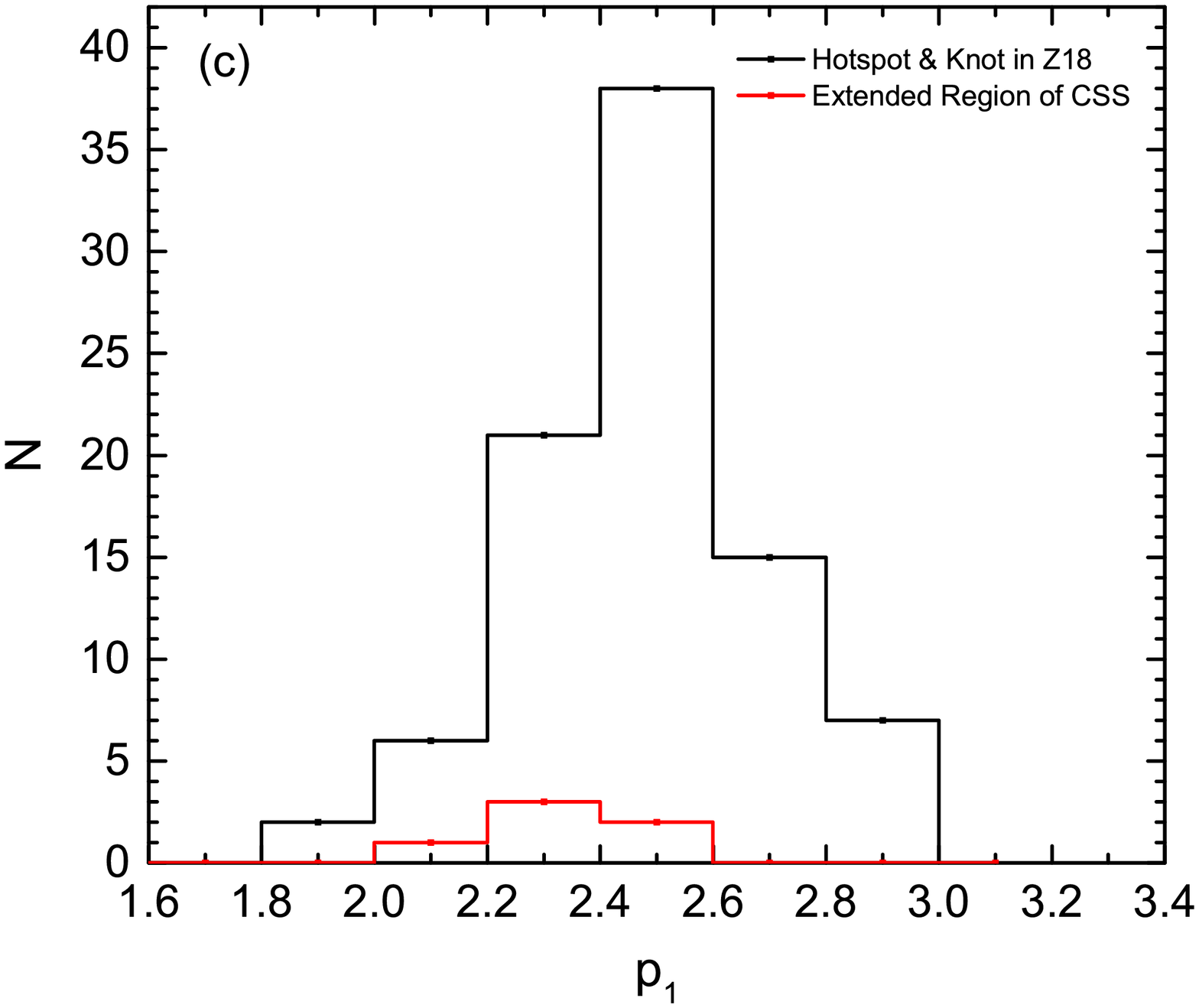}
\caption{Distributions of $\gamma_{\rm b}$, $B$, $p_1$ for the large-scale extended regions in the six CSSs (red lines). The large sample data of large-scale jet knots and hotspots (black lines) are taken from Zhang et al. (2018b).}\label{Para-LSJ}
\end{figure}

For the extended regions, the derived values of $\gamma_{\rm b}$, $B$, and $p_1$ are consistent with the values of substructures in large-scale jets, as displayed in Figure \ref{Para-LSJ}, where the data of these large-scale jet knots and hotspots are taken from Zhang et al. (2018b). The derived $\gamma_{\rm b}$ values of the large-scale extended regions in CSSs cluster at the lower end of the $\gamma_{\rm b}$ distribution. This is because we roughly take the large-scale extended region of CSS as a single-zone, which is not exactly like a knot or hotspot. The derived $\gamma_{\rm b}$ values are more similar to that in radio lobes (e.g., Takeuchi et al. 2012). $B$ of the large-scale extended regions in CSSs clusters at the large value end of the distribution, as displayed in Figure \ref{Para-LSJ}(b), but the $B$ distribution of the large-scale extended regions in CSSs is consistent with the distribution of knots and hotspots whose broadband SEDs can be well explained by the synchrotron radiation. The derived $p_1$ values of the large-scale extended regions in CSSs are consistent with that of those knots and hotspots, and are close or slightly larger than 2, which can be explained by the shock acceleration together with considering the cooling effect. This is consistent with the particle acceleration mechanisms in large-scale jets (e.g., Harris \& Krawczynski 2006).

\begin{figure}
 \centering
   \includegraphics[angle=0,scale=0.38]{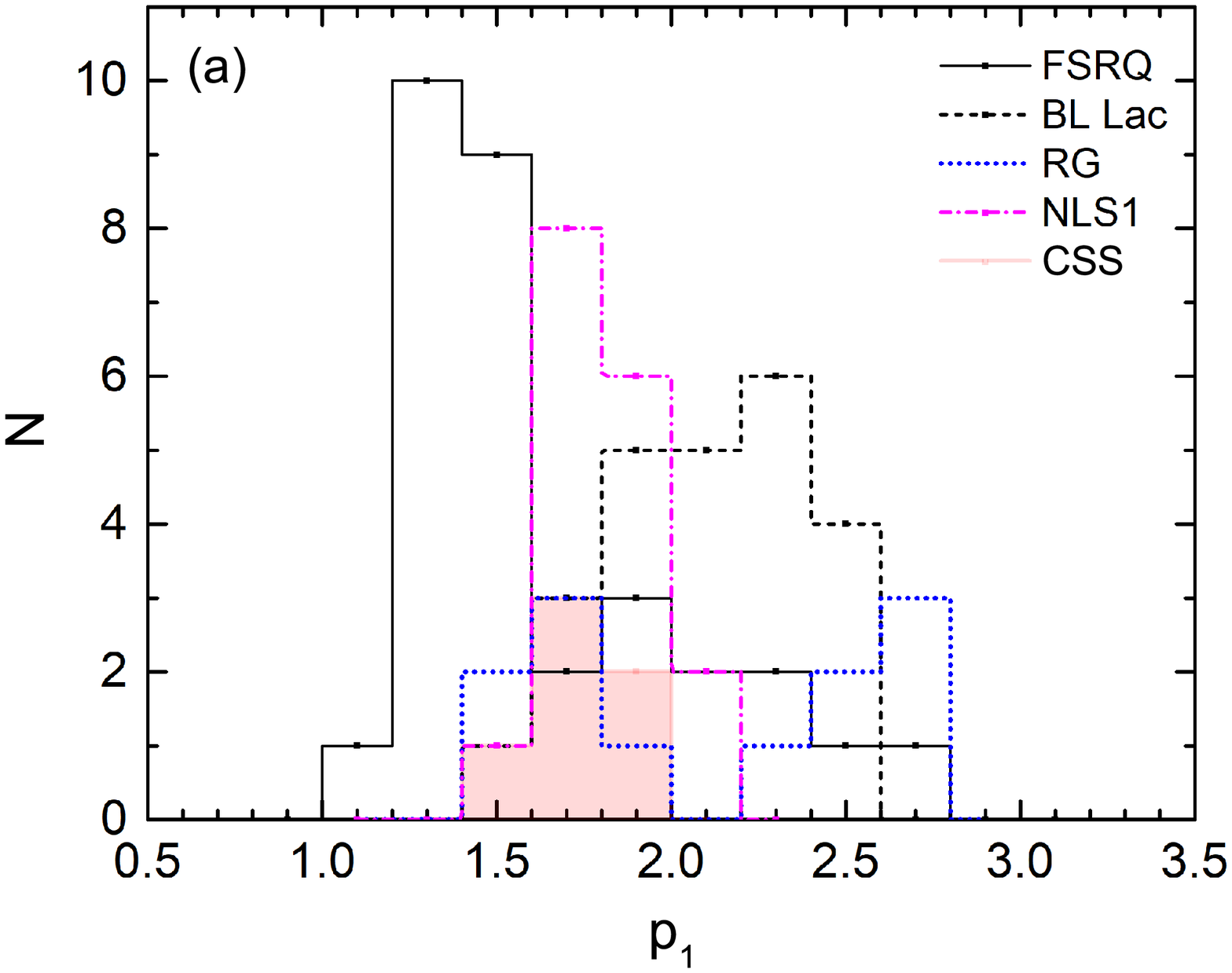}
   \includegraphics[angle=0,scale=0.38]{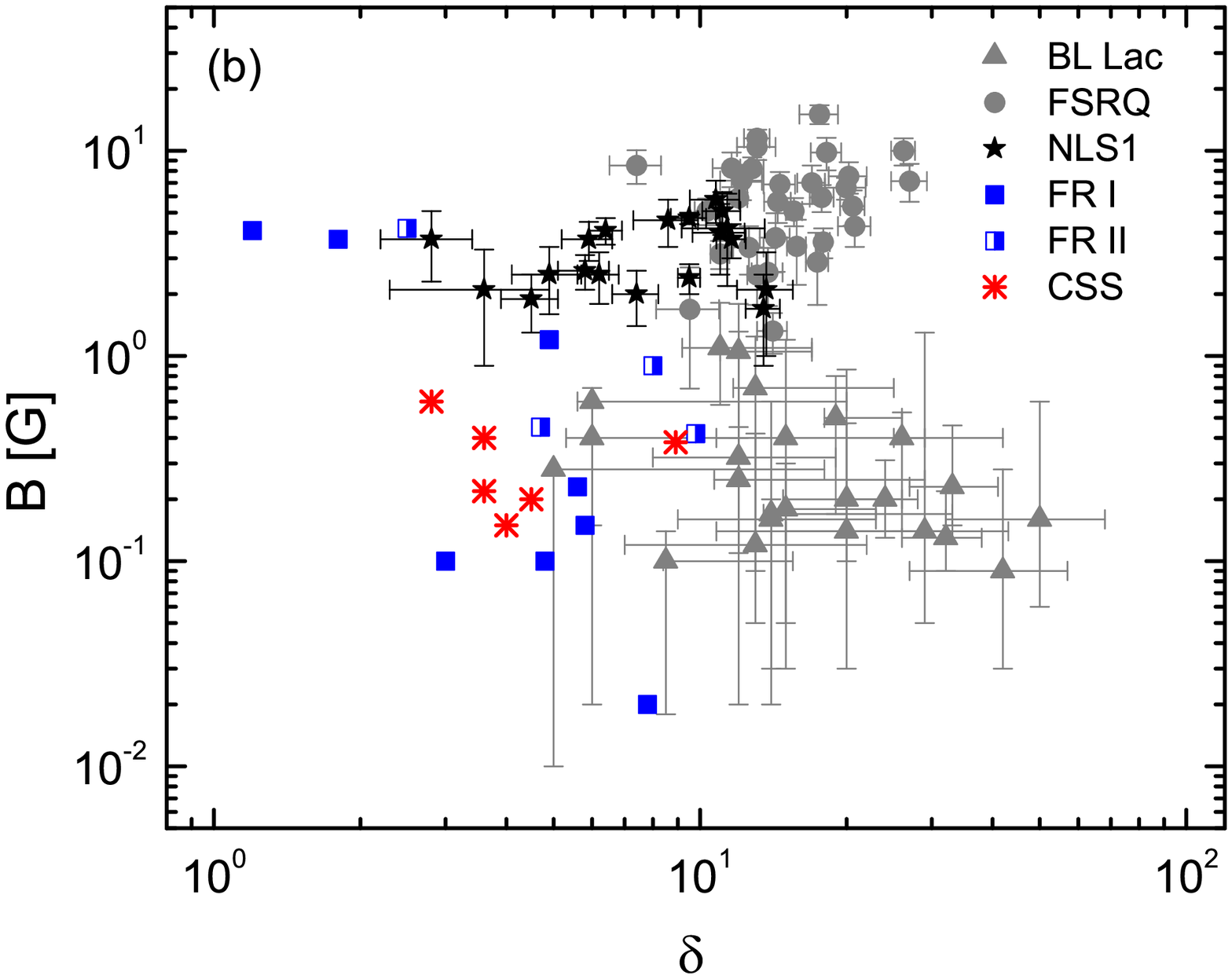}
   \includegraphics[angle=0,scale=0.38]{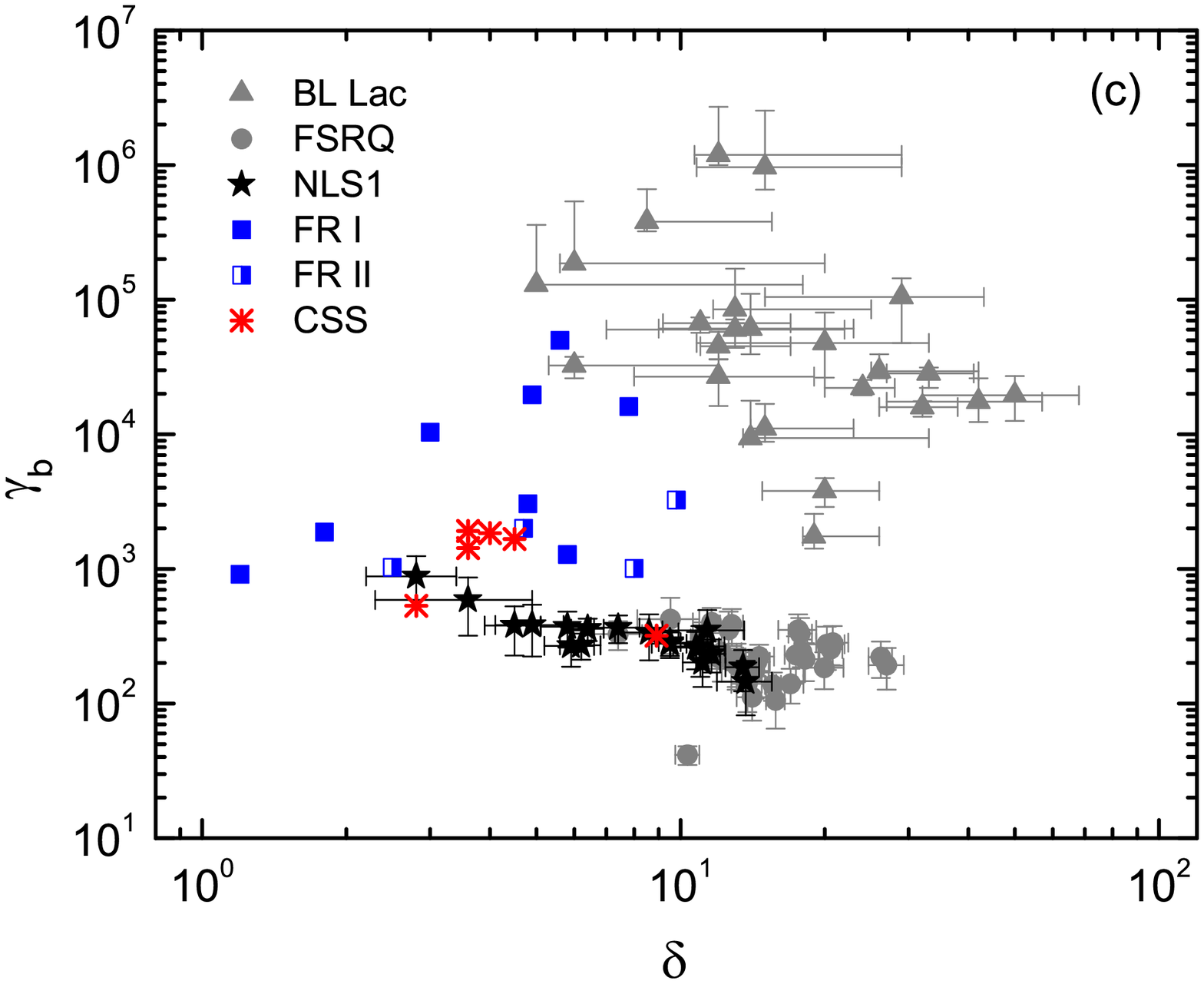}
   \includegraphics[angle=0,scale=0.38]{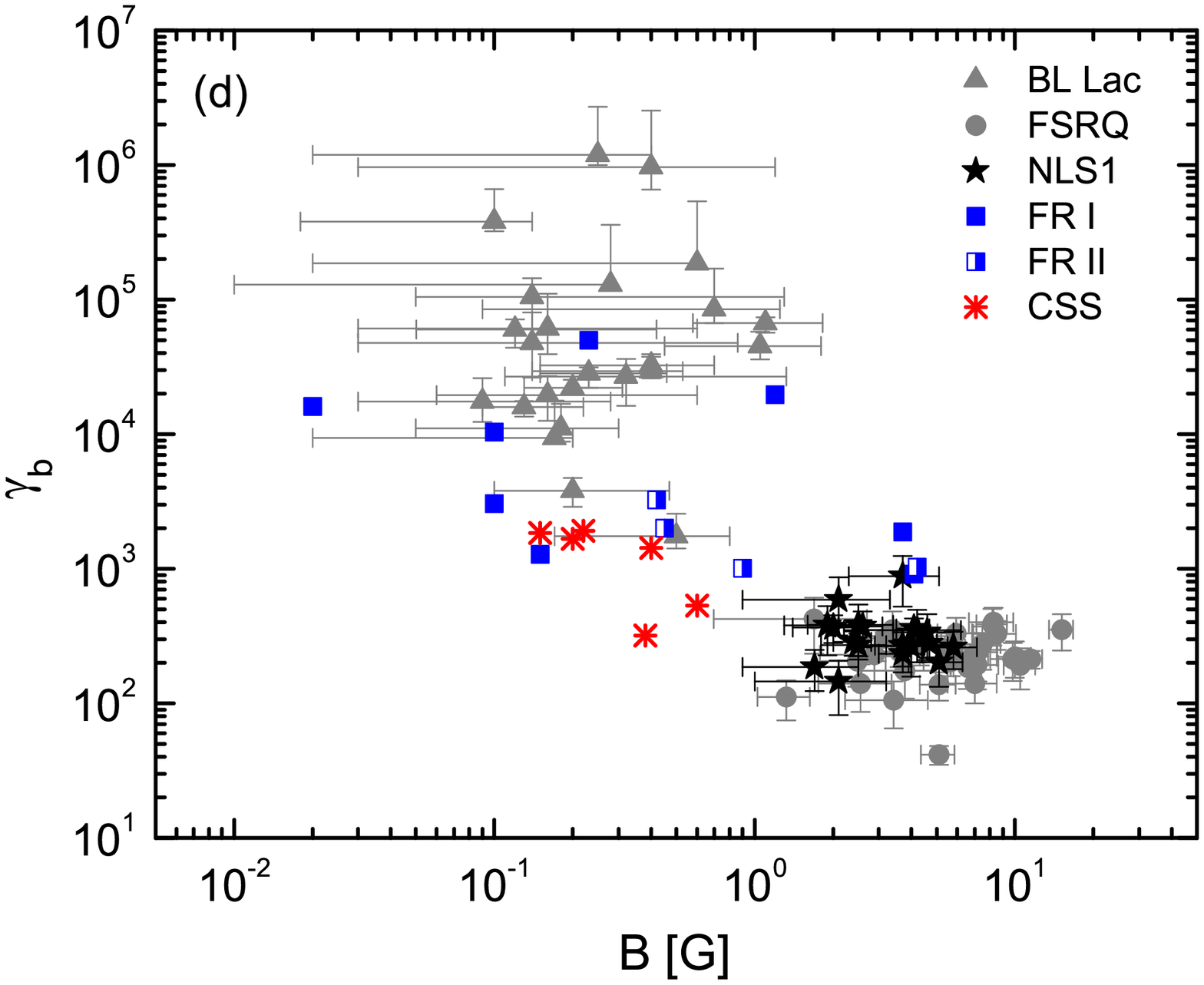}
\caption{Distributions of $p_1$ and correlations among $B$, $\delta$, and $\gamma_{\rm b}$ for these $\gamma$-ray emitting AGNs. $p_1$ of NLS1s are from Paliya et al. (2019). The more details of data for BL Lacs (Zhang et al. 2012), FSRQs (Zhang et al. 2014, 2015), NLS1s (Sun et al. 2015, Yang et al. 2018), and RGs (Xue et al. 2017) see Table 4.}\label{para-core}
\end{figure}

For the core regions, the derived Eddington ratio, i.e., $R_{\rm Edd}=L_{\rm disk}/L_{\rm Edd}$, where $L_{\rm Edd}$ and $L_{\rm disk}$ are the Eddington luminosity and accretion disk luminosity, respectively, ranges from 0.03 to 0.90, as listed in Table 3. $p_1$ ranges from 1.5 to 1.8; $\gamma_{\rm b}$ ranges within $\sim$320--2000; $B$ narrowly clusters at 0.15--0.60 G; $\delta$ ranges from 2.8 to 8.9 while $\Gamma$ narrowly clusters at 2.4--5.5. In order to compare the jet properties of core regions between these CSSs and other $\gamma$-ray emitting AGNs, we collect the jet parameter data together with $L_{\rm disk}$ and $M_{\rm BH}$ of other $\gamma$-ray emitting AGNs from the literature, as given in Table 4. As shown in Figure \ref{para-core}(a), the $p_1$ values of CSSs are smaller than the expected value of $p=2$ from the shock acceleration mechanism (e.g., Kirk et al. 2000; Achterberg et al. 2001; Virtanen \& Vainio 2005), which is similar to that of most FSRQs and NLS1s. Therefore, magnetic reconnection may be an effective process of particle acceleration in these jets, which can produce a flatter power-law particle spectrum (Zenitani \& Hoshino 2001; Sironi \& Spitkovsky 2014; Guo et al. 2015; Zhu et al. 2016). Correlations among $\gamma_{\rm b}$, $B$, and $\delta$ for these different kinds of $\gamma$-ray emitting AGNs are also illustrated in Figure \ref{para-core}. In the $\delta-B$ plane, the values of $\delta$ for CSSs are similar to these of RGs and slightly similar to these of NLS1s while $B$ of CSSs are similar to these of BL Lacs and RGs. In the $\delta-\gamma_{\rm b}$ plane, $\gamma_{\rm b}$ of CSSs are close to these of RGs. Hence, the values of $\gamma_{\rm b}$, $B$, and $\delta$ for CSSs are more closer to these of RGs than other $\gamma$-ray emitting AGNs. In the $B-\gamma_{\rm b}$ plane, no correlation between $B$ and $\gamma_{\rm b}$ is observed neither for BL Lacs nor FSRQs. However, it seems that there is an anti-correlation tendency between $B$ and $\gamma_{\rm b}$ for all the $\gamma$-ray emitting AGNs, RGs and CSSs are bridge between BL Lacs and FSRQs. The Pearson correlation analysis yields $r=-0.77$ and $p\sim0$.

GeV emission produced by the IC processes from CSS-like young AGNs was theoretically predicted (e.g., Stawarz et al. 2008; Migliori et al. 2014). Stawarz et al. (2008) and Migliori et al. (2014) suggested that the GeV emission of young sources may be dominated by the IC scattering of surrounding photon fields from the torus and/or UV-disk by the relativistic electrons in lobes/knots, which would be detectable with \emph{Fermi}/LAT if the lobes/knots are located at tens of pc from the central BH and the injected power is around $10^{46-47}$ erg s$^{-1}$. The IC processes of the accelerated electrons in shocked shells can also produce the detectable $\gamma$-ray emission with the \emph{Fermi}/LAT if the source is small, but for larger sources, producing the same predicted $\gamma$-ray fluxes requires a higher jet power and synchrotron emission dominating over IC emission (Ito et al. 2011). The IC/torus process of relativistic electrons in the mini shells on pc-scale may also produce VHE radiations, which may be detectable with the Cherenkov Telescope Array (Kino et al. 2013). By analyzing 4-yr \emph{Fermi}/LAT data, Migliori et al. (2014) did not find any statistically significant detections for a young AGN sample (including 3C 309.1) and presented the compatible upper limits with \emph{Fermi}/LAT flux threshold. Our analysis of $\sim$11-yr \emph{Fermi}/LAT data yields clear detections of CSSs in our sample. Our SED fitting results are broadly consistent with these previous analysis.

\section{Jet Power--Eddington Ratio Correlation and Unification of Jet Radiation for AGN Subclasses}

Jet power is essential to understand the jet physics. The conventional assumption is that the jet power is carried by electrons and protons with a one-to-one ratio,
magnetic fields, and radiation (e.g., Ghisellini et al. 2011, 2014; Zhang et al. 2012, 2014, 2015; Chen 2017, 2018). The energy density of protons cannot be constrained by observations, and using the one-to-one ratio of electron to proton to calculate the proton power seriously depends on the minimum energy of electrons, which is generally poor constrained for most AGNs (e.g., Zhang et al. 2014, 2015). Hence, we just consider the case of the electron--positron pair jet in this work. Based on the fitting parameters of core regions, we calculate the powers of the non-thermal electrons ($P_{\rm e}$) and magnetic fields ($P_{B}$) using $P_{\rm i}=\pi R^2\Gamma^2cU_{\rm i}$, where $U_{\rm i}$ could be the energy density of electrons or magnetic fields, and is measured in the comoving frame. The radiation power ($P_{\rm r}$) is estimated with the bolometric luminosity ($L_{\rm bol}$, the non-thermal radiation of compact core), i.e., $P_{\rm r}=\pi R^2\Gamma^2cU_{\rm r}=L_{\rm bol}\Gamma^2/4\delta^4$. The derived values of $P_{\rm e}$, $P_{B}$, $P_{\rm r}$, and $P^{e^{\pm}}_{\rm jet}=P_{\rm r}+P_{\rm e}+P_{B}$ for the six CSSs are given in Table 3.

\begin{figure}
 \centering
   \includegraphics[angle=0,scale=0.33]{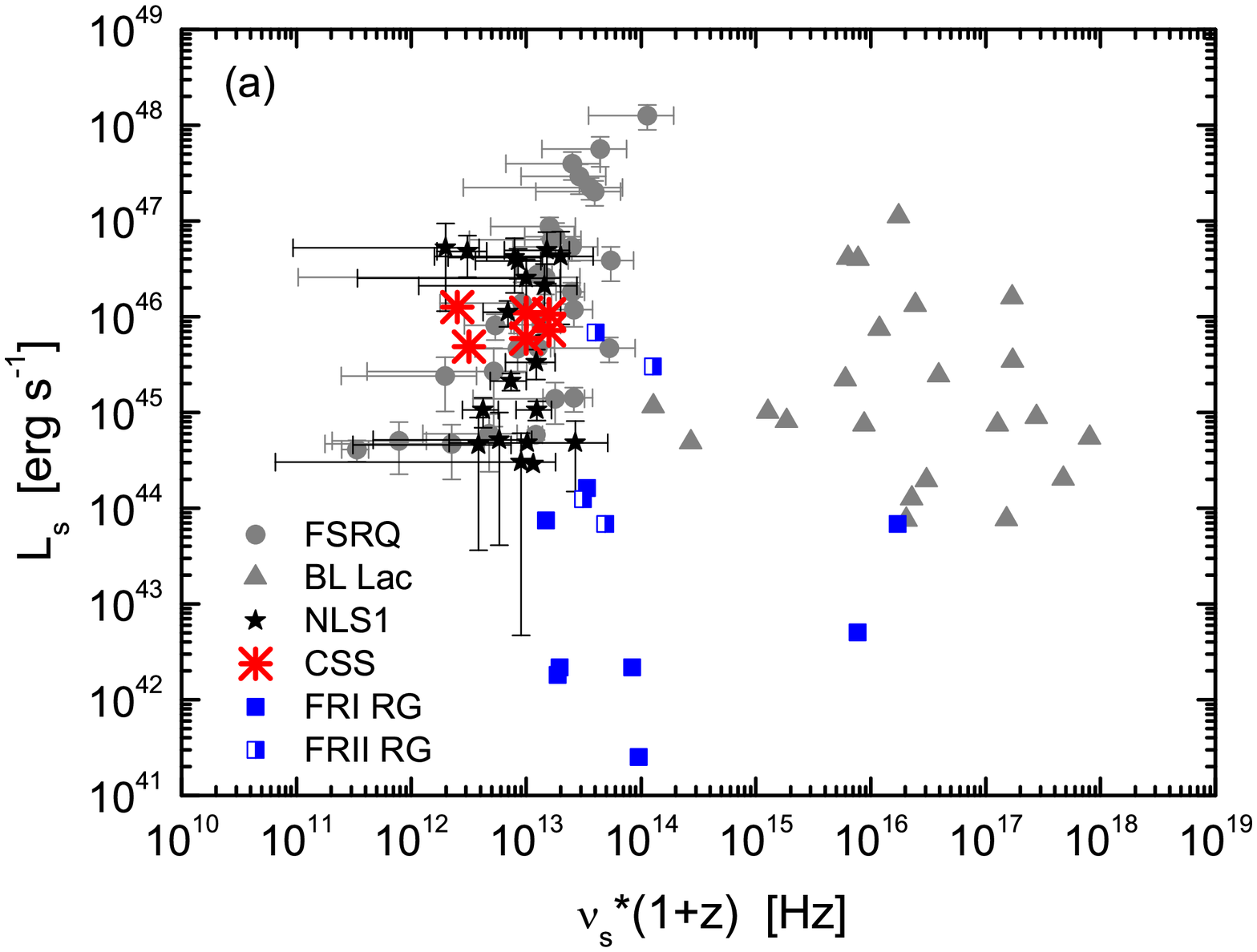}
   \includegraphics[angle=0,scale=0.33]{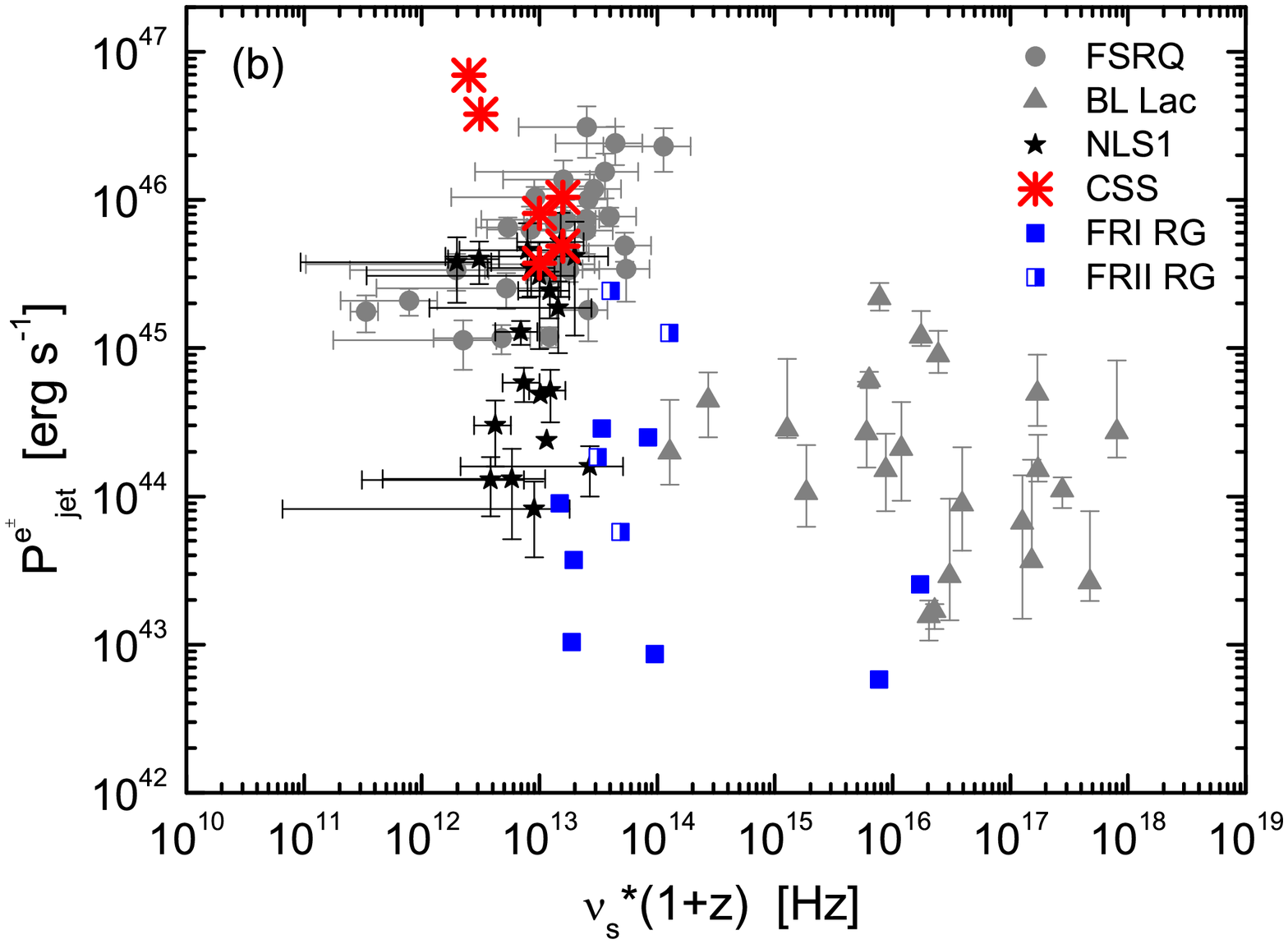}
\caption{The peak luminosity ($L_{\rm s}$) of synchrotron radiation and the jet power ($P^{e^{\pm}}_{\rm jet}$) as a function of the peak frequency ($\nu_{\rm s}$) of synchrotron radiation for these $\gamma$-ray emitting AGNs. The details of data see Tables 3 and 4.}\label{nus-Ls}
\end{figure}

We plot the peak luminosity of synchrotron radiation ($L_{\rm s}$) against the peak frequency of synchrotron radiation ($\nu_{\rm s}$) for these $\gamma$-ray emitting AGNs, as displayed in Figure \ref{nus-Ls}(a). No trend associated with the ``blazar sequence" is observed. CSSs cluster at within the distribution area of FSRQs and their $\nu_{\rm s}$ and $L_{\rm s}$ distribute within a narrow range. Replacing $L_{\rm s}$ with $P^{e^{\pm}}_{\rm jet}$ of sources, an anti-correlated tendency is presented for all the data points with $r=-0.54$ and $p=4.8\times10^{-8}$, as shown in Figure \ref{nus-Ls}(b). If we only consider the data of CSSs and blazars, the anticorrelations between them become stronger, i.e., $r=-0.77$ and $p=2.2\times10^{-12}$. This result may imply that the ``sequence" behavior of $\gamma$-ray emitting AGNs is dominated by the jet power (see also Xue et al. 2017; Fan \& Wu 2019). For these $\gamma$-ray emitting AGNs, $\nu_{\rm s}$ is related to jet radiation processes and thus should be correlated with the jet power.

\begin{figure}
 \centering
   \includegraphics[angle=0,scale=0.38]{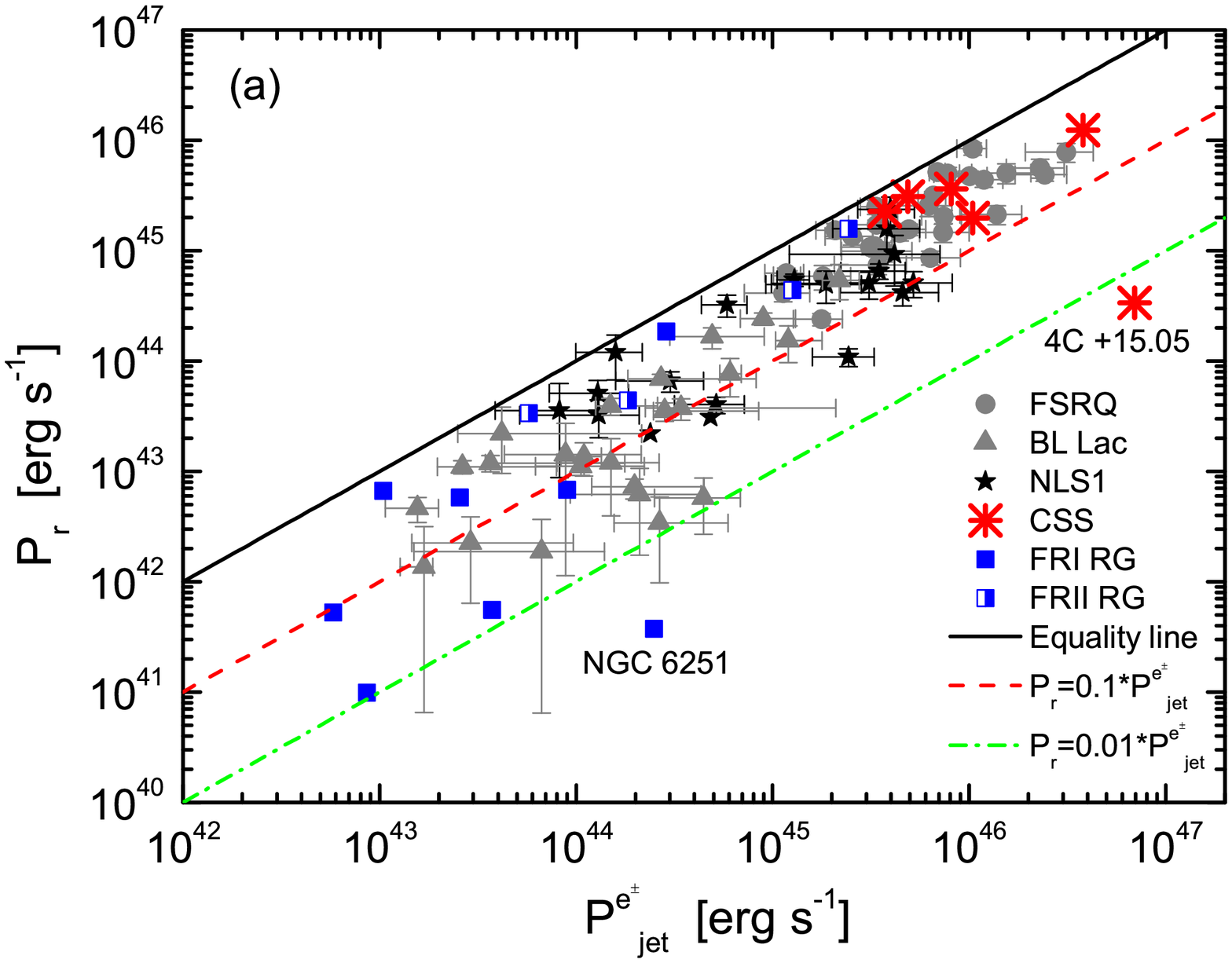}
   \includegraphics[angle=0,scale=0.38]{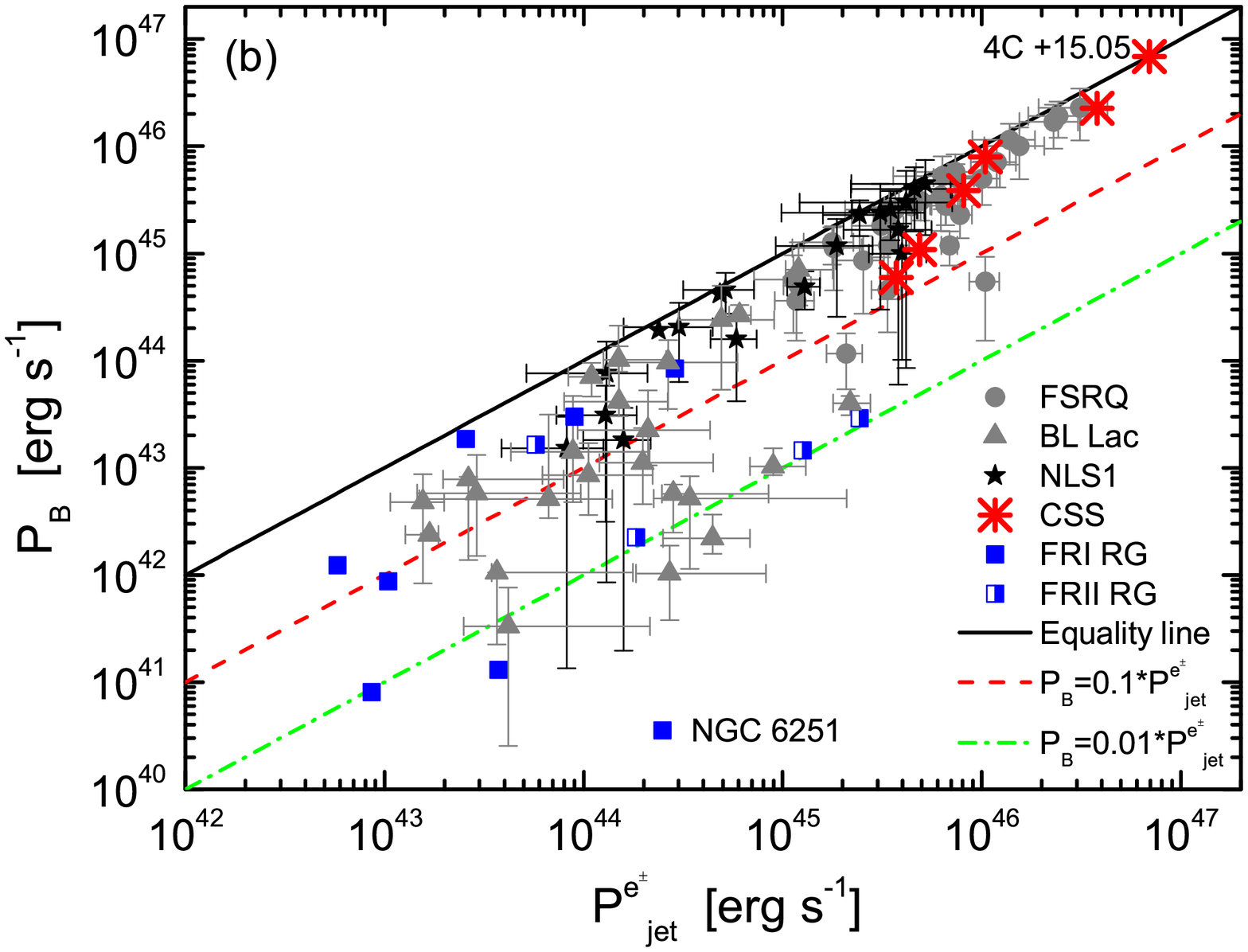}
\caption{$P_{\rm r}$ and $P_{B}$ as a function of $P^{e^{\pm}}_{\rm jet}$ for these $\gamma$-ray emitting AGNs. The linear fit in the log scale gives $\log P_{\rm r}=-(6.61\pm2.54)+(1.13\pm0.06)\log P^{e^{\pm}}_{\rm jet}$. The details of data see Tables 3 and 4.}\label{Pr-Pjet}
\end{figure}

In order to investigate the radiation efficiency and the magnetization of jet, we plot $P_{\rm r}$ and $P_{B}$ as a function of $P^{e^{\pm}}_{\rm jet}$ for these $\gamma$-ray emitting AGNs in Figure \ref{Pr-Pjet}. $P_{\rm r}$ is strongly correlated with $P^{e^{\pm}}_{\rm jet}$, and the Pearson correlation analysis gives $r=0.91$ and $p\sim0$. The linear fit in the log scale gives $\log P_{\rm r}=-(6.61\pm2.54)+(1.13\pm0.06)\log P^{e^{\pm}}_{\rm jet}$. Except for 4C 15.05, the other CSSs together with FSRQs are located at the high power end with high jet radiation efficiency ($P_{\rm r}$/$P^{e^{\pm}}_{\rm jet}$). Only 4C 15.05 and RG NGC 6251 have very low jet radiation efficiency ($<0.01$), even lower than that of some BL Lacs. Due to the lack of emission lines, the redshift of 4C 15.05 is still debated. As shown in Figure \ref{SED}, the two peak frequencies of its broadband SED could be well constrained, i.e., the synchrotron peak frequency ($\nu_{\rm s}\sim10^{12}$ Hz) and the IC peak frequency ($\nu_{\rm c}\sim10^{20}$ Hz). If the IC peak is produced by SSC process, it would give $B\delta\sim\frac{\nu_{\rm s}^2}{2.8\times10^6\nu_{\rm c}}\sim0.005$ (Tavecchio et al. 1998), even though $\delta=1$, then $B\sim0.005$ G. So the SED of 4C 15.05 cannot be well explained by the single-zone syn+SSC model as most of BL Lacs and RGs. As listed in Table 3, the jet power of 4C 15.05 is dominated by $P_{B}$. In this respect, it is similar to FSRQs.

$P_{B}$ is also correlated with $P^{e^{\pm}}_{\rm jet}$ for these $\gamma$-ray emitting AGNs, as presented in Figure \ref{Pr-Pjet}(b). The Pearson correlation analysis gives $r=0.88$ and $p\sim0$. We observe that there is a larger dispersion in the low power end for the $P_{B}$--$P^{e^{\pm}}_{\rm jet}$ relation. If only considering the data of CSSs, FSRQs, and NLS1s, the correlation between $P_{B}$ and $P^{e^{\pm}}_{\rm jet}$ becomes stronger with $r=0.93$ and $p\sim0$. On average the three kinds of $\gamma$-ray emitting AGNs have the larger $P_{B}$/$P^{e^{\pm}}_{\rm jet}$ than BL Lacs and RGs. As reported by Zhang et al. (2015), FSRQ jets and BL Lac jets have the different composition and radiation efficiency, and the jet properties of NLS1s are intermediate between them, but more analogous to FSRQ jets. It seems likely that CSS jets may be also high magnetized with high radiation efficiency.

\begin{figure}
   \includegraphics[angle=0,scale=0.4]{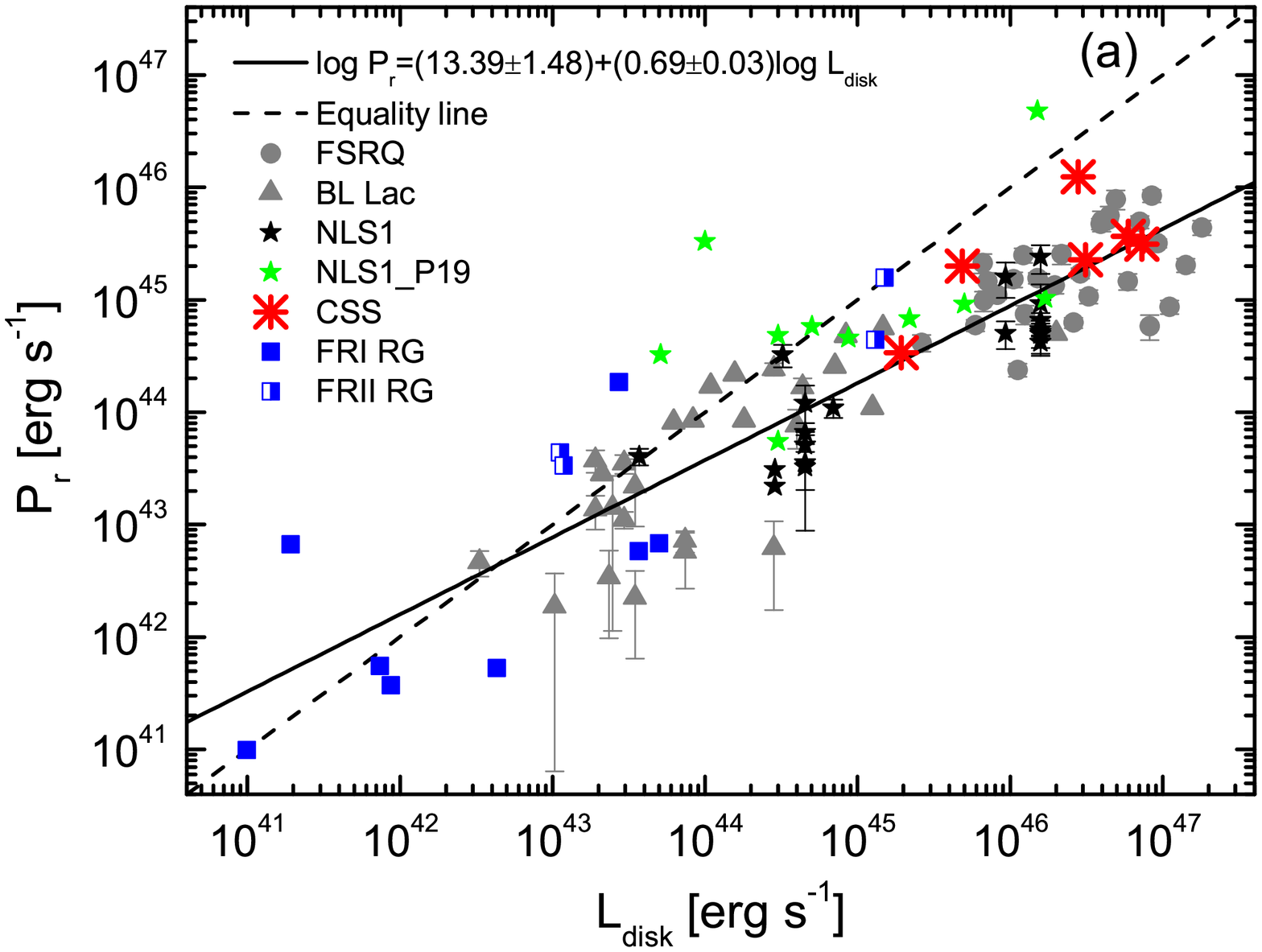}
   \includegraphics[angle=0,scale=0.4]{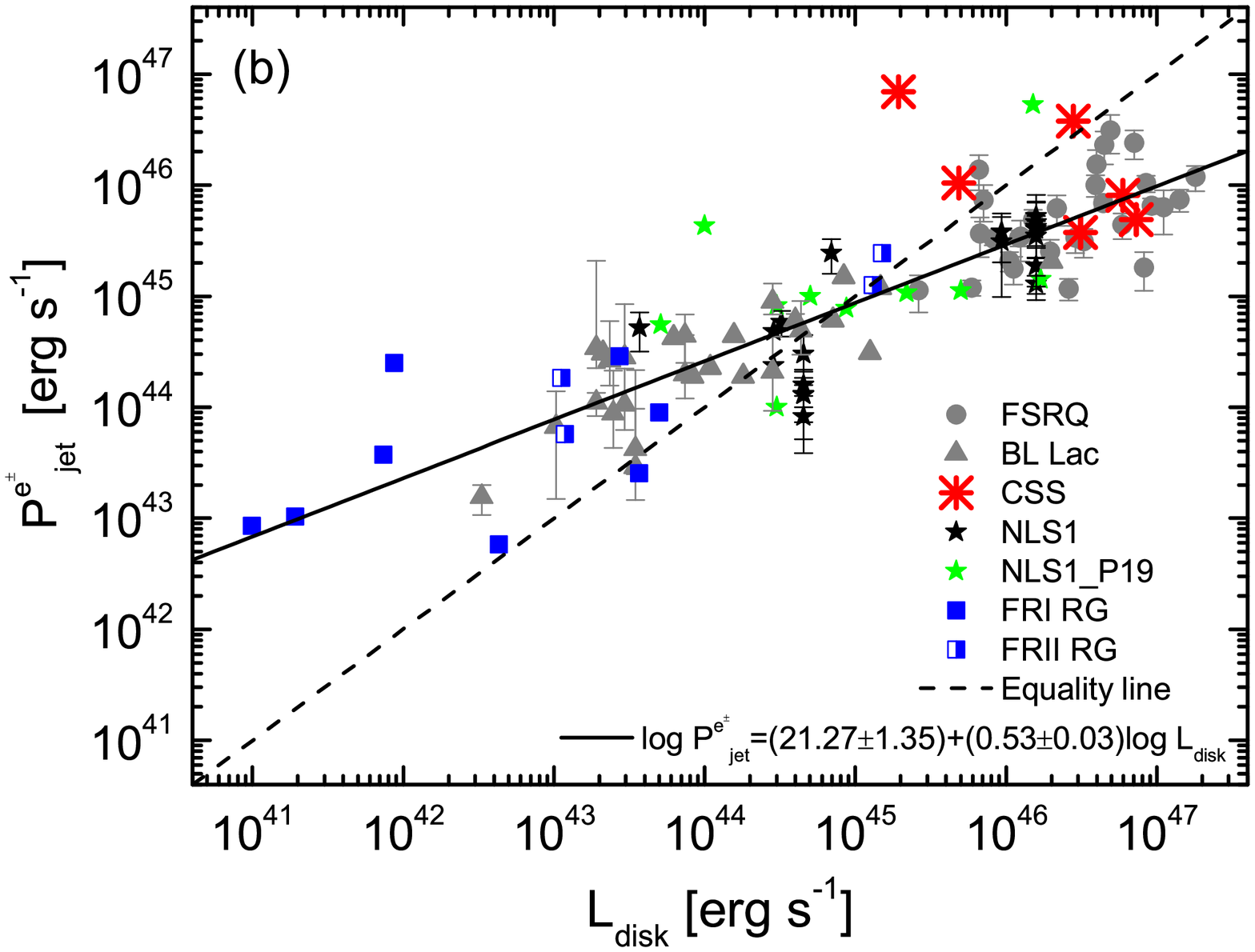}
   \includegraphics[angle=0,scale=0.4]{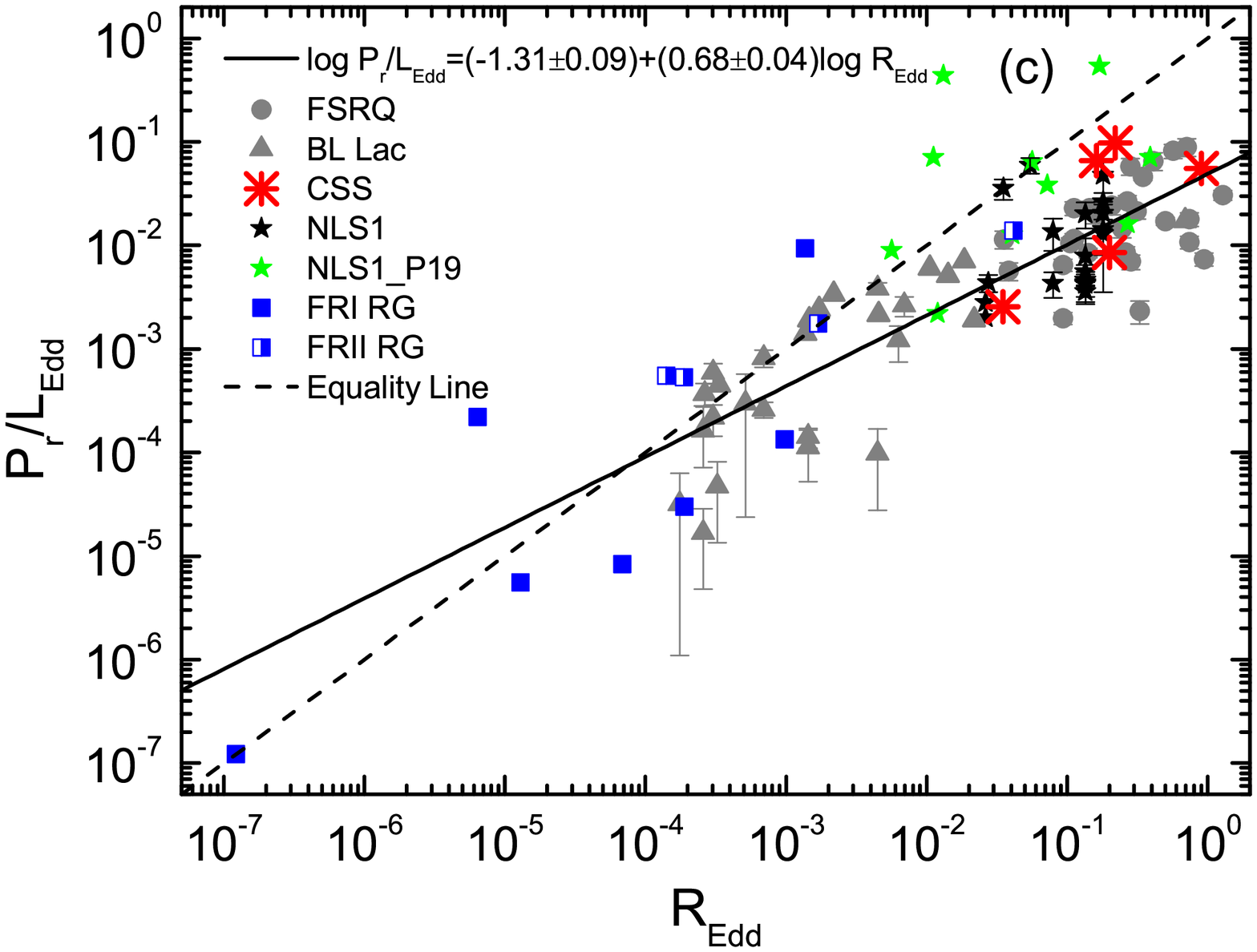}
   \includegraphics[angle=0,scale=0.4]{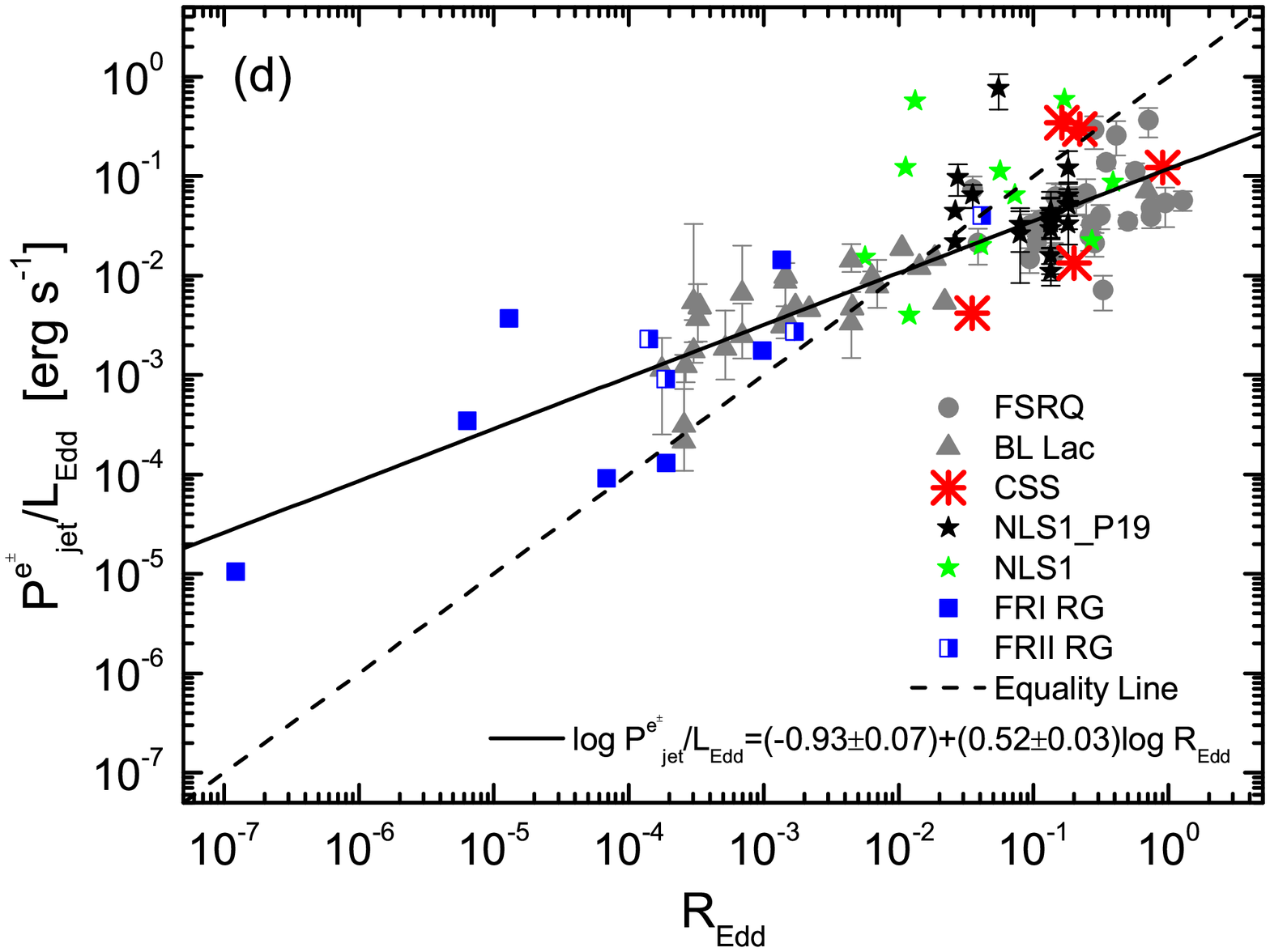}
\caption{$P_{\rm r}$ and $P^{e^{\pm}}_{\rm jet}$ as a function of $L_{\rm disk}$ (\emph{top-panels}) together with their relations in units of Eddington luminosity (\emph{bottom-panels}) for these $\gamma$-ray emitting AGNs. The solid lines are the linear regression fits for all the sources (except for the green stars). The details of data see Tables 3 and 4.}\label{R_edd-Pr}
\end{figure}

The emission of these $\gamma$-ray emitting AGNs is dominated by the jet radiation. In order to investigate the relation between jet and central engine, we plot $P_{\rm r}$ and $P^{e^{\pm}}_{\rm jet}$ as functions of $L_{\rm disk}$ and $M_{\rm BH}$, together with their relations in units of Eddington luminosity\footnote{The NLS1 data (green stars) from Paliya et al. (2019) are not included in correlation analysis and linear fits since $P_{\rm r}$ is estimated by slightly different method.}. In the $L_{\rm disk}$--$P_{\rm r}$ plane (Figure \ref{R_edd-Pr}(a)), the distributions of CSSs roughly overlap with these of FSRQs. Note that these $\gamma$-ray emitting AGNs form a sequence spanning six orders of magnitude in the $L_{\rm disk}$--$P_{\rm r}$ plane, and the similar feature has been observed for BL Lacs, NLS1s, and FSRQs in Zhang et al. (2015), but here FR I RGs further extend this sequence to the low power end. Except for several BL Lacs and RGs, and one NLS1, $P_{\rm r}$ of all the other sources are lower than their $L_{\rm disk}$. The Pearson correlation analysis yields $r=0.91$ and $p\sim0$. The linear fit in the log scale gives $\log P_{\rm r}=(13.49\pm1.49)+(0.68\pm0.03)\log L_{\rm disk}$, as shown in Figure \ref{R_edd-Pr}(a). The derived slope is flatter than that in Zhang et al. (2015), but is steeper than that in Paliya et al. (2017). However, only blazars and NLS1s are considered in Zhang et al. (2015) and both $\gamma$-ray loud and $\gamma$-ray quiet blazars are included in Paliya et al. (2017). Similar feature is also observed in the $L_{\rm disk}$--$P^{e^{\pm}}_{\rm jet}$ plane (Figure \ref{R_edd-Pr}(b)). The linear fit in the log scale yields $\log P^{e^{\pm}}_{\rm jet}=(21.32\pm1.35)+(0.53\pm0.03)\log L_{\rm disk}$ with $r=0.88$ and $p\sim0$. In this scenario, most of BL Lacs and RGs together with three CSSs have $P^{e^{\pm}}_{\rm jet}$ larger than $L_{\rm disk}$, and on average $P^{e^{\pm}}_{\rm jet}$ of CSSs is higher than that of other AGNs. The strong correlations are also observed when they are in units of Eddington luminosity, as illustrated in Figures \ref{R_edd-Pr}(c) and \ref{R_edd-Pr}(d). The linear fits in the log scale give $\log P_{\rm r}/L_{\rm Edd}=(-1.32\pm0.09)+(0.68\pm0.04)\log R_{\rm Edd}$ with $r=0.89$ and $p\sim0$ and $\log P^{e^{\pm}}_{\rm jet}/L_{\rm Edd}=(-0.94\pm0.07)+(0.52\pm0.03)\log R_{\rm Edd}$ with $r=0.87$ and $p\sim0$, respectively.

\begin{figure}
   \includegraphics[angle=0,scale=0.4]{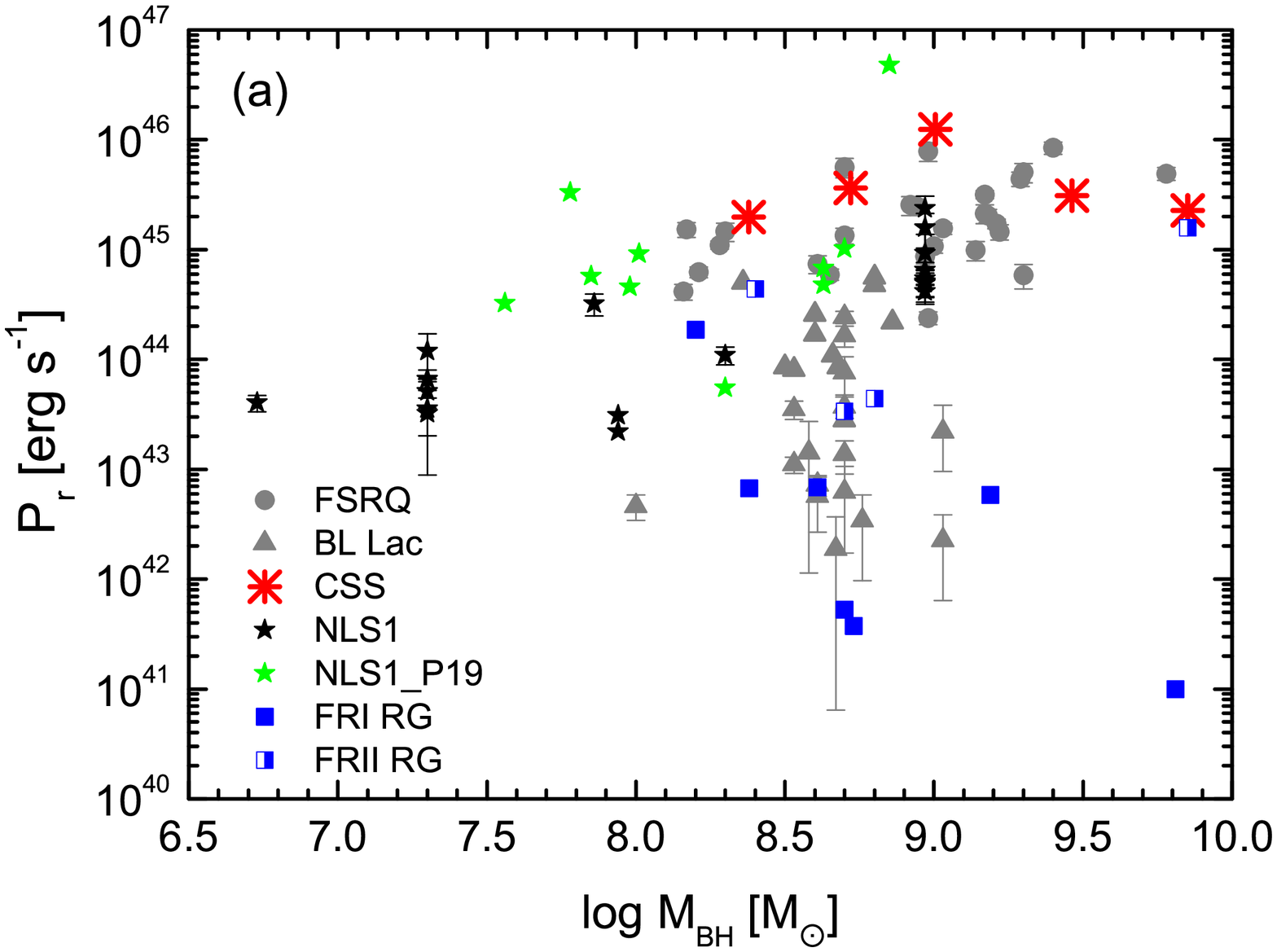}
   \includegraphics[angle=0,scale=0.4]{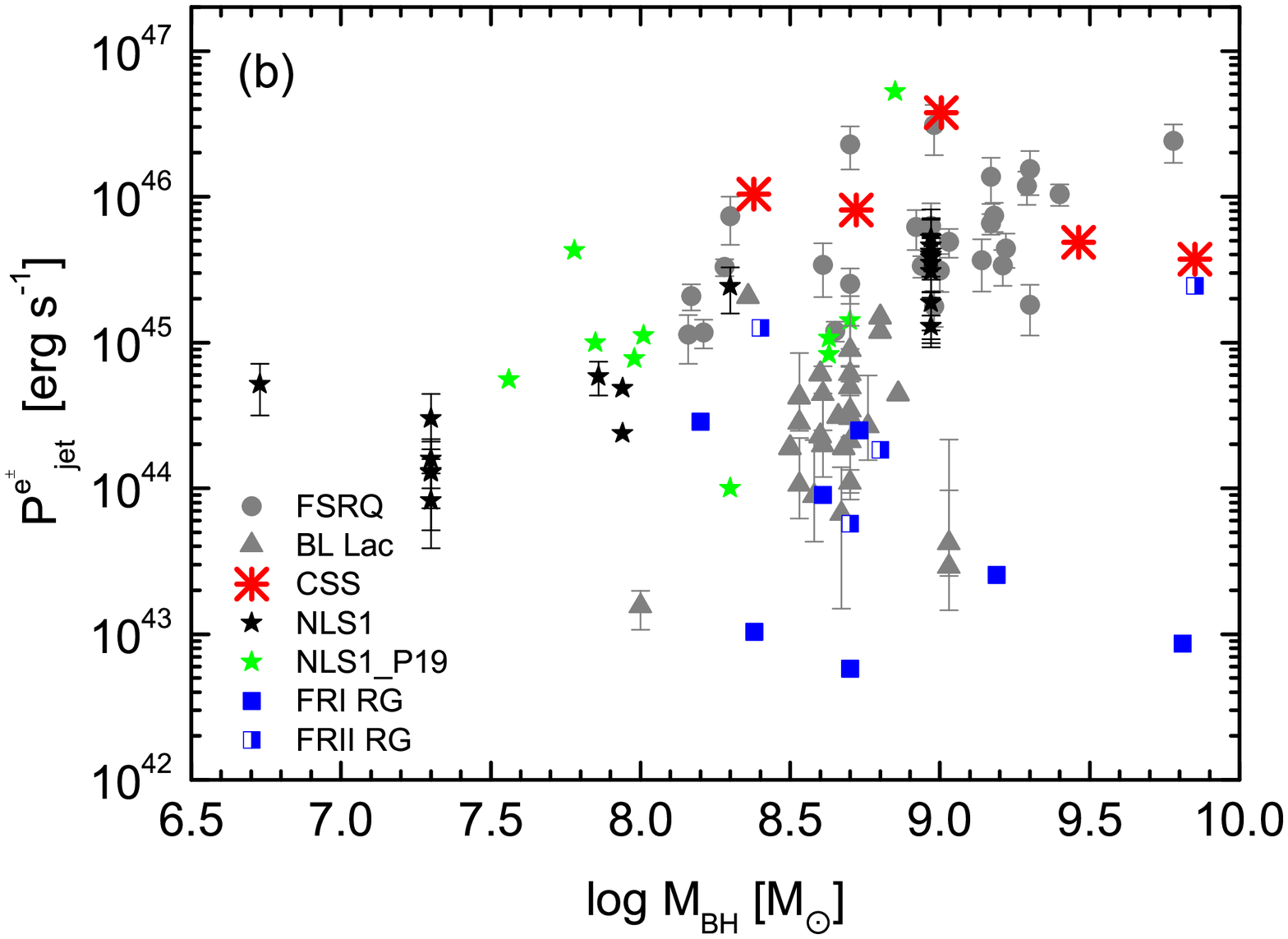}
\caption{$P_{\rm r}$ and $P^{e^{\pm}}_{\rm jet}$ as a function of $M_{\rm BH}$ for these $\gamma$-ray emitting AGNs. The details of data see Tables 3 and 4. }\label{M-Pr}
\end{figure}

No correlation between $P_{\rm r}$ and $P^{e^{\pm}}_{\rm jet}$ with $M_{\rm BH}$ is observed for these AGNs, as displayed in Figure \ref{M-Pr}. The range of $M_{\rm BH}$ for CSSs is similar to that of RGs and FSRQs, and NLS1s are the low-mass tail of $\gamma$-ray emitting AGNs (Sun et al. 2015; Zhang et al. 2015; Berton et al. 2016). However, $P_{\rm r}$ and $P^{e^{\pm}}_{\rm jet}$ would be strongly correlated with $M_{\rm BH}$ if only considering the data of FSRQs, NLS1s, and CSSs. The Pearson correlation analysis yields $r=0.79$ and $p=1.8\times10^{-11}$ for the $M_{\rm BH}$--$P_{\rm r}$ relation and $r=0.81$ and $p=1.3\times10^{-12}$ for the $M_{\rm BH}$--$P^{e^{\pm}}_{\rm jet}$ relation, respectively. This is consistent with that the jet power is connected to the BH mass (Heinz \& Sunyaev 2003). Hence, the lower jet power of NLS1s might be a consequence of the lower BH mass than FSRQs and CSSs. The predicted jet power should depend on the spin, mass, and horizon magnetic field of the BH (Blandford \& Znajek 1977; Ghisellini et al. 2014). The different relations between jet power and BH mass for the two kinds of blazars (FSRQs and BL Lacs) may indicate their different dominating jet formation mechanisms (Zhang et al. 2014).

Although the dominant mechanisms of jet launching may be different among these AGNs via either the Blandford--Payne (BP; Blandford \& Payne 1982) and/or Blandford--Znajek (BZ; Blandford \& Znajek 1977) mechanisms (Ghisellini \& Celotti 2001; Zhang 2013; Zhang et al. 2014; Ghisellini et al. 2014), the ejected jet power and jet radiation power are still connected with the disk luminosity and the Eddington ratio (e.g., Ghisellini et al. 2014; Zhang et al. 2015; Paliya et al. 2017). The different structures and accretion rates of accretion disks may result in the different dominating mechanisms of jet launching (Ghisellini \& Celotti 2001; Zhang 2013). Hence, the accretion rate (see also Shen \& Ho 2014) may be the fundamental parameter in the unified framework among different types of $\gamma$-ray emitting AGNs (e.g., Zhang et al. 2015). The $P^{e^{\pm}}_{\rm jet}/L_{\rm Edd}-R_{\rm Edd}$ (or $P_{\rm r}/L_{\rm Edd}-R_{\rm Edd}$) correlation likely signals a possible unification picture of jet radiation among subclasses of AGNs.

\begin{figure}
 \centering
   \includegraphics[angle=0,scale=0.4]{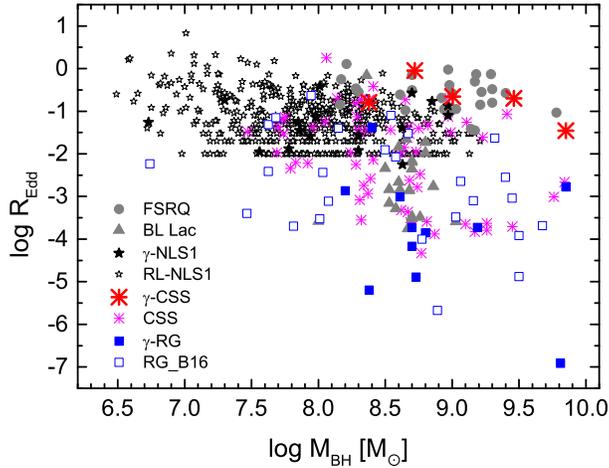}
\caption{Eddington ratio as a function of $M_{\rm BH}$. The RL-NLS1 sample data are from Viswanath et al. (2019), the CSS sample data are from Liao \& Gu et al. (2020), the RG sample data are from Berton et al. (2016), and the data of other $\gamma$-ray emitting AGNs see Tables 3 and 4. }\label{M-Redd}
\end{figure}

In order to further investigate the properties of $\gamma$-ray emitting CSSs, we collect the large sample data of CSSs and RL-NLS1s\footnote{It was suggested that $M_{\rm BH}$ of NLS1s derived from the emission line may be underestimated (Collin et al. 2006; Zhu et al. 2009; Calderone et al. 2013; Viswanath et al. 2019), hence the $M_{\rm BH}$ values of RL-NLS1s in this paper are estimated by fitting the big blue bump spectrum with the standard accretion disk model (Calderone et al. 2013; Viswanath et al. 2019).} together with a RG sample from the literature (Liao \& Gu et al. 2020; Viswanath et al. 2019; Berton et al. 2016). We plot $R_{\rm Edd}$ against $M_{\rm BH}$ for these sources together with the $\gamma$-ray emitting AGNs in Figure \ref{M-Redd}. It was proposed that CSSs may be the parent population of RL-NLS1s, but we find that the large sample of CSSs has higher $M_{\rm BH}$ and lower $R_{\rm Edd}$ than that of the large RL-NLS1 sample. We also find that on average the $\gamma$-ray detected NLS1s have high $M_{\rm BH}$ (but not high $R_{\rm Edd}$) among the large RL-NLS1 sample, likely indicating that the $\gamma$-ray emission is easier to detect in the RL-NLS1s with high BH mass. CSSs have higher $R_{\rm Edd}$ than that of these RGs on average. It is found that the $\gamma$-ray emitting CSSs have higher $R_{\rm Edd}$ and $M_{\rm BH}$ than others among the large CSS sample, which may be a helpful signal to find more $\gamma$-ray emitting CSS candidates in the future. The high $R_{\rm Edd}$ and $M_{\rm BH}$ feature of $\gamma$-ray emitting CSSs is similar to FSRQs. The high $R_{\rm Edd}$ of $\gamma$-ray emitting CSSs may imply that their BHs are located at an environment with rich gas to provide the high accretion.

\section{Discussion and Conclusions}

We analyzed the $\sim11$-year \emph{Fermi}/LAT data of six CSSs and all of them are located within the 95\% containment of the associating 4FGL point sources, confirming that they are spatially associated with these 4FGL point sources. By exploiting \emph{Fermi}/LAT observation data of the six CSSs, we obtained their long-term light curves, average spectra, and variability index (TS$_{\rm var}$). The derived TS$_{\rm var}$ values signal that five among six CSSs are obviously variable sources. This is consistent with the derived long-term light curves, and their variabilities are accompanied by the variations of photon spectral index. Even with the 180-day time bin size, we found that for four CSSs only several time bins in the long-term light curves have detections with TS$\geq$9, indicating that these CSSs are very weak $\gamma$-ray emitting AGNs.

We also compiled the broadband SEDs of the six CSSs from the literatures and ASDC to investigate their $\gamma$-ray emission properties. The broadband SEDs of $\gamma$-ray emitting CSSs can be well explained with the two-zone leptonic model. The steep radio spectra below 10 GHz are from the radiations of the extended regions in large-scale while the $\gamma$-ray emission should be dominated by the radiations of the compact cores. The derived values of $\gamma_{\rm b}$, $B$, and $p_1$ for the extended regions in CSSs are consistent with these of large-scale jet hotspots and knots in other AGNs. For the core regions, the derived beaming factors of these CSSs are smaller than that of the $\gamma$-ray emitting blazars. The flat electron spectra with $p_1\sim1.5-1.8$ likely imply that the magnetic reconnection may be an effective process of particle acceleration in these jets, similar to some FSRQs and NLS1s (Zhu et al. 2016).

Based on the fitting parameters, we also calculated $P_{\rm e}$, $P_{B}$, $P_{\rm r}$, and $P^{e^{\pm}}_{\rm jet}$ of the core regions for the $\gamma$-ray emitting CSSs and compared with other $\gamma$-ray emitting AGNs. Except for 4C +15.05, the other CSSs have high jet radiation power and jet radiation efficiency, but all of them have high ratio of $P_{B}$ to $P^{e^{\pm}}_{\rm jet}$, i.e., highly magnetized core-jets, similar to FSRQs and NLS1s. We found that for the different kinds of $\gamma$-ray emitting AGNs, $P_{\rm r}$ and $P^{e^{\pm}}_{\rm jet}$ in units of Eddington luminosity are strongly correlated with the Eddington ratio. In the $R_{\rm Edd}-P_{\rm r}/L_{\rm Edd}$ (or $R_{\rm Edd}-P^{e^{\pm}}_{\rm jet}/L_{\rm Edd}$) plane, CSSs, FSRQs, NLS1s, BL Lacs, and RGs form a sequence spanning more than six orders of magnitude. Hence, we proposed that the Eddington ratio may be a key physical driver for the unification scheme of AGN jet radiation. Comparing with a large CSS sample, the $\gamma$-ray emitting CSSs have higher Eddington ratio and $M_{\rm BH}$ than other CSSs, which may be a helpful signal to find more $\gamma$-ray emitting CSSs in the future.

We note that 3C 216 and 3C 380 may be not really CSSs and the compact size is due to the projection effects (O'Dea 1998). Both of them were reported to be embedded in a diffuse halo and more likely be oriented close to the line of sight, and then the halo may be radio lobes seen end-on (Barthel et al. 1988; Wilkinson et al. 1991; van Breugel et al. 1992; Taylor et al. 1995). For 3C 380, Polatidis \& Wilkinson (1998) also suggested that it is likely a powerful FR II source seen approximately end-on and the apparent acceleration with increasing radial distance along the jet is due to a bent jet and a varying angle to the line of sight. However, 3C 216 shows surprisingly weak ionization levels in its optical spectrum, i.e., the flux ratio [O~{\scriptsize II}]/[O~{\scriptsize III}]=1.2, which is also a fairly common property in the physically small CSSs, therefore its compact size may be a combination of projection effect and an intrinsically modest size (Pihlstr{\"o}m et al. 1999 and references therein). We find that the derived jet parameters, the radiation efficiency and the magnetization of jets for 3C 216 and 3C 380 do not show difference from other four $\gamma$-ray emitting CSSs, and both of them also follow the sequence of these $\gamma$-ray emitting AGNs in the $P^{e^{\pm}}_{\rm jet}/L_{\rm Edd}-R_{\rm Edd}$ (or $P_{\rm r}/L_{\rm Edd}-R_{\rm Edd}$) plane.

\acknowledgments
We thank the anonymous referee for his/her valuable suggestions. We appreciate helpful discussions with Shuang-Nan Zhang. This work is supported by the National Natural Science Foundation of China (grants 11973050, 11533003, 11863007, 11851304, U1731239 and U1831205), the National Key R\&D Program of China (2016YFA0400702), and Guangxi Science Foundation (grants AD17129006 and 2018GXNSFGA281007).

\clearpage

\begin{figure}
 \centering
   \includegraphics[angle=0,scale=0.3]{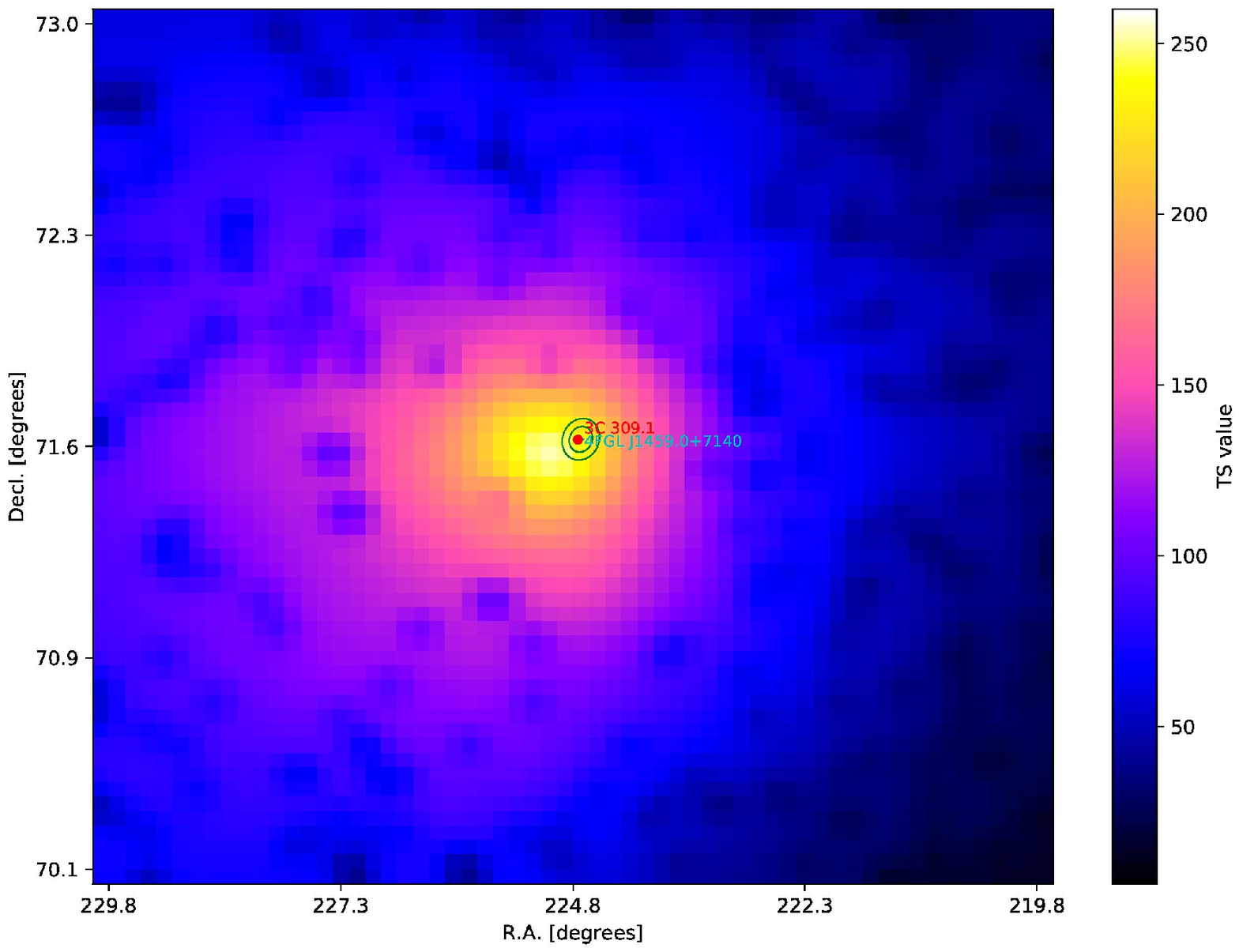}
   \includegraphics[angle=0,scale=0.3]{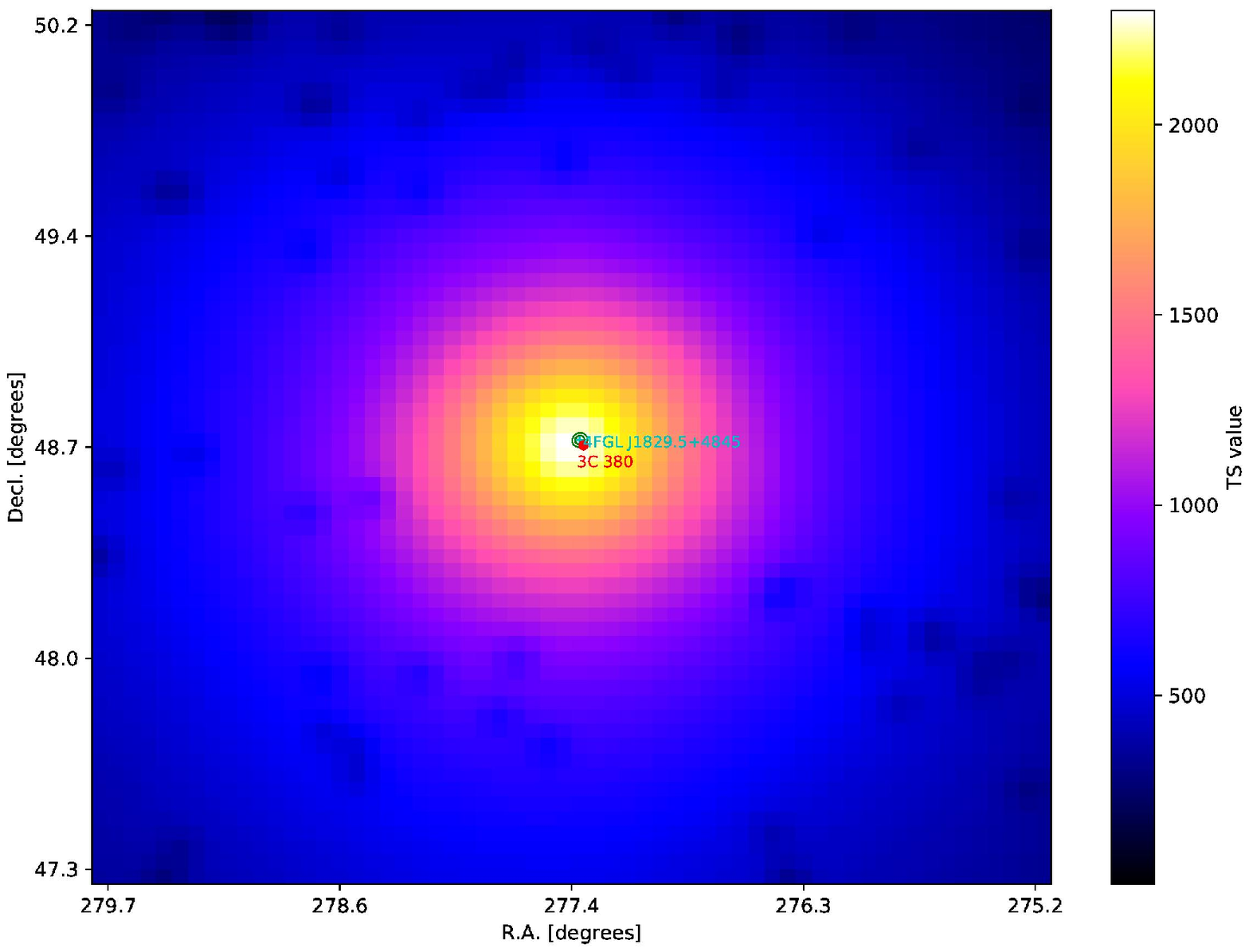}
   \includegraphics[angle=0,scale=0.3]{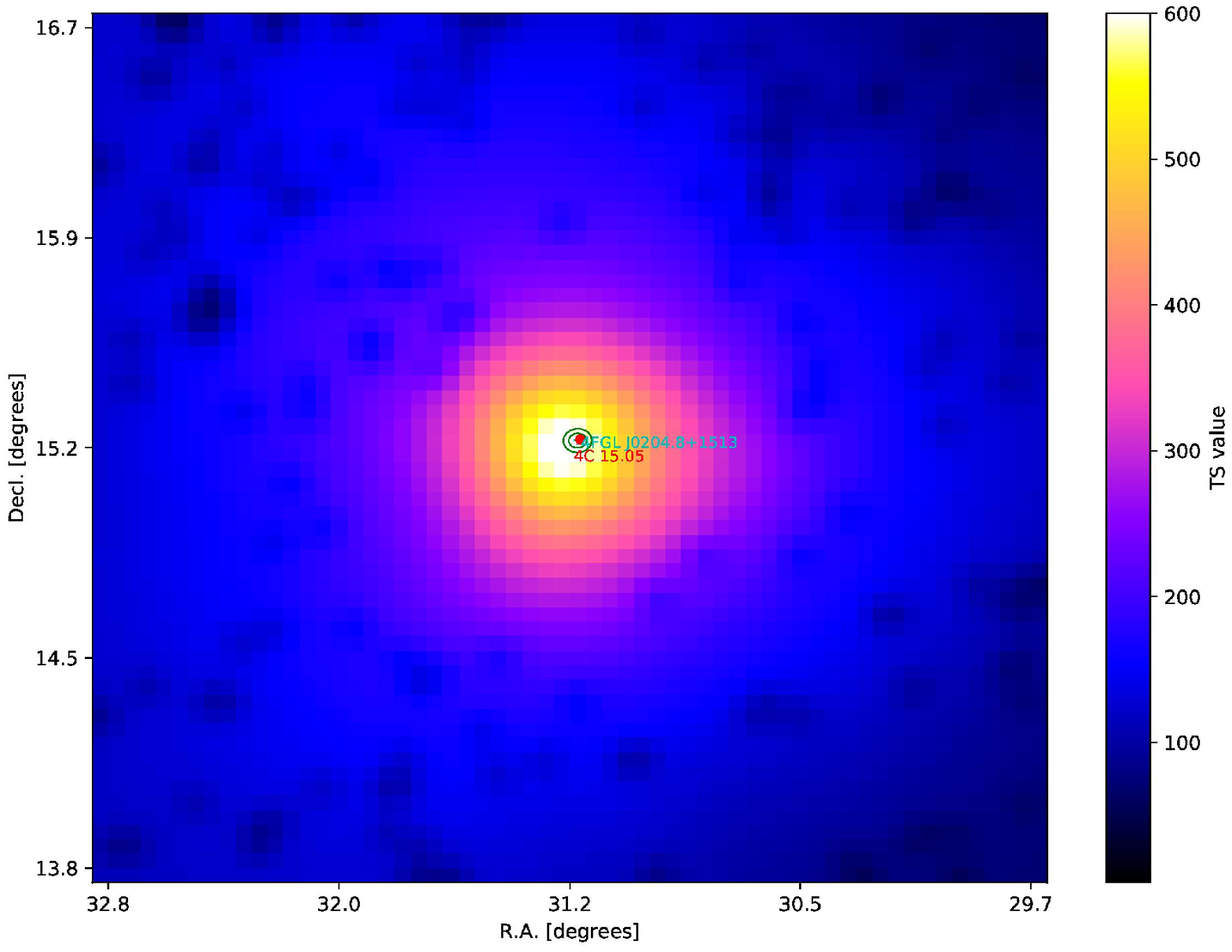}
\hfill\center{Fig. 1---  continued}
\end{figure}

\begin{figure}
 \centering
   \includegraphics[angle=0,scale=0.31]{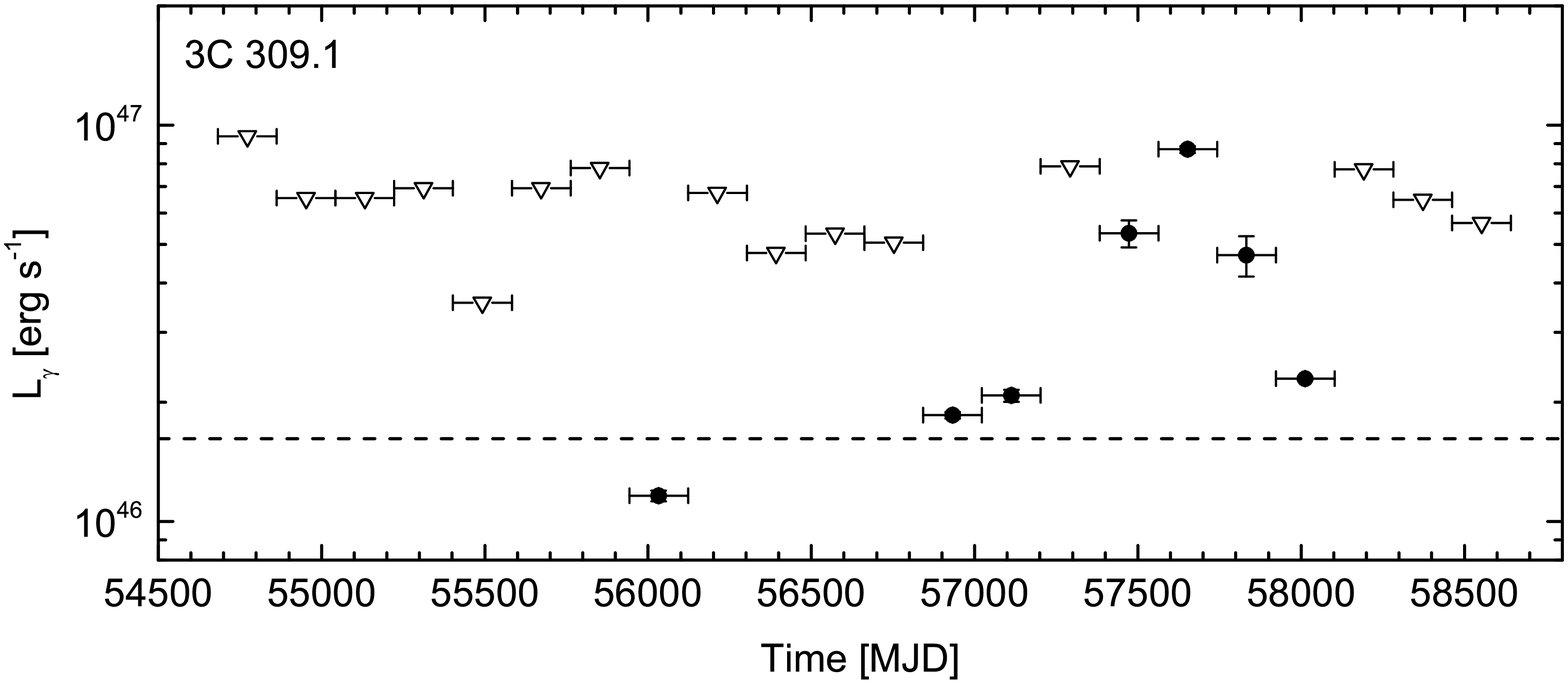}
   \includegraphics[angle=0,scale=0.27]{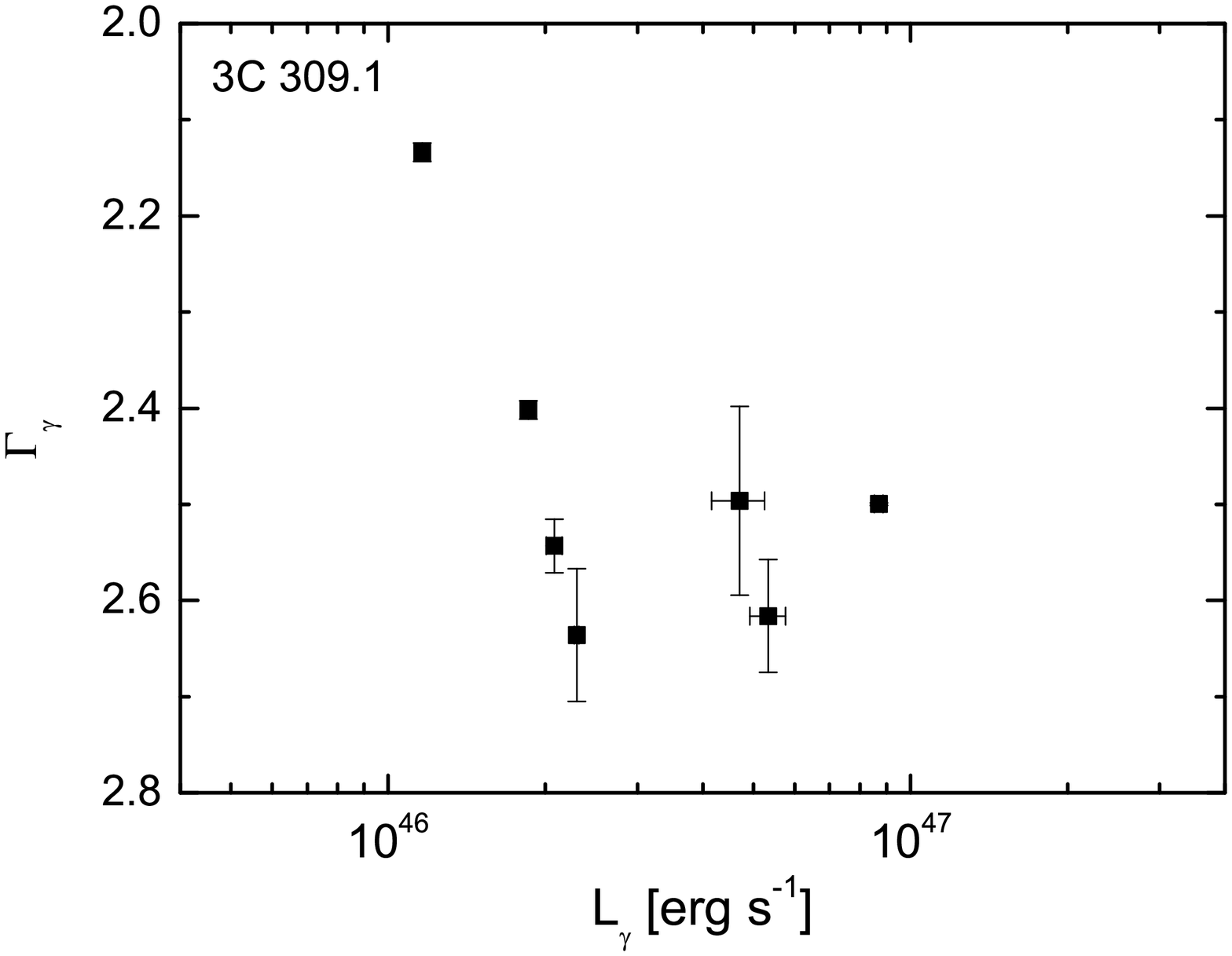}
   \includegraphics[angle=0,scale=0.31]{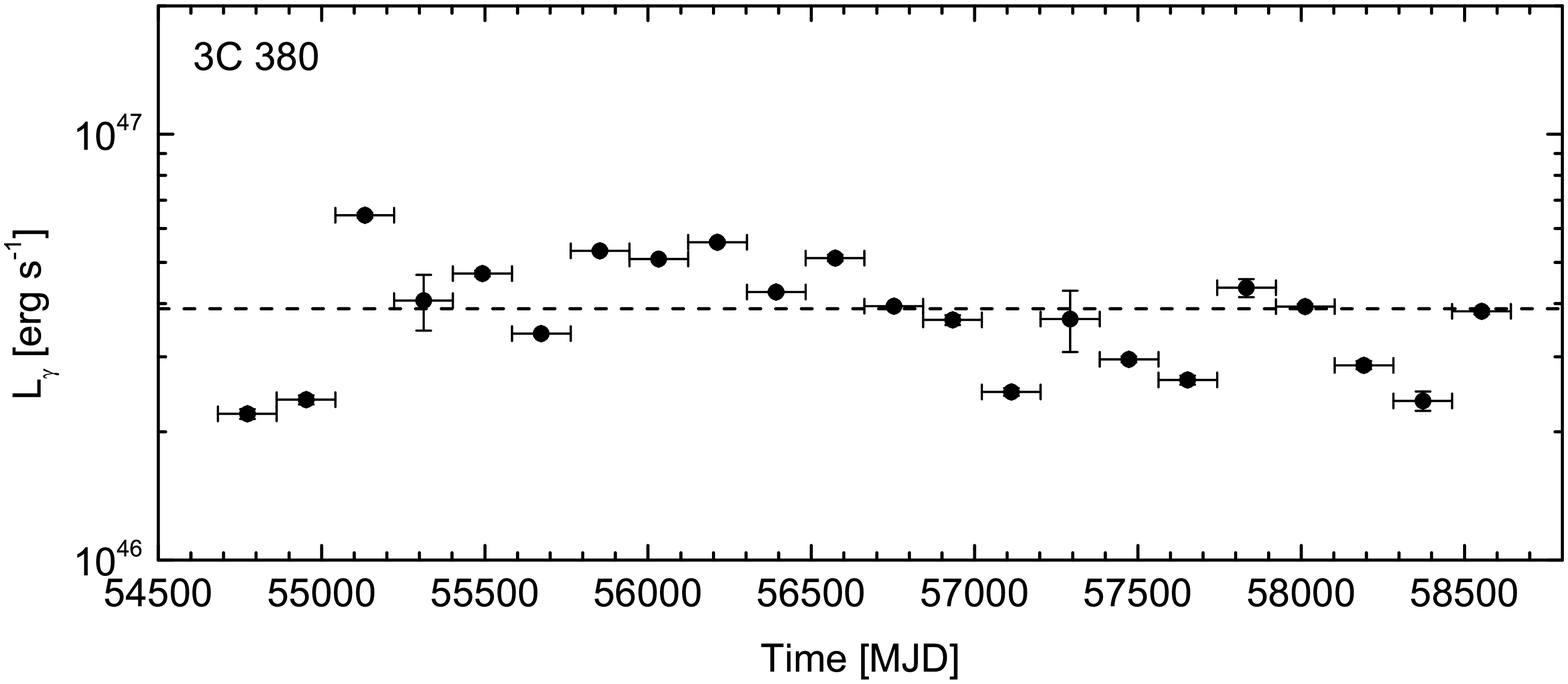}
   \includegraphics[angle=0,scale=0.27]{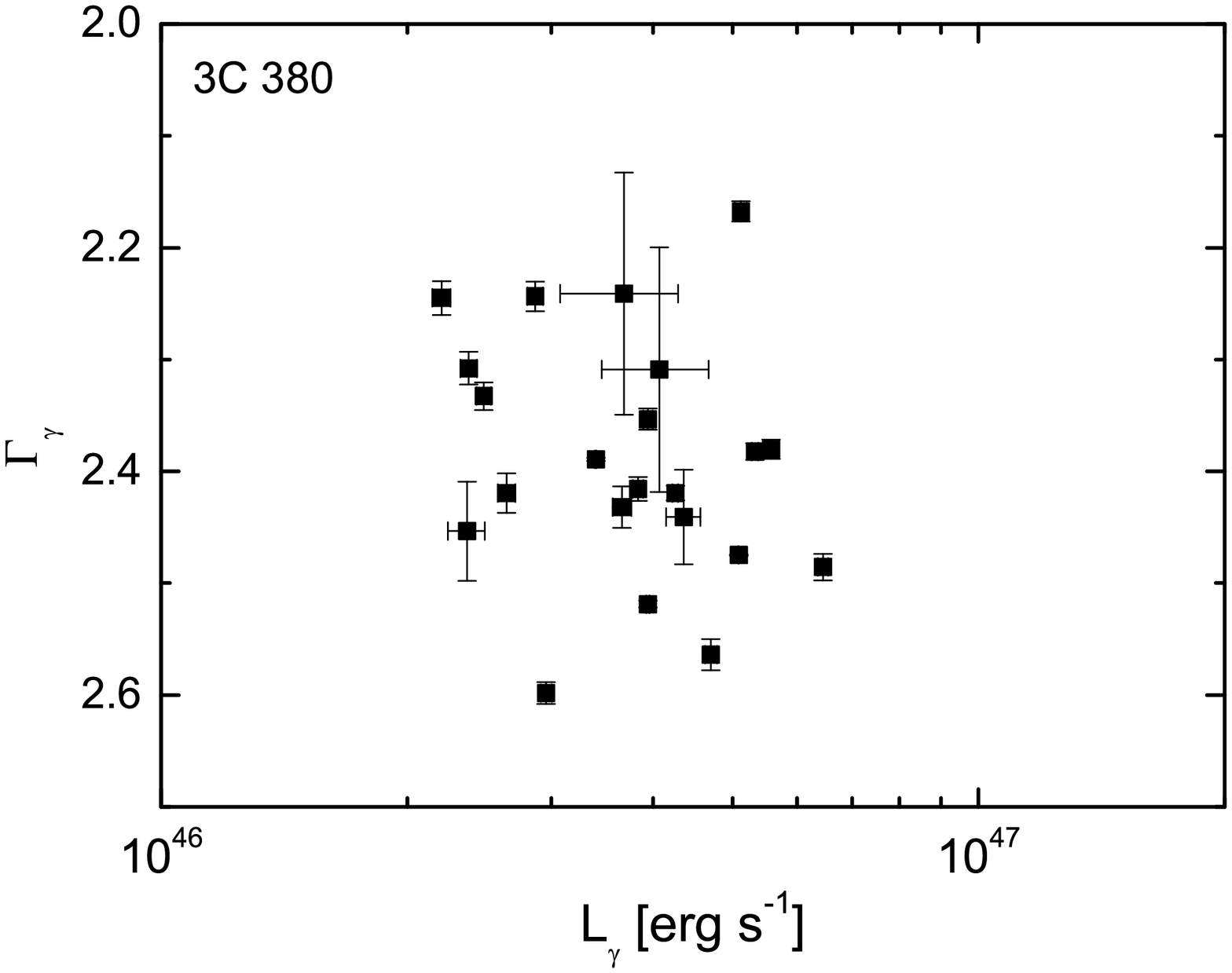}
   \includegraphics[angle=0,scale=0.31]{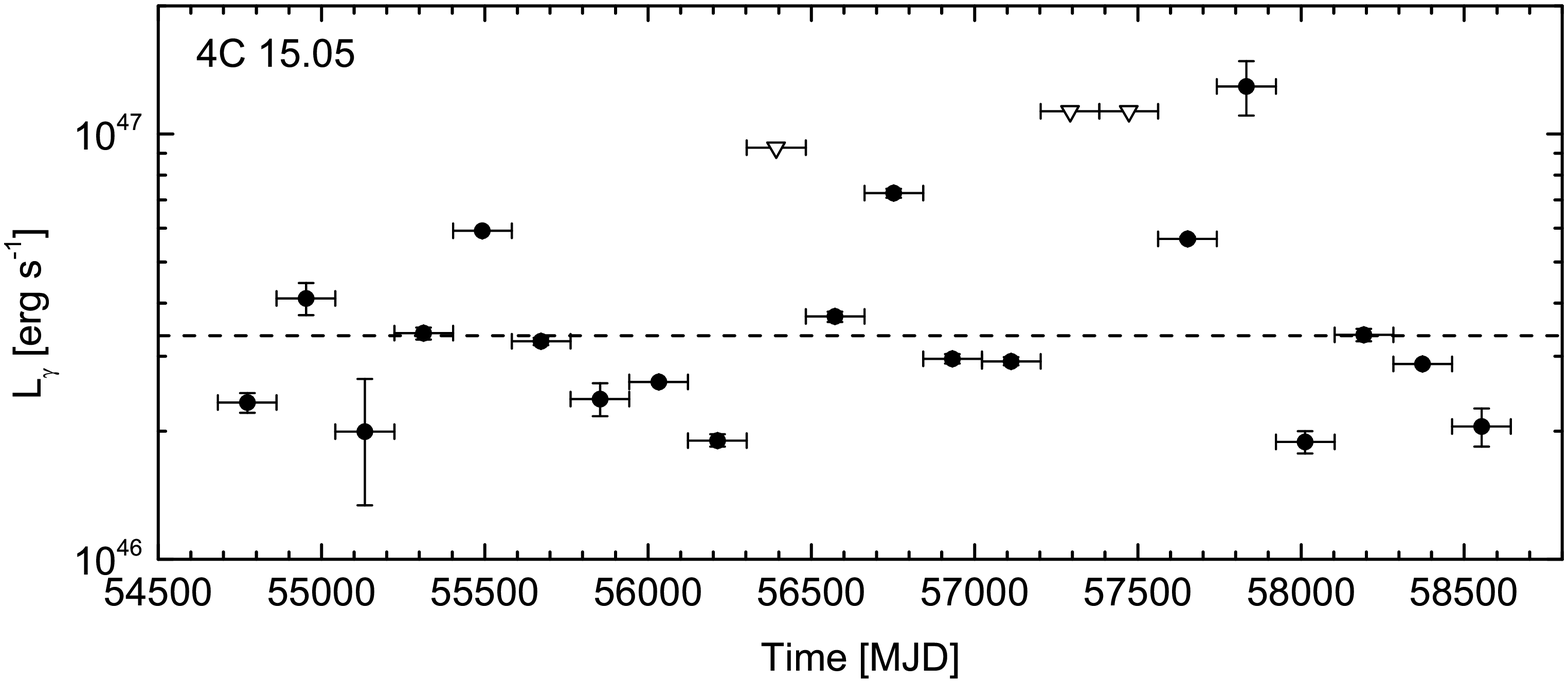}
   \includegraphics[angle=0,scale=0.27]{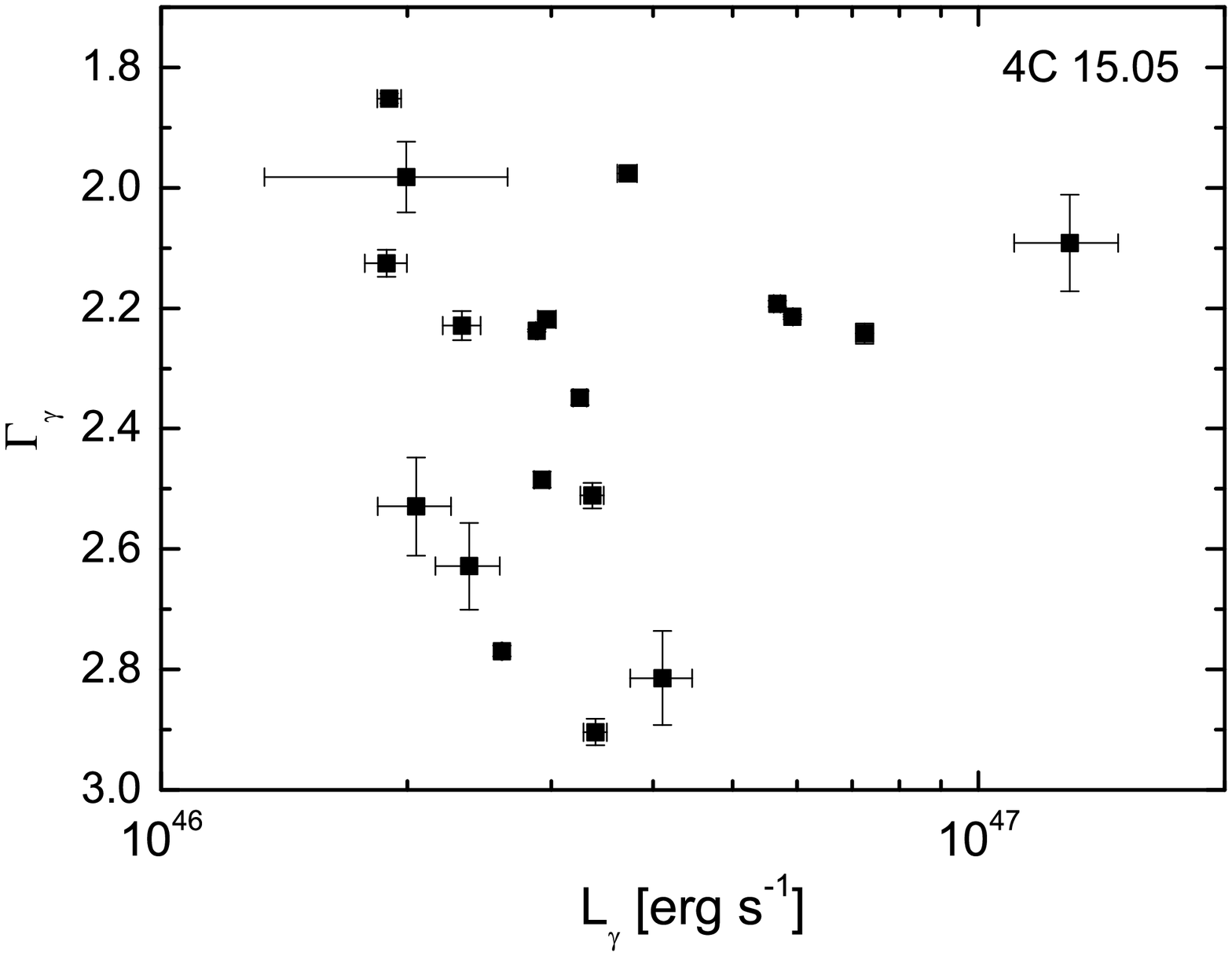}
\hfill\center{Fig. 2---  continued}
\end{figure}

\clearpage

\begin{deluxetable}{lccccccc}
\tabletypesize{\scriptsize} \tablecolumns{7} \tablewidth{0pc}
\tablecaption{\emph{Fermi}/LAT Analysis Results for the Six CSSs}\tablenum{1}
\tablehead{\colhead{Source Name} & \colhead{$R.A.$\tablenotemark{a}} &  \colhead{$Decl.$\tablenotemark{a}}&\colhead{TS}&\colhead{$F_{0.1-300~\rm GeV}$ } &\colhead{$\Gamma_{\rm \gamma}$}&\colhead{TS$_{\rm var}$}&\colhead{Association}\\
\colhead{}&\colhead{[deg]}&\colhead{[deg]}&\colhead{}&\colhead{[10$^{-12}$ erg cm$^{-2}$ s$^{-1}$]}&\colhead{}&\colhead{}&\colhead{}}
\startdata
3C 138&80.2912&16.6393&21.4&1.94$\pm$0.47&2.19$\pm$0.02&128.1&4FGL J0521.2+1637\\
3C 216&137.3897&42.8965&138.6&4.13$\pm$0.23&2.91$\pm$0.06&11.8&4FGL J0910.0+4257\\
3C 286&202.7845&30.5092&54.2&2.32$\pm$0.16&2.75$\pm$0.10&64.3&4FGL J1331.0+3032\\
3C 309.1&224.7816&71.6722&235.5&3.97$\pm$0.11&2.67$\pm$0.02&167.0&4FGL J1459.0+7140\\
3C 380&277.3824&48.7462&2184.2&18.44$\pm$0.10&2.50$\pm$0.01&149.1&4FGL J1829.5+4845\\
4C 15.05&31.2182&15.2326&611.5&10.05$\pm$0.52&2.47$\pm$0.05&211.4&4FGL J0204.8+1513\\
\enddata
\tablenotetext{a}{The coordinates of sources are from Abdollahi et al. (2020).}
\end{deluxetable}

\begin{deluxetable}{lccccccccccccccccccccc}
\tabletypesize{\tiny} \rotate \tablecolumns{22} \tablewidth{0pc}
\tablecaption{SED Fitting Parameters of the Six CSSs}\tablenum{2}
\tablehead{ \colhead{}  & \colhead{}  &\multicolumn{11}{c}{Compact Core} & \multicolumn{8}{c}{Extended Region} \\
\cline{3-13} \cline{15-22}\\
\colhead{Source} &\colhead{z}  & \colhead{$R$} &
\colhead{$B$} & \colhead{$\delta$}& \colhead{$\Gamma$} & \colhead{$\theta$} & \colhead{$\gamma_{\min}$} & \colhead{$\gamma_{\rm b}$} &\colhead{$\gamma_{\max}$}
&\colhead{$N_{0}$ } & \colhead{$p_1$} & \colhead{$p_2$} &\colhead{}& \colhead{$R$} & \colhead{$B$} & \colhead{$\gamma_{\min}$} & \colhead{$\gamma_{\rm b}$} &\colhead{$\gamma_{\max}$} &\colhead{$N_{0}$ } & \colhead{$p_1$} & \colhead{$p_2$} \\
\colhead{}& \colhead{}  &\colhead{[cm]} & \colhead{[G]} & \colhead{}& \colhead{} & \colhead{[deg]} & \colhead{} & \colhead{} &\colhead{} &\colhead{ [cm$^{-3}$]} & \colhead{} & \colhead{} &\colhead{} & \colhead{[kpc]}& \colhead{[$\mu$G]} & \colhead{} & \colhead{} & \colhead{} & \colhead{[cm$^{-3}$]}& \colhead{} & \colhead{}}
\startdata
3C 138&0.759&7.44E17&0.6&2.8&5.5&18&1&531&5E5&139.8&1.8&3.46&&2.95&277&5E2&5E3&2.5E5&6E-03&2.2&4\\
3C 216&0.67&1.01E18&0.4&3.6&3.6&15.9&1&1427&1E5&25.3&1.8&4.04&&17.54&66&1E2&2E4&1E6&8.8E-04&2.46&4\\
3C 286&0.849&1.13E18&0.2&4.5&4.5&12.7&1&1666&3E5&5.1&1.6&4.5&&19.18&72&1E3&1.9E4&9.5E5&4.7E-04&2.22&4\\
3C 309.1&0.904&8.75E17&0.22&3.6&2.8&15.5&50&1914&1E5&15&1.5&4.2&&15.65&77&5E2&1.5E4&7.5E5&4.3E-04&2.22&4\\
3C 380&0.692&1.11E18&0.15&4.0&2.4&9.5&1&1845&1E5&40.1&1.68&4.0&&7.11&190&1E2&1.5E4&7.5E5&5E-03&2.4&4\\
4C 15.05&0.833&2.26E18&0.38&8.9&5.0&3.9&20&320&6E4&0.36&1.7&3.6&&2.29&300&1E3&6E3&3E5&1.9E-03&2.0&4\\
\enddata
\end{deluxetable}

\begin{deluxetable}{lccccccccc}
\tabletypesize{\scriptsize} \tablecolumns{10} \tablewidth{0pc}
\tablecaption{BH Mass and Derived Parameters of Core Regions for the Six CSSs}\tablenum{3}
\tablehead{\colhead{Source} & \colhead{$P_{\rm e}$} &  \colhead{$P_{B}$}& \colhead{$P_{\rm r}$} &\colhead{$P^{e^{\pm}}_{\rm jet}$}&\colhead{$M_{\rm BH}$\tablenotemark{a}}&\colhead{$L_{\rm disk}$\tablenotemark{b}}&\colhead{$\nu_{\rm s}$}&\colhead{$L_{\rm s}$}&\colhead{$R_{\rm Edd}$}\\
\colhead{}&\colhead{[erg s$^{-1}$]}&\colhead{[erg s$^{-1}$]}&\colhead{[erg s$^{-1}$]}&\colhead{[erg s$^{-1}$]}&\colhead{[$M_{\bigodot}$]}&\colhead{[erg s$^{-1}$]}&\colhead{[Hz]}&\colhead{[erg s$^{-1}$]}&\colhead{}}
\startdata
3C 138&2.86E45&2.26E46&1.24E46&3.79E46&1.01E9&2.79E46&1.80E12&4.86E45&0.22\\
3C 216&4.92E44&7.93E45&1.99E45&1.04E46&2.39E8&4.87E45&9.49E12&7.19E45&0.16\\
3C 286&5.76E44&3.88E45&3.65E45&8.10E45&5.25E8&5.96E46&5.41E12&5.92E45&0.90\\
3C 309.1&6.77E44&1.09E45&3.11E45&4.88E45&2.9E9&7.28E46&8.32E12&1.06E46&0.20\\
3C 380&8.50E44&5.98E44&2.28E45&3.73E45&7.1E9&3.12E46&5.91E12&1.11E46&0.03\\
4C 15.05&5.14E43&6.91E46&3.39E44&6.95E46&&1.93E45&1.37E12&1.26E46&\\
\enddata
\tablenotetext{a}{$M_{\rm BH}$ of 3C 138, 3C 216, and 3C 309.1 are from Gu et al. (2001). $M_{\rm BH}$ of 3C 286 and 3C 380 are from Shen et al. (2011) and Woo et al. (2002), respectively. No $M_{\rm BH}$ of 4C 15.05 is found in the literature. }
\tablenotetext{b}{$L_{\rm disk}$ is estimated with the narrow blue bump of the accretion disk emission (green line) in SEDs. $L_{\rm disk}$ of 3C 216 is estimated by $10L_{\rm BLR}$, in which $L_{\rm BLR}$ is the BLR luminosity and taken from Celotti et al. (1997), and $L_{\rm disk}$ of 4C 15.05 is taken the luminosity at $10^{15}$ Hz of the model-fitting line.}
\end{deluxetable}

\clearpage
\begin{deluxetable}{lccccccccccccc}
\tabletypesize{\tiny} \rotate \tablecolumns{14} \tablewidth{51pc}\setlength{\tabcolsep}{1.2mm}
\tablecaption{Data of other $\gamma$-ray Emitting AGNs}\tablenum{4}
\tablehead{ \colhead{Source\tablenotemark{a}} &\colhead{z}  & \colhead{$\delta$} & \colhead{$B$}& \colhead{$\gamma_{\rm b}$} & \colhead{$\nu_{\rm s}$} &\colhead{$L_{\rm s}$} &\colhead{$P_{\rm r}$ } & \colhead{$P_{\rm e}$} & \colhead{$P_{B}$} & \colhead{$P^{e^{\pm}}_{\rm jet}$} & \colhead{$L_{\rm disk}$\tablenotemark{b}} &\colhead{$M_{\rm BH}$\tablenotemark{c}} &\colhead{$R_{\rm Edd}$} \\
\colhead{}& \colhead{}  &\colhead{} & \colhead{[G]} & \colhead{}& \colhead{[Hz]} & \colhead{[erg s$^{-1}$]} & \colhead{[erg s$^{-1}$]} & \colhead{[erg s$^{-1}$]} &\colhead{[erg s$^{-1}$]} &\colhead{[erg s$^{-1}$]} & \colhead{[erg s$^{-1}$]} & \colhead{[$M_{\bigodot}$]} &\colhead{} }
\startdata
\scriptsize{BL Lacs}\\
\hline
BL Lacertae$^L$&0.069&19$^{+7}_{-1}$&0.5$^{+0.3}_{-0.33}$&1.74$^{+0.81}_{-0.33}$E3&1.2E14&1.16E45&7.25$^{+1.23}_{-1.61}$E42&1.79$^{+2.48}_{-0.78}$E44&1.12$^{+1.23}_{-0.66}$E43&1.98$^{+2.48}_{-0.78}$E44&7.38E43&8.61&1.44E-03\\
BL Lacertae$^H$&\nodata&20$^{+6}_{-5.2}$&0.2$^{+0.27}_{-0.1}$&3.80$^{+0.91}_{-0.92}$E3&2.54E14&4.9E44&5.73$^{+3.00}_{-3.05}$E42&4.36$^{+2.41}_{-1.94}$E44&2.20$^{+1.47}_{-0.62}$E42&4.44$^{+2.41}_{-1.94}$E44&7.38E43&8.61&1.44E-03\\
Mkn 421$^L$&0.031&29$^{+14}_{-14}$&0.14$^{+1.16}_{-0.09}$&1.05$^{+0.39}_{-0.57}$E5&1.23E17&7.49E44&1.88$^{+1.81}_{-1.81}$E42&5.96$^{+6.73}_{-5.16}$E43&5.12$^{+26.35}_{-1.73}$E42&6.66$^{+7.23}_{-5.17}$E43&1.03E43&8.67&1.75E-04\\
Mkn 501$^L$&0.034&14$^{+9}_{-5}$&0.16$^{+0.44}_{-0.13}$&6.12$^{+4.91}_{-2.18}$E4&2.95E16&1.96E44&2.26$^{+1.61}_{-1.61}$E42&2.09$^{+6.74}_{-1.36}$E43&5.78$^{+7.44}_{-4.28}$E42&2.90$^{+6.78}_{-1.43}$E43&3.47E43&9.03&2.57E-04\\
Mkn 501$^H$&\nodata&15$^{+14}_{-4.2}$&0.4$^{+0.8}_{-0.37}$&9.65$^{+15.69}_{-3.08}$E5&1.8E19&2.91E45&2.19$^{+1.62}_{-1.23}$E43&1.97$^{+17.3}_{-1.15}$E43&3.31$^{+4.31}_{-3.05}$E41&4.19$^{+17.40}_{-1.69}$E43&3.47E43&9.03&2.57E-04\\
PKS 2005--489$^H$&0.071&42$^{+15}_{-15}$&0.09$^{+0.19}_{-0.06}$&1.75$^{+0.85}_{-0.51}$E4&5.63E15&2.19E45&3.41$^{+2.44}_{-2.44}$E42&1.66$^{+3.22}_{-0.91}$E44&9.59$^{+6.02}_{-6.03}$E43&2.66$^{+3.28}_{-1.10}$E44&2.35E43&8.76&3.24E-04\\
1ES 1218+30.4&0.182&20$^{+13}_{-9.2}$&0.14$^{+0.72}_{-0.11}$&4.78$^{+3.26}_{-2.15}$E4&3.29E16&2.44E45&1.42$^{+1.31}_{-1.31}$E43&6.00$^{+12.08}_{-4.24}$E43&1.41$^{+3.24}_{-0.93}$E43&8.83$^{+12.57}_{-4.53}$E43&2.47E43&8.58&5.17E-04\\
W Com$^L$&0.102&15$^{+8}_{-1}$&0.18$^{+0.12}_{-0.13}$&1.10$^{+0.59}_{-0.22}$E4&1.68E15&8.13E44&1.11$^{+0.19}_{-0.19}$E43&8.63$^{+11.6}_{-4.33}$E43&8.49$^{+8.43}_{-4.88}$E42&1.06$^{+1.16}_{-0.44}$E44&2.95E43&8.53$^{\rm P17}$&6.93E-04\\
W Com$^H$&\nodata&14$^{+19}_{-0.5}$&0.17$^{+0.03}_{-0.15}$&9.36$^{+8.41}_{-0.57}$E3&1.16E15&1.01E45&3.51$^{+0.66}_{-0.66}$E43&2.41$^{+5.66}_{-0.34}$E44&5.75$^{+1.25}_{-3.26}$E42&2.82$^{+5.66}_{-0.34}$E44&2.95E43&8.53&6.93E-04\\
PKS 2155--304$^L$&0.116&50$^{+18}_{-18}$&0.16$^{+0.44}_{-0.1}$&1.94$^{+0.78}_{-0.69}$E4&1.07E16&7.47E45&6.20$^{+4.47}_{-4.46}$E42&1.81$^{+2.20}_{-1.16}$E44&2.24$^{+3.07}_{-1.19}$E43&2.10$^{+2.22}_{-1.17}$E44&2.82E44&8.7$^{\rm P17}$&4.47E-03\\
PKS 2155--304$^H$&\nodata&26$^{+16}_{-1}$&0.4$^{+0.13}_{-0.26}$&2.95$^{+0.97}_{-0.34}$E4&2.2E16&1.33E46&2.42$^{+0.31}_{-0.31}$E44&6.40$^{+4.11}_{-2.08}$E44&1.03$^{+0.49}_{-0.17}$E43&8.93$^{+4.13}_{-2.10}$E44&2.82E44&8.7&4.47E-03\\
1ES 1959+650$^L$&0.048&11$^{+6}_{-1.8}$&1.1$^{+0.72}_{-0.52}$&6.67$^{+0.72}_{-1.00}$E4&2.64E17&8.91E44&1.37$^{+0.45}_{-0.46}$E43&2.55$^{+0.10}_{-0.76}$E43&7.04$^{+2.44}_{-2.44}$E43&1.10$^{+0.25}_{-0.26}$E44&1.91E43&8.7$^{\rm P17}$&3.02E-04\\
1ES 1959+650$^H$&\nodata&12$^{+17}_{-1.3}$&0.25$^{+0.15}_{-0.23}$&1.19$^{+1.52}_{-0.19}$E6&1.48E19&1.81E45&3.73$^{+0.83}_{-0.83}$E43&3.00$^{+17.46}_{-1.17}$E44&5.15$^{+3.16}_{-4.02}$E42&3.42$^{+17.46}_{-1.17}$E44&1.91E43&8.7&3.02E-04\\
PG 1553+113&0.3&32$^{+6}_{-6}$&0.13$^{+0.09}_{-0.04}$&1.59$^{+0.16}_{-0.23}$E4&4.87E15&4.11E46&7.64$^{+2.90}_{-2.91}$E43&2.69$^{+0.38}_{-0.63}$E44&2.64$^{+0.68}_{-0.10}$E44&6.09$^{+0.83}_{-0.70}$E44&3.98E44&8.7$^{\rm P17}$&6.32E-03\\
1ES 1011+496&0.212&13$^{+12}_{-1.3}$&0.7$^{+0.55}_{-0.61}$&8.48$^{+8.58}_{-1.79}$E4&1.4E17&1.59E46&1.66$^{+0.34}_{-0.36}$E44&8.72$^{+31.47}_{-4.13}$E43&2.40$^{+2.61}_{-1.86}$E44&4.93$^{+4.11}_{-1.94}$E44&4.37E44&8.7$^{\rm P17}$&6.93E-03\\
Mkn 180&0.045&6$^{+6}_{-0.7}$&0.4$^{+0.3}_{-0.25}$&3.25$^{+0.50}_{-0.64}$E4&1.96E16&7.54E43&4.64$^{+1.20}_{-1.20}$E42&6.16$^{+1.33}_{-2.62}$E42&4.77$^{+3.94}_{-3.94}$E42&1.56$^{+0.43}_{-0.49}$E43&3.31E42&8$^{\rm P17}$&2.63E-04\\
RGB J0152+017&0.08&5$^{+13}_{-0}$&0.28$^{+0}_{-0.27}$&1.29$^{+2.31}_{-0.00}$E5&1.4E17&7.69E43&1.19$^{+0.20}_{-0.20}$E43&2.38$^{+14.02}_{-0.00}$E43&1.06$^{+0.00}_{-0.83}$E42&3.67$^{+14.02}_{-0.22}$E43&\nodata&\nodata&\nodata\\
H1426+428&0.129&8.5$^{+7}_{-0.1}$&0.1$^{+0.04}_{-0.082}$&3.79$^{+2.82}_{-0.57}$E5&7.16E17&5.45E44&6.88$^{+0.69}_{-0.69}$E43&2.01$^{+5.55}_{-0.87}$E44&1.03$^{+0.84}_{-0.65}$E42&2.71$^{+5.55}_{-0.88}$E44&\nodata&\nodata&\nodata\\
PKS 0548--322&0.069&6$^{+14}_{-0.4}$&0.6$^{+0}_{-0.58}$&1.85$^{+3.52}_{-0.00}$E5&4.43E17&2.03E44&1.10$^{+0.15}_{-0.15}$E43&7.54$^{+53.13}_{-0.49}$E42&7.79$^{+0.50}_{-6.41}$E42&2.63$^{+5.32}_{-0.66}$E43&\nodata&\nodata&\nodata\\
1ES 2344+514&0.044&13$^{+9}_{-6}$&0.12$^{+0.3}_{-0.07}$&6.01$^{+1.15}_{-1.63}$E4&2.18E16&1.26E44&1.37$^{+1.79}_{-1.30}$E42&1.31$^{+0.03}_{-0.37}$E43&2.37$^{+0.08}_{-0.08}$E42&1.68$^{+0.18}_{-0.39}$E43&\nodata&\nodata&\nodata\\
1ES 1101--232&0.186&12$^{+5}_{-1}$&1.05$^{+0.75}_{-0.6}$&4.51$^{+1.28}_{-0.91}$E4&1.45E17&3.48E45&3.91$^{+0.67}_{-0.67}$E43&8.57$^{+4.84}_{-3.84}$E42&1.02$^{+1.09}_{-0.23}$E44&1.50$^{+1.10}_{-0.24}$E44&\nodata&\nodata&\nodata\\
3C 66A&0.44&24$^{+4}_{-4}$&0.2$^{+0.11}_{-0.07}$&2.21$^{+0.33}_{-0.26}$E4&5.38E15&4.01E46&5.40$^{+2.08}_{-1.82}$E44&1.60$^{+0.53}_{-0.34}$E45&4.02$^{+0.64}_{-0.92}$E43&2.18$^{+0.57}_{-0.39}$E45&\nodata&\nodata&\nodata\\
PKS 1424+240&0.5&33$^{+8}_{-6}$&0.23$^{+0.23}_{-0.08}$&2.83$^{+0.31}_{-0.62}$E4&1.16E16&1.12E47&1.52$^{+0.56}_{-0.56}$E44&3.47$^{+0.43}_{-1.55}$E44&7.01$^{+5.76}_{-0.13}$E44&1.20$^{+0.58}_{-0.16}$E45&\nodata&\nodata&\nodata\\
1ES 0806+524&0.138&12$^{+7}_{-4}$&0.32$^{+1}_{-0.21}$&2.68$^{+0.95}_{-1.07}$E4&7.73E15&7.49E44&1.19$^{+0.80}_{-0.80}$E43&9.71$^{+6.48}_{-6.94}$E43&4.12$^{+9.42}_{-1.10}$E43&1.50$^{+1.15}_{-0.71}$E44&\nodata&\nodata&\nodata\\
PKS 0521--36&0.055&&&&&&1.7E44&5.01E43&7.59E42&2.28E44&1.09E44&8.6&2.18E-03\\
PKS 0829+46&0.174&&&&&&8.51E43&5.25E43&5.13E43&1.89E44&8.31E43&8.68&1.38E-03\\
PKS 0851+202&0.306&&&&&&2.19E44&1.95E44&3.02E43&4.44E44&1.57E44&8.86&1.72E-03\\
TXS 0954+658&0.367&&&&&&8.13E43&3.31E44&8.91E42&4.21E44&6.21E43&8.53&1.46E-03\\
PMN 1012+063&0.727&&&&&&8.51E43&5.25E43&5.13E43&1.89E44&1.81E44&8.5&4.55E-03\\
PKS 1057--79&0.581&&&&&&5.62E44&6.03E44&2.4E43&1.19E45&1.47E45&8.8&1.85E-02\\
PKS 1519--273&1.294&&&&&&4.79E44&1.26E44&8.91E44&1.5E45&8.4E44&8.8&1.06E-02\\
PKS 1749+096&0.322&&&&&&1.1E44&1.07E44&9.33E43&3.1E44&1.26E45&8.66&2.19E-02\\
S5 1803+78&0.68&&&&&&5.01E44&1.15E44&1.45E45&2.06E45&2E46&8.36&6.94E-01\\
TXS 1807+698&0.051&&&&&&2.82E43&2.69E44&8.71E42&3.06E44&2.12E43&8.7&3.37E-04\\
PKS 2240--260&0.774&&&&&&2.57E44&2.51E44&9.77E43&6.06E44&7.12E44&8.6&1.42E-02\\
\hline
\scriptsize{RGs}\\
\hline
NGC 1218&0.02865&5.6&0.23&4.98E4&7.45E15&5.06E42&5.3E41&4.03E42&1.24E42&5.8E42&4.3E42&8.7$^{\rm R05}$&6.82E-05\\
NGC 1275&0.01756&5.8&0.15&1.27E3&1.45E13&7.49E43&6.83E42&5.28E43&3.01E43&8.97E43&5E43&8.61$^{\rm B03}$&9.77E-04\\
NGC 6251&0.02471&7.8&0.02&1.6E4&8.13E13&2.19E42&3.75E41&2.49E44&3.54E40&2.5E44&8.72E41&8.73$^{\rm M03}$&1.29E-05\\
3C 120&0.03301&1.8&3.7&1.88E3&3.27E13&1.62E44&1.86E44&1.66E43&8.47E43&2.88E44&2.72E43&8.2$^{\rm B03}$&1.36E-03\\
PKS 0625--35&0.05459&4.9&1.2&1.96E4&1.62E16&6.85E43&5.81E42&1.08E42&1.87E43&2.56E43&3.69E43&9.19$^{\rm B03}$&1.89E-04\\
M 87&0.00428&3&0.1&1.04E4&9.52E13&2.51E41&9.94E40&8.42E42&8.1E40&8.6E42&9.92E40&9.81$^{\rm G09}$&1.22E-07\\
Cen A&0.00183&1.2&4.1&9.09E2&1.87E13&1.81E42&6.7E42&2.84E42&8.72E41&1.04E43&1.92E41&8.38$^{\rm S05}$&6.37E-06\\
Cen b&0.01292&4.8&0.1&3.04E3&1.93E13&2.17E42&5.54E41&3.66E43&1.31E41&3.73E43&7.39E41&&\\
3C 111&0.0485&4.7&0.45&2.01E3&2.96E13&1.25E44&4.4E43&1.37E44&2.26E42&1.84E44&1.12E43&8.8$^{\rm K11}$&1.4E-04\\
3C 207&0.6808&9.8&0.42&3.26E3&7.51E13&3.02E45&4.4E44&8.08E44&1.45E43&1.26E45&1.31E45&8.4$^{\rm S11}$&4.13E-02\\
3C 380&0.692&8&0.9&1.01E3&2.38E13&6.86E45&1.58E45&8.33E44&2.89E43&2.44E45&1.51E45&9.851$^{\rm W02}$&1.69E-03\\
Pictor A&0.03506&2.5&4.2&1.03E3&4.73E13&6.89E43&3.36E43&7.38E42&1.64E43&5.74E43&1.19E43&8.7$^{\rm K11}$&1.88E-04\\
\hline
\scriptsize{NLS1s}\\
\hline
1H 0323+342$^1$&0.0629&2.8$\pm$0.6&3.7$\pm$1.4&883$\pm$361&2.51$\pm$2.31E13&4.83$\pm$3.34E44&1.19$\pm$0.52E44&2.09$\pm$2.00E43&1.82$\pm$1.80E43&1.59$\pm$0.59E44&4.54E44&7.3$^{\rm Z07}$&0.181\\
1H 0323+342$^2$&\nodata&3.6$\pm$1.3&2.1$\pm$1.2&591$\pm$273&8.51$\pm$8.45E12&3.05$\pm$3.00E44&3.56$\pm$2.68E43&3.16$\pm$3.10E43&1.51$\pm$1.50E43&8.24$\pm$4.36E43&4.54E44&7.3&0.181\\
1H 0323+342$^3$&\nodata&4.9$\pm$0.8&2.5$\pm$0.9&383$\pm$160&5.50$\pm$5.06E12&5.17$\pm$4.77E44&3.24$\pm$1.21E43&2.24$\pm$2.20E43&7.59$\pm$7.50E43&1.31$\pm$0.79E44&4.54E44&7.3&0.181\\
1H 0323+342$^4$&\nodata&4.5$\pm$0.6&1.9$\pm$0.6&378$\pm$151&3.63$\pm$3.34E12&4.61$\pm$4.25E44&5.12$\pm$1.52E43&4.68$\pm$4.60E43&3.09$\pm$2.78E43&1.29$\pm$0.56E44&4.54E44&7.3&0.181\\
1H 0323+342$^5$&\nodata&6.2$\pm$0.6&2.5$\pm$0.7&271$\pm$60&3.98$\pm$1.37E12&1.06$\pm$0.36E45&6.61$\pm$1.41E43&3.16$\pm$2.05E43&2.04$\pm$1.41E44&3.02$\pm$1.43E44&4.54E44&7.3&0.181\\
PMN J0948+0022$^1$&0.5846&11$\pm$1.35&4$\pm$1.5&267$\pm$109&6.31$\pm$6.10E12&2.51$\pm$1.45E46&5.03$\pm$1.39E44&1.95$\pm$1.84E44&2.38$\pm$2.08E45&3.08$\pm$2.09E45&9.32E45&8.97$^{\rm V19}$&0.079\\
PMN J0948+0022$^2$&\nodata&10.77$\pm$1.25&5.8$\pm$1.35&260$\pm$80&9.55$\pm$5.49E12&4.92$\pm$2.72E46&5.11$\pm$1.38E44&2.24$\pm$1.70E44&4.47$\pm$2.98E45&5.20$\pm$2.99E45&1.58E46&8.97&0.135\\
PMN J0948+0022$^3$&\nodata&8.6$\pm$1.3&4.6$\pm$1.2&336$\pm$126&9.12$\pm$8.39E12&2.09$\pm$1.44E46&4.94$\pm$1.62E44&2.04$\pm$1.75E44&1.17$\pm$0.92E45&1.87$\pm$0.95E45&1.58E46&8.97&0.135\\
PMN J0948+0022$^4$&\nodata&11.1$\pm$1&5.1$\pm$1.2&202$\pm$69&5.01$\pm$3.69E12&4.17$\pm$2.40E46&4.18$\pm$1.00E44&1.86$\pm$1.37E44&3.98$\pm$2.38E45&4.59$\pm$2.39E45&1.58E46&8.97&0.135\\
PMN J0948+0022$^5$&\nodata&11.6$\pm$0.8&3.7$\pm$0.7&234$\pm$64&5.37$\pm$3.09E12&3.80$\pm$1.31E46&6.54$\pm$1.06E44&2.57$\pm$1.48E44&2.57$\pm$1.24E45&3.48$\pm$1.26E45&1.58E46&8.97&0.135\\
PMN J0948+0022$^6$&\nodata&9.5$\pm$0.5&2.4$\pm$0.4&285$\pm$59&4.37$\pm$1.71E12&1.12$\pm$0.34E46&5.43$\pm$0.70E44&2.57$\pm$1.25E44&4.90$\pm$1.92E44&1.29$\pm$0.24E45&1.58E46&8.97&0.135\\
PMN J0948+0022$^7$&\nodata&13.5$\pm$1.1&1.7$\pm$0.8&186$\pm$63&1.95$\pm$0.90E12&4.79$\pm$2.21E46&2.38$\pm$0.68E45&6.03$\pm$6.00E44&1.00$\pm$0.90E45&3.98$\pm$1.27E45&1.58E46&8.97&0.135\\
PMN J0948+0022$^8$&\nodata&13.7$\pm$1.8&2.1$\pm$1.1&145$\pm$63&1.26$\pm$1.20E12&5.25$\pm$4.11E46&1.59$\pm$0.55E45&5.56$\pm$5.50E44&1.66$\pm$1.60E45&3.80$\pm$1.78E45&9.32E45&8.97&0.079\\
PMN J0948+0022$^9$&\nodata&11.37$\pm$2.2&4.2$\pm$2&350$\pm$147&1.26$\pm$1.16E13&4.27$\pm$3.44E46&9.28$\pm$4.47E44&2.45$\pm$2.40E44&2.99$\pm$2.90E45&4.16$\pm$2.94E45&1.58E46&8.97&0.135\\
PKS 1502+036&0.409&9.5$\pm$0.8&4.7$\pm$0.3&277$\pm$60&8.71$\pm$4.01E12&3.36$\pm$1.16E45&1.09$\pm$0.20E44&2.95$\pm$1.50E43&2.29$\pm$0.84E45&2.43$\pm$0.84E45&6.92E44&8.3$^{\rm C13}$&0.028\\
SBS 0846+513&0.5835&7.4$\pm$0.8&2$\pm$0.6&366$\pm$84&4.68$\pm$1.61E12&2.13$\pm$0.44E45&3.23$\pm$0.73E44&1.05$\pm$0.65E44&1.58$\pm$1.17E44&5.86$\pm$1.52E44&3.22E44&7.86$^{\rm V19}$&0.035\\
PKS 2004--447&0.24&6.4$\pm$0.5&4.1$\pm$0.6&359$\pm$69&1.00$\pm$0.35E13&1.06$\pm$0.25E45&4.03$\pm$0.68E43&1.95$\pm$0.81E43&4.57$\pm$2.00E44&5.17$\pm$2.00E44&3.7E43&6.73$^{\rm O01}$&0.055\\
TXS 2116--077$^H$&0.26&5.9$\pm$0.7&3.7$\pm$0.8&269$\pm$81&8.10$\pm$0.00E12&4.81$\pm$0.00E44&3.08$\pm$0.00E43&3.46$\pm$0.00E43&4.18$\pm$0.00E44&4.83$\pm$0.00E44&2.88E44&7.94$^{\rm V19}$&0.026\\
TXS 2116--077$^L$&\nodata&5.8$\pm$0.7&2.6$\pm$0.5&372$\pm$111&9.11$\pm$0.00E12&2.89$\pm$0.00E44&2.19$\pm$0.00E43&2.44$\pm$0.00E43&1.92$\pm$0.00E44&2.38$\pm$0.00E44&2.88E44&7.94&0.026\\
TXS 0929+533&0.6&&&&&&9.12E44&9.77E43&1.1E44&1.12E45&5.01E45&8.01$^{\rm V19}$&3.9E-01\\
GB6 J0937+5008&0.28&&&&&&3.24E44&2.24E44&8.13E42&5.56E44&5.13E43&7.56$^{\rm P19}$&1.12E-02\\
TXS 0955+326&0.53&&&&&&1.02E45&1.55E44&2.34E44&1.41E45&1.7E46&8.7$^{\rm P19}$&2.7E-01\\
FBQS J1102+2239&0.45&&&&&&3.31E45&3.72E44&6.03E44&4.29E45&1E44&7.78$^{\rm P19}$&1.32E-02\\
PMN J1222+0413&0.97&&&&&&4.79E46&1.32E45&3.63E45&5.28E46&1.51E46&8.85$^{\rm P19}$&1.7E-01\\
SDSS J1246+0238&0.36&&&&&&4.79E44&6.46E43&2.82E44&8.25E44&3.02E44&8.63$^{\rm V19}$&5.63E-03\\
TXS 1419+391&0.49&&&&&&6.76E44&1.74E44&2.24E44&1.07E45&2.19E45&8.63$^{\rm V19}$&4.08E-02\\
TXS 1518+423&0.48&&&&&&5.75E44&1.74E44&2.45E44&9.95E44&5.01E44&7.85$^{\rm P19}$&5.63E-02\\
RGB J1644+263&0.14&&&&&&5.5E43&1.78E43&2.69E43&9.97E43&3.02E44&8.3$^{\rm V19}$&1.2E-02\\
PMN J2118+0013&0.46&&&&&&4.57E44&6.31E43&2.57E44&7.77E44&8.71E44&7.98$^{\rm V19}$&7.26E-02\\
\hline
\scriptsize{FSRQs}\\
\hline
3C 273&0.158&7.41$\pm$0.9&8.5$\pm$1.6&328$\pm$79&2.24$\pm$1.03E13&1.42$\pm$0.41E45&5.85$\pm$1.47E44&1.00$\pm$0.58E44&1.12$\pm$0.67E45&1.81$\pm$0.69E45&8.23E46&9.3&0.328\\
3C 454.3&0.859&15.6$\pm$0.6&5.1$\pm$0.8&137$\pm$32&2.88$\pm$1.33E12&8.14$\pm$2.44E45&3.18$\pm$0.28E45&5.63$\pm$2.60E44&2.81$\pm$0.98E45&6.55$\pm$1.05E45&9.27E46&9.17&0.498\\
PKS 0208--512&1.003&15.8$\pm$0.7&3.42$\pm$1.2&105$\pm$40&9.84$\pm$8.61E11&2.41$\pm$1.39E45&1.72$\pm$0.18E45&4.60$\pm$4.25E44&1.19$\pm$0.82E45&3.38$\pm$0.94E45&2.88E46&9.21&0.141\\
PKS 0420--01&0.916&12.8$\pm$0.7&8.193$\pm$1.1&385$\pm$118&2.75$\pm$1.90E13&4.71$\pm$1.36E45&1.56$\pm$0.18E45&2.69$\pm$1.24E44&3.09$\pm$1.07E45&4.92$\pm$1.09E45&1.51E46&9.03&0.112\\
PKS 0528+134&2.07&17.42$\pm$0.9&2.88$\pm$1.1&230$\pm$86&3.00$\pm$2.42E12&1.39$\pm$0.90E46&8.43$\pm$1.09E45&1.44$\pm$1.36E45&5.44$\pm$3.90E44&1.04$\pm$0.18E46&8.46E46&9.4&0.268\\
B3 0650+453&0.933&14.1$\pm$1&1.325$\pm$0.3&111$\pm$37&4.04$\pm$2.97E11&5.08$\pm$2.81E44&1.52$\pm$0.23E45&4.43$\pm$3.49E44&1.17$\pm$0.63E44&2.08$\pm$0.42E45&1.05E46&8.17&0.566\\
PKS 0727--11&1.589&20.6$\pm$1.2&5.38$\pm$1.05&254$\pm$61&1.00$\pm$0.46E13&1.19$\pm$0.41E46&4.74$\pm$0.52E45&3.16$\pm$1.75E44&5.01$\pm$2.19E45&1.01$\pm$0.23E46&3.92E46&\nodata&\nodata\\
PKS 1127--145&1.184&13.1$\pm$0.8&11.5$\pm$1.2&213$\pm$32&1.15$\pm$0.32E13&1.81$\pm$0.46E46&2.04$\pm$0.28E45&2.66$\pm$0.98E44&5.11$\pm$1.67E45&7.42$\pm$1.70E45&1.41E47&9.18&0.741\\
1Jy 1308+326&0.997&12.6$\pm$0.9&3.39$\pm$0.9&353$\pm$129&8.91$\pm$7.18E12&1.40$\pm$0.64E45&2.50$\pm$0.38E45&3.98$\pm$3.21E44&4.57$\pm$2.74E44&3.36$\pm$0.57E45&1.22E46&8.94&0.112\\
PKS 1508--055&1.185&16.96$\pm$1.1&7$\pm$1.5&141$\pm$41&3.86$\pm$2.22E12&4.63$\pm$2.03E45&8.65$\pm$1.23E44&1.49$\pm$0.93E44&5.30$\pm$2.72E45&6.32$\pm$2.72E45&1.11E47&8.97&0.943\\
PKS 1510--089&0.36&11$\pm$0.5&3.15$\pm$0.5&305$\pm$32&8.91$\pm$1.23E12&5.83$\pm$0.67E44&5.94$\pm$0.68E44&1.02$\pm$0.28E44&5.01$\pm$1.73E44&1.20$\pm$0.19E45&5.92E45&8.65&0.105\\
TXS 1846+322&0.798&13.1$\pm$0.6&2.48$\pm$0.7&206$\pm$72&2.66$\pm$1.96E12&6.00$\pm$3.59E44&6.26$\pm$0.69E44&1.82$\pm$1.42E44&3.63$\pm$2.09E44&1.17$\pm$0.26E45&2.61E46&8.21&1.278\\
PKS 2123--463&1.67&17.9$\pm$0.6&3.6$\pm$0.6&243$\pm$56&5.11$\pm$2.35E12&8.27$\pm$2.86E45&5.17$\pm$0.41E45&5.30$\pm$2.53E44&1.18$\pm$0.41E45&6.88$\pm$0.64E45&4.42E46&\nodata&\nodata\\
TXS 2141+175&0.213&10.34$\pm$0.6&5.1$\pm$0.75&42$\pm$6&2.78$\pm$0.74E11&4.09$\pm$1.03E44&2.39$\pm$0.31E44&2.63$\pm$1.05E44&1.27$\pm$0.48E45&1.77$\pm$0.49E45&1.12E46&8.98&0.094\\
PKS 2144+092&1.113&14.28$\pm$1&3.77$\pm$1.2&175$\pm$66&2.48$\pm$2.28E12&2.69$\pm$1.99E45&1.34$\pm$0.21E45&3.26$\pm$2.82E44&8.62$\pm$5.88E44&2.53$\pm$0.69E45&1.96E46&8.7$^{\rm P17}$&0.312\\
PKS 2204--54&1.215&14.45$\pm$0.9&5.66$\pm$1.2&205$\pm$46&5.71$\pm$2.24E12&4.75$\pm$1.53E45&1.07$\pm$0.15E45&2.42$\pm$1.30E44&1.81$\pm$0.89E45&3.13$\pm$0.91E45&3.27E46&9$^{\rm P17}$&0.26\\
PMN 2345--1555&0.621&13.778$\pm$1&2.555$\pm$0.8&141$\pm$55&1.38$\pm$1.27E12&4.71$\pm$2.71E44&4.15$\pm$0.70E44&1.43$\pm$1.20E44&5.72$\pm$3.90E44&1.13$\pm$0.41E45&2.63E45&8.16&0.145\\
S4 0133+47&0.859&13.1$\pm$1.2&10.48$\pm$1.45&191$\pm$65&8.91$\pm$7.18E12&6.35$\pm$1.83E46&1.46$\pm$0.27E45&1.20$\pm$0.64E44&5.75$\pm$2.65E45&7.34$\pm$2.66E45&7.06E45&8.3&0.281\\
PKS 0227--369&2.115&17.8$\pm$1&5.93$\pm$0.9&331$\pm$102&1.26$\pm$0.87E13&2.03$\pm$0.58E47&5.02$\pm$0.66E45&4.47$\pm$2.06E44&2.29$\pm$0.90E45&7.75$\pm$1.14E45&\nodata&\nodata&\nodata\\
4C 28.07&1.213&14.6$\pm$1.1&6.88$\pm$1&223$\pm$51&8.13$\pm$3.74E12&6.87$\pm$2.69E46&1.46$\pm$0.23E45&2.04$\pm$1.08E44&2.75$\pm$1.14E45&4.41$\pm$1.17E45&5.91E46&9.22&0.283\\
PKS 0347--211&2.944&26.2$\pm$1.5&10$\pm$1.5&222$\pm$68&1.12$\pm$0.78E13&5.61$\pm$1.94E47&4.92$\pm$0.64E45&1.95$\pm$1.03E44&1.91$\pm$0.70E46&2.42$\pm$0.70E46&7.08E46&9.78$^{\rm P17}$&0.093\\
PKS 0454--234&1.003&19.98$\pm$1.9&6.6$\pm$0.8&184$\pm$56&7.94$\pm$5.49E12&8.72$\pm$2.21E46&2.14$\pm$0.41E45&1.32$\pm$0.67E44&1.15$\pm$0.48E46&1.37$\pm$0.48E46&6.59E45&9.17&0.035\\
S4 0917+44&2.19&18.2$\pm$1.3&9.82$\pm$1.8&213$\pm$66&9.12$\pm$6.30E12&2.92$\pm$1.01E47&4.41$\pm$0.66E45&3.47$\pm$2.00E44&7.08$\pm$2.93E45&1.18$\pm$0.30E46&1.81E47&9.29&0.739\\
4C 29.45&0.729&11.6$\pm$1&8.25$\pm$1.6&400$\pm$112&3.16$\pm$1.82E13&3.85$\pm$1.51E46&7.43$\pm$1.40E44&1.10$\pm$0.68E44&2.57$\pm$1.36E45&3.42$\pm$1.37E45&1.25E46&8.61&0.245\\
3C 279&0.536&12$\pm$0.5&5.9$\pm$0.35&219$\pm$37&8.13$\pm$2.81E12&2.74$\pm$0.28E46&1.10$\pm$0.09E45&1.91$\pm$0.48E44&2.00$\pm$0.41E45&3.29$\pm$0.43E45&8.26E45&8.28&0.345\\
PKS 1454--354&1.424&20.2$\pm$1.8&7.5$\pm$1.3&276$\pm$99&1.48$\pm$1.36E13&2.24$\pm$0.57E47&5.07$\pm$1.02E45&3.63$\pm$2.01E44&1.00$\pm$0.51E46&1.54$\pm$0.52E46&3.98E46&9.3$^{\rm P17}$&0.159\\
PKS 1502+106&1.839&27$\pm$2.3&7.14$\pm$1.5&192$\pm$66&8.91$\pm$6.57E12&3.96$\pm$1.28E47&7.86$\pm$1.51E45&2.45$\pm$1.47E44&2.29$\pm$1.16E46&3.10$\pm$1.17E46&4.91E46&8.98&0.409\\
B2 1520+31&1.487&20.8$\pm$1.6&4.3$\pm$0.9&283$\pm$93&1.00$\pm$0.69E13&5.34$\pm$1.54E46&2.54$\pm$0.49E45&2.04$\pm$1.22E44&3.47$\pm$1.84E45&6.22$\pm$1.90E45&2.16E46&8.92&0.207\\
4C 66.20 &0.657&12.2$\pm$1.2&7.16$\pm$1.4&240$\pm$95&8.91$\pm$8.85E12&2.60$\pm$0.90E46&9.85$\pm$1.97E44&1.17$\pm$0.73E44&2.57$\pm$1.42E45&3.67$\pm$1.44E45&6.68E45&9.14&0.039\\
PKS 2325+093&1.843&17.6$\pm$1.6&15.1$\pm$1.6&354$\pm$107&3.98$\pm$2.75E13&1.26$\pm$0.36E48&5.64$\pm$1.11E45&3.09$\pm$1.49E44&1.70$\pm$0.74E46&2.29$\pm$0.75E46&4.46E46&8.7&0.709\\
\enddata
\tablenotetext{a}{The superscripts denote the different SEDs in the literatures of Zhang et al. (2012), Sun et al. (2015), and Yang et al. (2018). BL Lacs that are not included in Zhang et al. (2012) are taken from Ghisellini et al. (2011). FSRQs are from Zhang et al. (2014, 2015). RGs are from Xue et al. (2017). NLS1s that are not included in Sun et al. (2015) and Yang et al. (2018) are taken from Paliya et al. (2019).}
\tablenotetext{b}{Except for five BL Lacs (PKS 2155--304, 1ES 1959+650, PG 1553+113, 1ES 1011+496, Mkn 180), $L_{\rm disk}$ are taken from Paliya et al. (2017), $L_{\rm disk}$ of other BL Lacs is taken from Zhang et al. (2015), which is estimated with the luminosity of BLR. $L_{\rm disk}$ of two FSRQs (PKS 0347--211 and PKS 1454--354) are taken from Paliya et al. (2017) while others are from Zhang et al. (2015). For NLS1s, $L_{\rm disk}$ is from Sun et al. (2015) and Yang et al. (2018) while $L_{\rm disk}$ of that are not included in the two literatures are from Paliya et al. (2019). $L_{\rm disk}$ of RGs is taken the luminosity at $10^{15}$ Hz of the model-fitting lines in Xue et al. (2017).}
\tablenotetext{c}{ The superscripts denote the references--P17: Paliya et al. (2017), P19: Paliya et al. (2019), Z07: Zhou et al. (2007), C13: Calderone et al. (2013), O01: Oshlack et al. (2001), V19: Viswanath et al. (2019), R05: Rinn et al. (2005), B03: Bettoni et al. (2003), M03: Merloni et al. (2003), G09: Gebhardt et al. (2009), S05: Silge et al. (2005), K11: Kataoka et al. (2011), S11: Shen et al. (2011), W02: Woo \& Urry (2002).  For BL Lacs, except for five sources, $M_{\rm BH}$ of others are from  Ghisellini et al. (2011) and Sbarrato et al. (2012). For FSRQs, except for four sources, $M_{\rm BH}$ of others are taken from  Zhang et al. (2015 and reference therein). }
\end{deluxetable}

\end{document}